\documentclass{jfm-nobanner} 

\usepackage{graphicx}
\usepackage{newtxtext}
\usepackage{newtxmath}
\usepackage{natbib}
\usepackage{hyperref}
\hypersetup{
    colorlinks = true,
    urlcolor   = blue,
    citecolor  = black,
}

\newcommand{\RomanNumeralCaps}[1]
\linenumbers

\usepackage[utf8]{inputenc}
\usepackage{textcomp}

\usepackage[final]{changes} 

\usepackage{graphicx}
\usepackage{epstopdf, epsfig}
\usepackage{amsmath}  
\usepackage{amssymb}
\usepackage{lscape}
\usepackage{mathtools}
\usepackage{pgfplots} 
\pgfplotsset{compat=newest}
\pgfplotsset{plot coordinates/math parser=false}
\newlength\figureheight
\newlength\figurewidth 
\usepackage{tikz}
\usetikzlibrary{decorations.pathreplacing,angles,quotes}
\usepackage{xifthen}
\usetikzlibrary{arrows}
\newcommand{\defeq}{\vcentcolon=}

\DeclareMathOperator\erf{erf}

\title{Non-iterative vortex-based smearing correction for the actuator line method}

\author{{Vitor G. Kleine\aff{1,2}
  \corresp{\email{vitok@mech.kth.se}}},
  Ardeshir Hanifi\aff{1}
 \and Dan S. Henningson\aff{1}}

\affiliation{\aff{1}KTH Royal Institute of Technology, FLOW, SE-10044, Stockholm, Sweden
\aff{2}Instituto Tecnol\'{o}gico de Aeron\'{a}utica, Pra\c{c}a Marechal Eduardo Gomes, 50, Vila das Ac\'{a}cias, 12228-900, S\~{a}o Jos\'{e} dos Campos - SP, Brazil}

\begin{document}
\maketitle

\begin{abstract}
The actuator line method (ALM) is extensively used in wind turbine and rotor simulations. However, its original uncorrected formulation overestimates the forces near the tip of the blades and does not reproduce well forces on translating wings. The recently proposed vortex-based smearing correction for the ALM is a correction based on physical and mathematical properties of the simulation that allows for a more accurate and general ALM. So far, to correct the forces on the blades, the smearing correction depended on an iterative process at every time step, which is usually slower, less stable and less deterministic than direct methods. In this work, a non-iterative process is proposed and validated. First, we propose a formulation of the non-linear lifting line that is equivalent to the ALM with smearing correction, showing that their results are practically identical for a translating wing. Then, by linearizing the lifting line method, the iterative process of the correction is substituted by the direct solution of a small linear system. No significant difference is observed in the results of the iterative and non-iterative corrections, both in wing and rotor simulations. Additional contributions of the present work include the use of a more accurate approximation for the velocity induced by a smeared vortex segment and the implementation of a free-vortex wake model to define the vortex sheet, that contribute to the accuracy and generality of the method. The results present here may motivate the adoption of the ALM by other communities, for example, in fixed-wing applications.
\end{abstract}


\section{Introduction}
The actuator line method (ALM) was developed to represent blades or wings as lifting lines in numerical simulations of the Navier-Stokes equations \citep{sorensen2002numerical}. It allows the reduction of computational costs by replacing the geometry of the blades by lines carrying body forces calculated using the local velocity and airfoil data \citep{sorensen2015simulation}.

In order to accurately calculate the forces, the local velocity needs to be corrected near the tip of the blades. This correction has been a mildly controversial issue in the past, with different proposed models \citep{shen2005tip,sorensen2016general} and even arguments that it should not be needed \citep{martinez2012comparison,sorensen2016general} due to the fact that the ALM creates tip vortices that should lead to lower loads near the tip. However, the overestimation of loads near the tip on the numerical results indicates that a correction is needed \citep{sorensen2016general,dag2017combined}.

Some recent discoveries have shed some light on this apparent controversy. \citet{dag2020new} (originally in \citep{dag2017combined}) observed that the bound vortex created by the actuator line showed a Gaussian vorticity distribution equals to a Lamb-Ossen vortex model \citep{lamb1932hydrodynamics,oseen1911uber,saffman1992vortex}. The authors then conjectured that this pattern would be extended to the vortex sheet and proposed a correction based on this model to approximate the velocity in the ALM to the velocity induced by the singular vorticity distribution predicted by a discretized Prandtl's lifting line. The simulations of rotors showed clear improvements compared to the uncorrected ALM. The method brings the forces smoothly to zero near the tip and the hub of the blades. The results for a translating planar wing show that the results approximate a lifting line method for this case. \cite{meyer2019vortex} compared the results of a similar correction with the results of a lifting line technique for rotating blades with and without a viscous core model. They showed that the forces of the ALM without correction agree with a vortex method with finite core size, while the ALM with a vortex-based smearing correction agrees with the vortex method using ideal vortices.

The mathematical connection between the vortices generated by the ALM and a Lamb-Ossen vortex model has been proven by \cite{forsythe2015coupled} for the bound vortex and \cite{martinez2019filtered} (originally in \citep{martinez2017large}) for the vortex sheet of a translating wing. They showed that it is a consequence of the convolution of the discrete forces with a Gaussian kernel, necessary to distribute the forces and avoid numerical instabilities. The correction of \citep{martinez2019filtered}, termed ``subfilter-scale velocity correction'' or ``filtered actuator line model'', is, in essence, a variant of a vortex-based smearing correction. It was applied to a wind turbine by \citep{stanly2022large}.

\cite{caprace2019lifting} modified Prandtl's lifting line by distributing the vorticity by using a Gaussian kernel. Even though this work does not concern directly the ALM, the connections between the mollified (smeared) lifting line and the ALM were clearly stated by the authors. They studied the effect of three-dimensional, two-dimensional (normal to the line) and one-dimensional (normal to the line and streawise direction) regularizations. Both the three-dimensional and two-dimensional regularizations are used in the ALM~\citep{mikkelsen2003actuator}. In that work, a variable smoothing parameter approaching zero or a very low finite value near the tip leads to improved results, a result also observed in ALM works \citep{shives2013mesh,jha2014guidelines,jha2018actuator}. However, in the ALM, a low value of smoothing parameter $\varepsilon$ may lead to numerical instabilities. Since the minimum $\varepsilon$ is usually around 2 to 3 times the grid spacing \citep{troldborg2009actuator}, a variable smoothing parameter may impose a very refined grid near the tips.

Other methods may allow a more compact representation of the effect of the forces. For example, one strategy in finite volume methods to represent an actuator disk or surface is to model the effects of the forces as pressure jumps, allowing it to be restricted to one grid element~\citep{rethore2012discrete,troldborg2015consistent}. Also, for vortex particle methods, which sheds vortex elements in the wake, the smoothing parameter may be equal to the grid spacing~\citep{caprace2020immersed}. Nevertheless, in numerical methods, the compactness of the representation of the effects of the forces is limited by the grid spacing and is guided by numerical reasons, not by the physical properties of the flow.

Depending on the numerical method or application, a larger $\varepsilon$ may be necessary or desirable in the ALM. For example, \citet{kleusberg2019parametric}, using a spectral element method, observed spatially growing oscillations for $\varepsilon=2 \Delta x$ (where $\Delta x$ is the average grid spacing), while the amplitude of oscillations was bounded for $3 \Delta x \leq \varepsilon \leq 4 \Delta x$. The oscillations for $\varepsilon=2 \Delta x$ were not sufficient to cause numerical instabilities on force calculations, however, they may affect applications that require low numerical disturbances in the wake. For this reason, \citet{kleine2022stability} used $\varepsilon=3.5 \Delta x$ for a vortex stability study. Also, \citet{shives2013mesh} recommended a smearing parameter around $\varepsilon=4 \Delta x$ if lower errors in the angle of attack are desirable (see also~\citep{forsting2020generalised}).

For this reason, corrections that take into account the difference between the vorticity created by the convoluted forces and the vorticity created by discrete singular forces, such as proposed by \citep{dag2020new}, \citep{martinez2019filtered} and \citep{meyer2019vortex}, seem to be more cost-effective than reducing $\varepsilon$. Some techniques to further reduce the computational cost of the smearing correction are explored by \cite{meyer2020brief}.

\citet{meyer2019wake} investigated the wake created by the actuator line with smearing correction, confirming that the smearing parameter continues to have an influence on the wake, despite its influence on the forces at the blades being greatly reduced.

The corrections of~\citep{dag2020new} and~\citep{meyer2019vortex} apply an iterative process at each time step. The circulation is dependent on the local induced velocity, which depends on the circulation. For this reason, at each iteration, the velocity or circulation is updated using a relaxation iterative process. From our experience, the choice of a relaxation factor close to unity can lead to numerical instability while a low relaxation factor increases the number of iterations and can increase the run time of the simulation.

The correction of~\citep{martinez2019filtered} avoids an iterative procedure at each time step by using the circulation of the previous time step to calculate the induced velocity. Then, a weighted average of the correction velocity of the previous time step and the correction velocity of the current time step is used as the correction term. Hence, the induced velocities are calculated from the values of circulation of the previous two time steps, without taking into consideration the current value of circulation. Conceptually it is equivalent to a first iteration of the iterative method of other works. For steady simulations, the values converge to the steady solution. However, for unsteady problems, this procedure does not guarantee compatibility between the current circulation and local velocity at each time step. This approach can be justified due to the small difference between the circulations between time steps, for most simulations. Nevertheless, an error is introduced, which is dependent on the difference of the circulation between time steps and the weighting factor (in that work, called ``relaxation factor'').

In this work, a method is introduced that avoids the iterative method while maintaining the compatibility between the current circulation and velocities. We propose a direct way of computing the smearing correction, based on a linearized version of the lifting line. Using this method, the correction at each time step is found by solving a linear system of equations of the order $N$, where $N$ is the number of actuator line control points. In order to achieve this, two formulations of the discretized lifting line method based on the actuator line are presented: the non-linear formulation and its linearized version.

Additionally, some other aspects of the smearing correction are discussed, such as the choice of correction function and formation of the vortex sheet. Regarding these aspects, we aim to keep the method as general as possible. Previous works~\citep{dag2020new,martinez2019filtered,meyer2019vortex} already showed that the smearing correction can make the forces calculated by the ALM closely match the results of the lifting line method. By introducing fewer assumptions, avoiding especially assumptions related to rotating blades, the method could be applied to other problems beyond rotor simulations, such as simulations of fixed-wing aircrafts.

This work is structured as follows. First, we present the general idea of the actuator line method with smearing correction in Section \ref{sec:ACL}. Then, in Section \ref{sec:vorticity}, we draw on the theoretical work of \citep{martinez2019filtered} to develop the specific correction for velocities induced by vortex segments and describe the formation of the vortex sheet. The iterative lifting line method based on the actuator line is presented in Section \ref{sec:NLLL}. The linearization of this lifting line method (Appendix \ref{sec:LLL}) is the basis for the non-iterative smearing correction detailed in Section \ref{sec:nonitesmearcorr}. Results of the simulations of a translating wing and the NREL 5-MW wind turbine are shown in Section \ref{sec:results}. Finally, the main conclusions are summarized in Section \ref{sec:conclusions}.
\section{The actuator line method} \label{sec:ACL}

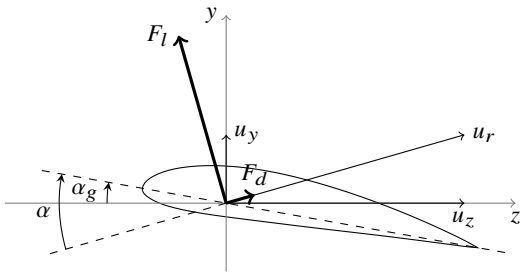
\begin{figure}
  \centering
  \begin{tikzpicture}
  \def\c{4.5}
  \def\r{0.35*\c}
  \draw[scale=\c,rotate=-10] plot file{Figures/naca4415.dat} -- cycle;
  \draw[rotate=-10,dashed] (-0.55*\c,0)--(0.85*\c,0);
  \draw[->,gray] (-0.65*\c,0)--(0.85*\c,0) node[below,black]{$z$};
  \draw[->,gray] (0,-0.2*\c)--(0,0.55*\c) node[left,black]{$y$};
  \draw[->] (0,0)--(0.7*\c,0) node[below]{$u_z$};
  \draw[->] (0,0)--(0,0.2*\c) node[right]{$u_y$};
  \draw[->] (0,0)--(0.7*\c,0.2*\c) node[right]{$u_{r}$};
  \draw[dashed] (0,0)--(-0.7*0.75*\c,-0.2*0.75*\c);
  \draw[->,very thick] (0,0)--(0.7*0.12*\c,0.2*0.12*\c) node[above]{$F_{d}$};
  \draw[->,very thick] (0,0)--(-0.2*0.7*\c,0.7*0.7*\c) node[left]{$F_{l}$};
  \draw[-stealth] (0,0) +(180:\r) arc(180:170:\r) node[left,midway]{$\alpha_g$};
  \draw[-stealth] (0,0) +(195.945:1.4*\r) arc(195.945:170:1.4*\r) node[left,midway]{$\alpha$};
\end{tikzpicture}
  \caption{Local reference system defined by a cross-section of the blade.}
  \label{fig:2dref}
\end{figure}

The incompressible Navier-Stokes equation written in primitive variables (pressure $p$ and velocity $\mathbf{u}$) are:
\begin{equation}
  \frac{\partial \mathbf{u}}{\partial t} + \mathbf{u} \cdot \nabla \mathbf{u} = -\frac{1}{\rho} \nabla p + \nu \nabla^2 \mathbf{u} + \mathbf{f}
\end{equation}
where $\rho$ and $\nu$ are the density and kinematic viscosity of the fluid, respectively. The body force term, $\mathbf{f}$, in the case of the actuator line method, is used to model the turbine \citep{sorensen2002numerical,mikkelsen2003actuator,troldborg2009actuator}. The body forces are based on the two-dimensional force per spanwise unit length, $\mathbf{F}_{2D}$, given by
\begin{equation}
  \mathbf{F}_{2D} = (F_l,F_d) = \left( \frac{1}{2}\rho \, u_r^2 \, c \, C_l,\frac{1}{2}\rho \, u_r^2 \, c \, C_d \right)
\end{equation}
where $F_l$ and $F_d$ are the lift and drag forces (lift is perpendicular to the relative velocity and drag is parallel to the relative velocity, see figure \ref{fig:2dref}), calculated from the relative velocity $u_r=\sqrt{u_y^2+u_z^2}$, the local chord $c$ and the two-dimensional lift and drag coefficients, $C_l$ and $C_d$, which are obtained from the airfoil data at the local Reynolds number and angle of attack $\alpha$ (calculated using the local relative velocity).

The dimensions of $\mathbf{F}_{2D}$ are not consistent with the dimensions of the body force $\mathbf{f}$. As an intermediate step, an idealized three-dimensional body force $\mathbf{f}^i$ is defined based on $\mathbf{F}_{2D}$, where at each cross-section the body force is given by
\begin{equation}
  \label{eq:f2d}
  \mathbf{f}^i = \frac{1}{\rho} \mathbf{F}_{2D} \delta(y) \delta(z) ,
\end{equation}
where $\delta$ is the Dirac delta function, which can be interpreted as the limit of the Gaussian function
\begin{equation}
  \delta(y) = \lim_{\varepsilon \to 0} \frac{1}{\pi^{1/2}\varepsilon} e^{-\frac{y^2}{\varepsilon^2}} .
\end{equation}
This body force $\mathbf{f}^i$, where the superscript $^i$ indicates the idealized case, is concentrated at the origin of the local reference system (figure \ref{fig:2dref}). Considering all cross-sections of a wing, the space of non-zero $\mathbf{f}^i$ defines a line, which can be interpreted as the limit of the actuator line method, when the smearing parameter $\varepsilon$ goes to zero. It is easy to see that $\mathbf{f}^i$ is singular at the origin but its integration in the two-dimensional plane gives $\mathbf{F}_{2D}/\rho$.

In the ALM, to avoid numerical problems related to singularities, the forces need to be distributed smoothly on several mesh points. This is usually performed by convolving the force with a regularization kernel. A common choice for the regularization kernel is a three-dimensional Gaussian kernel.  Considering the same constant Gaussian width in the three directions:
\begin{equation}
  \label{eq:eta3d}
  \eta_3(x,y,z) \defeq \frac{1}{\pi^{3/2}\varepsilon^3} \exp{\left(-\frac{x^2+y^2+z^2}{\varepsilon^2}\right)}=\eta(x)\eta(y)\eta(z)
\end{equation}
where
\begin{equation}
  \eta(x) \defeq \frac{1}{\pi^{1/2}\varepsilon} \exp{\left(-\frac{x^2}{\varepsilon^2}\right)}
\end{equation}
and symbol $\defeq$ denoting equal to by definition. The smeared force is
\begin{equation}
  \mathbf{f} = \mathbf{f}^i * \eta_3 .
  \label{eq:fconvoluted}
\end{equation}
Non-uniform and anisotropic kernels have also been proposed \citep{mikkelsen2003actuator,shives2013mesh,martinez2017optimal,churchfield2017advanced,jha2018actuator,cormier2021evaluation}, but these are not investigated in the present work.

This presentation of the convolution operation in the actuator line method using an idealized body force formed by Dirac delta functions as an intermediate step may seem like an unusual approach, but is mathematically equivalent to the method of classical works~\citep{sorensen2002numerical,mikkelsen2003actuator} (save for possible typos regarding the density term, which is not relevant for incompressible flows), which employ one-dimensional integrals. This procedure, which was already applied by \cite{martinez2019filtered}, formalizes the connection between the two-dimensional forces and the convolution in three dimensions, which is relevant for Section \ref{sec:vorticity}.

Similarly to previous studies \citep{martinez2019filtered,dag2020new,meyer2019vortex}, we focus on the lift force, which has a greater influence on the forces of the turbine, and leave the effects related to the smearing of the drag force for future studies (a two-dimensional correction for drag has been proposed by~\citep{martinez2017optimal} and an outline for a three-dimensional drag force correction can be found in~\citep{kleine2022onstability}).

\subsection{Lift force in the basic formulation of the actuator line method} \label{sec:basicACL}

In the general formulation of the actuator line \citep{sorensen2002numerical}, the lift force $F_l$ at each spanwise section $j$ of the wing (force per spanwise length) is obtained from airfoil data from:
\begin{equation}
  F_{lj} = \frac{1}{2}\rho u_r^2 \, c_j \, C_l(\alpha_j)
  \label{eq:NLlift}
\end{equation}
where $c_j$ is the local chord of the airfoil and $C_l(\alpha_j)$ is the local lift coefficient, obtained from airfoil data at an angle of attack $\alpha_j$. The local reference system of figure \ref{fig:2dref} is used here. The velocities $u_z$ and $u_y$, relative to the actuator line, are obtained from the CFD (computational fluid dynamics) simulation at the control point position, $\mathbf{x}_j$ of the actuator line segment. The lift coefficient $C_l(\alpha_j)$ is calculated by the interpolating tabulated airfoil coefficients using the effective local angle of attack
\begin{equation}
  \alpha_j=\alpha_{gj} + \arctan{\left(\frac{u_y}{u_z}\right)}
  \label{eq:effalpha}
\end{equation}
where $\alpha_{gj}$ is the geometric angle of attack given by the twist and incidence of the wing (or, in the case of rotating blades, the opposite of the local pitch angle \citep{troldborg2009actuator}).

If needed, the circulation at each spanwise section can be calculated from the Kutta–Joukowski theorem
\begin{equation}
  \Gamma_j = \frac{F_{lj}}{\rho u_r} = \frac{1}{2} u_r \, c_j \, C_l(\alpha_j) .
  \label{eq:GammaNLL}
\end{equation}
Usually, some form of tip or smearing correction is applied to this basic formulation. The smearing corrections employed in the present work are described in Sections \ref{sec:itesmearcorr} and \ref{sec:nonitesmearcorr}.

\subsection{Vortex-based smearing correction} \label{sec:vortexsmearcorr}

In the actuator line method with smearing correction, the velocity $\mathbf{u}^s$, sampled from the CFD simulation at control point $\mathbf{x}_j$, is summed to the ``missing velocity'' $\mathbf{u}^m$ to arrive at the corrected velocity $\mathbf{u}^c$:
\begin{equation}
  \mathbf{u}^{c}(\mathbf{x}_j) = \mathbf{u}^{s}(\mathbf{x}_j) + \mathbf{u}^{m}(\mathbf{x}_j) .
  \label{eq:ucsmear}
\end{equation}
The ``missing velocity'' is defined as the difference between the velocities induced by the vortices created by the actuator line and ``reference'' vortices. The concept of what is considered a ``reference'' vortex varies between the works. \citet{dag2020new} and \citet{meyer2019vortex} consider a vortex filament with an infinitesimal core, while \citet{martinez2019filtered} considers the vortex created using the optimum smearing parameter developed in \citep{martinez2017optimal}. A viscous core model with the radius of the vortex core evolving in time could also be easily implemented. If there is the possibility of the vortices impinging the actuator lines, it might be useful to employ some desingularization method.

In the present work, we adopt the definition of reference vortices as vortex filaments with an infinitesimal core. In this method, the missing velocity is calculated from the difference of velocities induced by ideal, singular, vortices (superscript $^{vi}$) and vortices with finite core created by the actuator lines (superscript $^v$)
\begin{equation}
  \mathbf{u}^{m}(\mathbf{x}_j) = \mathbf{u}^{vi}(\mathbf{x}_j) - \mathbf{u}^{v}(\mathbf{x}_j) .
  \label{eq:um}
\end{equation}
Using this definition of missing velocity, the aim of the method is to reproduce the results of a lifting line method. The idea behind the vortex-based smearing correction is that the velocity sampled from the numerical simulation, in a linear approximation, is given by the sum of the local undisturbed velocity, $\mathbf{U}$, and the velocity induced by the vortices created by the actuator line
\begin{equation}
  \mathbf{u}^{s}(\mathbf{x}_j) \approx \mathbf{U}(\mathbf{x}_j) + \mathbf{u}^{v}(\mathbf{x}_j) .
  \label{eq:us}
\end{equation}
From equation~\eqref{eq:ucsmear}, the corrected velocity becomes
\begin{equation}
  \mathbf{u}^{c}(\mathbf{x}_j) \approx \mathbf{U}(\mathbf{x}_j) + \mathbf{u}^{vi}(\mathbf{x}_j) ,
  \label{eq:ucmodel}
\end{equation}
which reproduces the results of a lifting line method. It should be noted that the local undisturbed velocity $\mathbf{U}$ is not known in ALM simulations (except for simple cases, used mostly as validation). Only the velocity sampled from the simulations $\mathbf{u}^{s}$ is known. From this, comes the need to model the velocity induced by the vorticity created by the actuator lines, $\mathbf{u}^{v}$, described in section~\ref{sec:vorticity}.

On top of the results from the vortex-based smearing correction, other corrections can be introduced, based on the known limitations of the lifting line method. For example, \citet{dag2017combined} combined the smearing correction with the decambering correction of~\citep{sorensen2016refined}. Limitations and further corrections of the lifting line method are out of the scope of the present work, and we consider it as the reference for validation of the vortex-based smearing correction.

\subsection{Iterative smearing correction} \label{sec:itesmearcorr}

The missing velocity is calculated from the circulation of the bound and wake vortices. At the same time, the circulation is calculated from the corrected velocity, using equation~\eqref{eq:GammaNLL}. For this reason, an iterative procedure was used by~\citet{dag2020new} and \citet{meyer2019vortex}. The approach of \citep{meyer2019vortex} applies the relaxation factor to the velocity. In this work, we apply the relaxation factor to the circulation, because we solve for the circulation in Section \ref{sec:nonitesmearcorr}.

The steps of the iterative smearing correction are, for each time-step:
\begin{enumerate}
  \item Start from a guess of the circulation distribution, usually the circulation from the previous time-step;
  \item \label{item:formvortexsm} Form the vortex sheet by:
  \begin{itemize}
      \item prescribing the vortex sheet, for example, by assuming helical or horseshoe vortices; or
      \item employing a free-vortex wake method, for example, by advecting the vortices with the CFD velocity or by a combination of the CFD velocities and the velocities induced by the free vortices.
  \end{itemize}
  \item Calculate the missing velocity $\mathbf{u}^m(\mathbf{x}_j)$ at every control point. Find the local corrected velocities according to equation \eqref{eq:ucsmear}; \label{item:velocitysm}
  \item Calculate the effective angle of attack using equation \eqref{eq:effalpha};
  \item Find the local lift coefficient $C_l$, interpolating from the airfoil data table;
  \item Calculate the new value of circulation $\Gamma_j^{new}$ using equation \eqref{eq:GammaNLL};
  \item Update the current value of circulation using a relaxation factor $r$:
  \begin{equation}
    \Gamma_j=r\Gamma_j^{new} + (1-r) \Gamma_j^{old}
  \end{equation}
  \item Restart from step (\ref{item:formvortexsm}) if the local velocity affects the formation of the vortex sheet or from step (\ref{item:velocitysm}) otherwise, using the value of circulation calculated in the previous step. Iterate until a chosen convergence criterion is reached.
\end{enumerate}

The forces are calculated after convergence. There is an ambiguity regarding the choice of velocity used to calculate the forces. This ambiguity is resolved for the lift force in~\citep{kleine2022simulating}. For the present simulations, the difference is negligible, because the error is of second order (see \citep{kleine2022simulating} for further discussion on this topic). In the present work, all forces are calculated using the corrected velocities.

\section{Vorticity created by body forces and the missing velocity} \label{sec:vorticity}

\subsection{Vorticity generated by body forces} \label{sec:vorticitybody}

Some of the development of \citep{martinez2019filtered} is reproduced in this section for completeness. Neglecting viscous effects in a steady flow, the vorticity equation in steady flow becomes
\begin{equation}
  \mathbf{u} \cdot \nabla \boldsymbol{\omega} = \boldsymbol{\omega} \cdot \nabla \mathbf{u} + \nabla \times \mathbf{f} .
  \label{eq:vorticityeq}
\end{equation}
Equation \eqref{eq:vorticityeq} can be linearized considering an uniform baseflow $\mathbf{U} = const.$, arriving at
\begin{equation}
  \mathbf{U} \cdot \nabla \boldsymbol{\omega} =  \nabla \times \mathbf{f} .
  \label{eq:vorticitylineareq}
\end{equation}
Considering uniform flow aligned with the $z$-axis, $\mathbf{U} = (0,0,U_z)$, for a straight wing with only lift forces, that can be considered to act aligned with the $y$-axis, $\mathbf{f}=(0,f_y,0)$:
\begin{equation}
  \label{eq:vortxsteady}
  U_z \frac{\partial\omega_x}{\partial z} = -\frac{\partial f_y}{\partial z} \implies \omega_x=-\frac{f_y}{U_z}
\end{equation}
\begin{equation}
  \label{eq:vortzsteady}
  U_z \frac{\partial\omega_z}{\partial z} = \frac{\partial f_y}{\partial x} \implies \omega_z=\int_{-\infty}^{z} \frac{1}{U_z} \frac{\partial f_y}{\partial x} dz
\end{equation}
while $\omega_x,\omega_z \gg \omega_y \approx 0$ (for a more detailed derivation, the reader is referred to~\citep{kleine2022onstability}).

The bound vorticity $\omega_x^i$ generated by an ideal concentrated force $f_y^i = - F_l(x)/\rho \delta(y)\delta(z)$ is
\begin{equation}
  \label{eq:vortxideal}
  \omega_x^i = -\frac{f_y^i}{U_z} = \frac{1}{U_z \rho} F_l(x) \delta(y)\delta(z) ,
\end{equation}
where it is relevant to note that a positive lift force $F_l$ applies a negative force to the fluid. The integral of the vorticity in the spanwise direction is equal to the circulation, $\Gamma$, at the section. For the case with concentrated force
\begin{equation}
  \label{eq:circulation}
  \Gamma (x) = \int_{-\infty}^{+\infty} \int_{-\infty}^{+\infty} \omega_x^i dy dz = \frac{1}{U_z \rho} F_l(x) \delta(y)\delta(z) = \frac{1}{U_z \rho} F_l(x) ,
\end{equation}
which agrees with the Kutta-Joukowski theorem.

\subsection{Vorticity generated by a discontinuous distribution of force}
\label{sec:vorticityshed}

In numerical methods, usually, the force is discretized and represented by interpolating functions inside segments. The theory for the semi-infinite wing of~\citep{martinez2019filtered} can be used to construct a discretized method. In the current implementation of the actuator line method, for each segment $j$, the force is calculated at control point $x_{j}$ and considered constant for $x$-positions between the boundaries of the segment, $x_{j_-} \le x \le x_{j_+}$. The two-dimensional force can be defined based on the 
\begin{equation}
  \mathbf{F}_{2D}(x)=\sum_{j=1}^{N}(H(x-x_{j_-})-H(x-x_{j_+})) \mathbf{F}_{2D}(x_{j}) .
\end{equation}
where $H(z)$ is the Heaviside step function with the half-maximum convention. From equation~\eqref{eq:f2d}, the ideal body force is then given by
\begin{equation}
  \mathbf{f}^i(x,y,z) =  \sum_{j=1}^{N}(H(x-x_{j_-})-H(x-x_{j_+})) \delta(y)\delta(z) \frac{\mathbf{F}_{2D}(x_{j})}{\rho} .
\end{equation}
which, in most cases, will be discontinuous at the boundary of the segments (the average is considered at the boundary of the segments). The convolution with a Gaussian function (equation~\eqref{eq:fconvoluted}) transforms the body forces to
\begin{equation}
  \mathbf{f}(x,y,z) =   \sum_{j=1}^{N}(H_{\varepsilon}(x-x_{j_-})-H_{\varepsilon}(x-x_{j_+})) \eta(y)\eta(z) \frac{\mathbf{F}_{2D}(x_{j})}{\rho} ,
  \label{eq:fcode}
\end{equation}
where the function
\begin{equation}
  H_{\varepsilon}(x)=\frac{\erf(\frac{x}{\varepsilon})+1}{2}
\end{equation}
is defined as a smeared (mollified) Heaviside step function \citep{caprace2019lifting}.

Equation~\eqref{eq:fcode} is employed directly in the current simulations. As far as the authors are aware, this may be the first implementation of the analytical convolution in an ALM code. However, it should be noted that most works do not describe how the convolution operation is performed in practice. We are aware, however, that many implementations, including the previous version of our code~\citep{kleusberg2019wind}, perform the convolution operation numerically.

Considering only lift force, the bound vortex is (from equation \eqref{eq:vortxsteady})
\begin{equation}
  \label{eq:vortxidealheav}
  \begin{split}
    \omega_x^i= \sum_{j=1}^{N}(H(x-x_{j_-})-H(x-x_{j_+})) \delta(y)\delta(z) \Gamma_j 
  \end{split}
\end{equation}
\begin{equation}
  \label{eq:vortxheav}
  \begin{split}
    \omega_x = \sum_{j=1}^{N} (H_{\varepsilon}(x-x_{j_-})-H_{\varepsilon}(x-x_{j_+})) \eta(y) \eta(z) \Gamma_j ,
  \end{split}
\end{equation}
where $\Gamma_j=F_l(x_j)/(U_z \rho) $. We can notice that the vorticity not only spread as a Gaussian function in the directions normal to the actuator line but also spread outside the boundaries of the line in the spanwise direction (terms $H_{\varepsilon}$). From equation \eqref{eq:vortzsteady}, the shed vorticity is:
\begin{equation}
  \label{eq:vortzidealheav}
  \omega_z^i= - \sum_{j=1}^{N} \bigg[ \delta(x-x_{j_-}) \delta(y) H(z)  - \delta(x-x_{j_+}) \delta(y) H(z)  \bigg] \Gamma_j
\end{equation}
\begin{equation}
  \label{eq:vortzheav}
  \omega_z= - \sum_{j=1}^{N} \bigg[   \eta(x-x_{j_-}) \eta(y) H_{\varepsilon}(z) - \eta(x-x_{j_+}) \eta(y) H_{\varepsilon}(z) \bigg] \Gamma_j  .
\end{equation}

The first two terms of equation \eqref{eq:vortzidealheav} are the singular semi-infinite vortices generated by the discontinuities in the circulation at the boundary of the segment, in positions $x_{j_-}$ and $x_{j_+}$. These terms in equation \eqref{eq:vortzheav} become semi-infinite Lamb-Oseen vortices centered in $x_{j_-}$ and $x_{j_+}$. This case regresses to a discretized Prandtl's lifting line in the case of equation \eqref{eq:vortzidealheav} and to a lifting line with vortices with Gaussian core in the case of equation \eqref{eq:vortzheav}

The vorticity of equations~\eqref{eq:vortxidealheav} and~\eqref{eq:vortzidealheav} is identical to the vorticity considered in the lifting line method, under the assumption that the velocity $U_z$ is approximately constant inside each segment. For rotating blades, this assumption is an approximation, since the local velocity changes along the blades. Nevertheless, this assumption is consistent with the linear theory and is a good approximation if the length of the segment is small compared to the radius.

The missing velocity is defined as the velocity needed to recover the velocity induced by singular vortices, as explained in Section~\ref{sec:vortexsmearcorr}. If the implementation of the ALM considers constant circulation inside each segment, the correction proposed by \cite{dag2020new} is equivalent to a lifting line method, according to the linear approximation, as proposed by \cite{dag2020new} and \cite{meyer2019vortex} (putting aside other simplifications, as discussed in Sections \ref{sec:lamboseenvelocity} and \ref{sec:formvortexsheet}). If the implementation of the ALM considers a non-constant distribution of circulation inside each segment, the correction by \cite{dag2020new} is not identical to the discrete lifting line method, because equations \eqref{eq:vortzidealheav} and \eqref{eq:vortzheav} would have extra terms. This case is not treated here but could be a topic for further research. Interestingly, also the discretized implementation of \citep{martinez2019filtered} is not exactly identical (although very similar) to the classical lifting line method, because it implicitly considers the vortices located at the control points (equation 5.7 of that reference), not at the boundaries of segments, as would be usual in a classical discretized lifting line.

The bound vortex is also affected by the convolution with the Gaussian kernel, as seen in equation \eqref{eq:vortxheav}. However, it does not affect the forces in most cases, since most geometries have straight wings or blades and a straight bound vortex does not induce velocity on itself. For multi-blade configurations, the blades are usually too distant for the correction to make any difference (distance is several times $\varepsilon$). A possible exception may be the hub, however, the circulation at the hub is low and the forces are not as relevant for performance. As a consequence, the effect of the convolution on the vorticity generated and the implementation of corrections for bound vorticity have been, justifiably, neglected.

\subsection{Velocity induced by a smeared vortex segment} \label{sec:lamboseenvelocity}

In the method of \citep{dag2020new,meyer2019vortex}, the vortex wake is discretized by segments, and the missing velocity is calculated for each finite-length segment of vortex with a Gaussian distribution of vorticity. Hence, it is relevant to study the velocity induced by a finite-length segment with a cross-section of Gaussian vorticity distribution.

\begin{figure}
    \centering
    \sbox0{\includegraphics[width=0.65\textwidth]{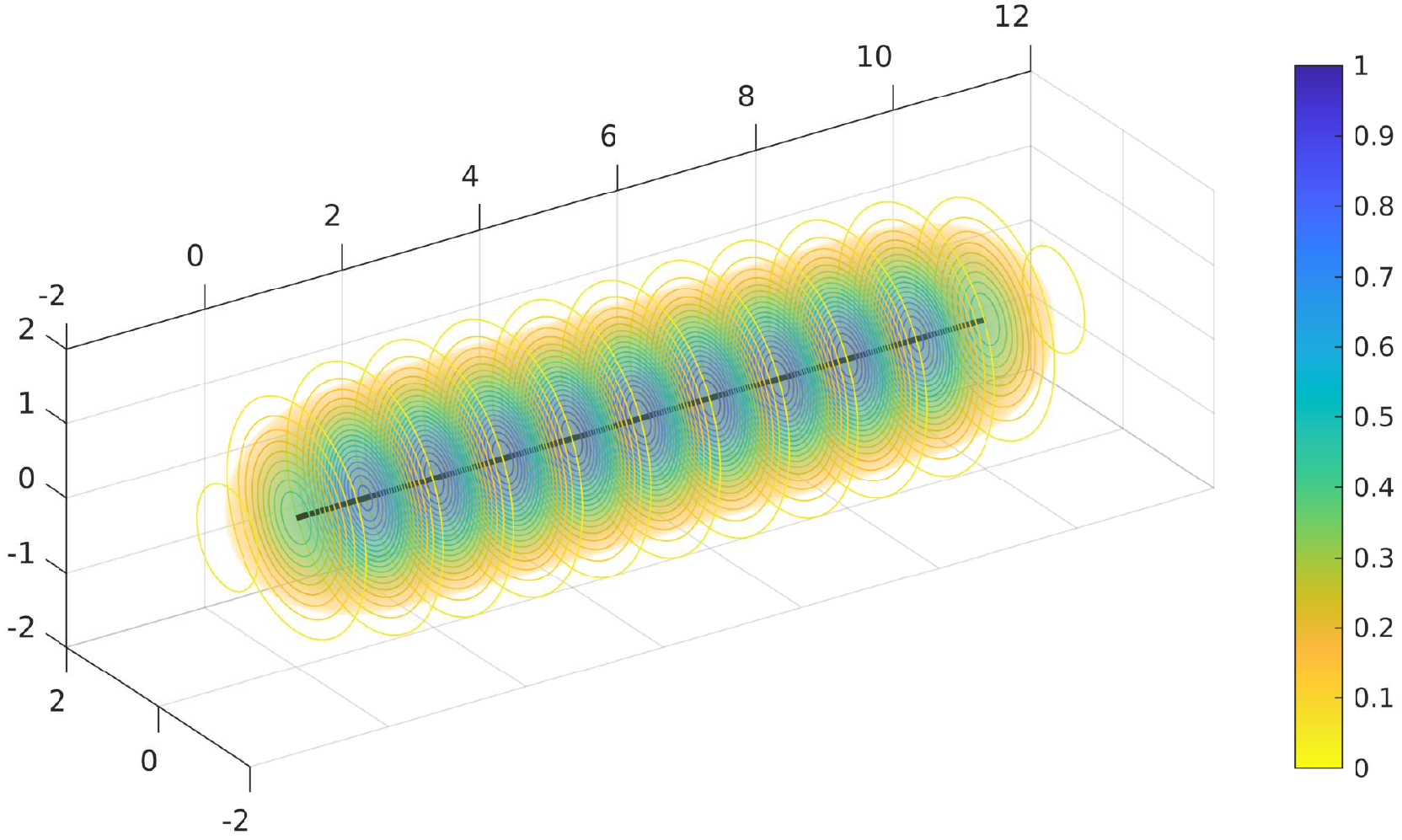}}
    \begin{tikzpicture}
      \node[anchor=south west,inner sep=0] (image) at (0,0) {\includegraphics[clip,trim={.0\wd0} {.0\ht0} {.0\wd0} {.0\ht0},width=0.65\textwidth]{Figures/smearedv.pdf}};
      \node at (0.06\wd0,0.06\ht0) {\small{$\frac{x}{\varepsilon}$}};
      \node at (0.0\wd0,0.43\ht0) {\small{$\frac{y}{\varepsilon}$}};
      \node at (0.38\wd0,0.85\ht0) {\small{$\frac{z-z_{j_-}}{\varepsilon}$}};
    \end{tikzpicture}
    \caption{Smeared vortex segment of length $z_{j_+}-z_{j_-} = 10 \varepsilon$. Isocontours and volume rendering indicate normalized vorticity $\omega_z \pi\varepsilon^2/\Gamma_j$, given by equation \eqref{eq:vortzsegm}. Black line indicates the segment of vortex filament with concentrated vorticity.}
    \label{fig:smearedvortex}
\end{figure}

Without loss of generality, we consider a straight segment of vortex filament with constant circulation $\Gamma_j$, located at $(x,y)=(0,0)$ and aligned with the $z$-direction, from $z_{j_-}$ to $z_{j_+}$, given by
\begin{equation}
  \label{eq:vortzidealsegm}
  \omega_z^i = \delta(x) \delta(y) \left( H(z-z_{j_-})-H(z-z_{j_+}) \right) \Gamma_j .
\end{equation}
A ``smeared vortex segment'' is defined as the smeared equivalent of the straight segment of vortex filament, formed by the convolution of equation \eqref{eq:vortzidealsegm} with a Gaussian function in three dimensions. In order to compute the velocity induced by a smeared vortex segment, the approach of \citet{leonard1980vortex} is taken (see also~\citep{leonard1985computing,winckelmans1989topics}). The vorticity of a segment of vortex filament convoluted with a 3-d Gaussian function is given by
\begin{equation}
  \label{eq:vortzsegm}
  \omega_z = \Gamma_j \int^{z_{j_+}}_{z_{j_-}} \eta_3(x,y,z-z') dz' = \eta(x) \eta(y) \left( H_{\varepsilon}(z-z_{j_-})-H_{\varepsilon}(z-z_{j_+}) \right) \Gamma_j .
\end{equation}
Figure \ref{fig:smearedvortex} shows a vortex filament of length $10 \varepsilon$ with concentrated (infinite) vorticity in black and the vorticity distribution of the corresponding smeared vortex segment.

The velocity induced by a smeared vortex segment, is given by~\citep{leonard1980vortex}:
\begin{equation}
  \label{eq:velocitysegm}
    u^{v}(\mathbf{x}) = \frac{\Gamma_j}{4 \pi} \int^{z_{j_+}}_{z_{j_-}} 
    \frac{\hat{\mathbf{e}}_x \times (\mathbf{x}-\mathbf{x'})}{|\mathbf{x}-\mathbf{x'}|^3} g\left( |\mathbf{x}-\mathbf{x'}| \right) d z' ,
\end{equation}
where function $g(s)$ is defined as
\begin{equation}
  \label{eq:functiong}
    g(s) = 4 \pi \int^s_0 \eta_3(s') s'^2 ds' = \erf{\left(\frac{s}{\varepsilon}\right)} - 2 s \eta(s),
\end{equation}
in which the 3-d Gaussian kernel $\eta_3(s')$ (equation~\eqref{eq:eta3d}) is rewritten in terms of the spherical coordinate $s'=\sqrt{x^2+y^2+z^2}$. Hence, the induced velocity, in the azimuthal direction,  is 
\begin{equation}
  \label{eq:velocitythetasegm}
  \begin{split}
    u_{\theta}^{v}(r,z) = & \frac{\Gamma_j}{4 \pi} r \int^{z_{j_+}}_{z_{j_-}} 
    \frac{\erf{\left(\frac{\sqrt{r^2+(z-z')^2}}{\varepsilon}\right)} - 2 \sqrt{r^2+(z-z')^2} \, \eta\left(\sqrt{r^2+(z-z')^2}\right)}{(r^2+(z-z')^2)^{3/2}} d z' \\
    = & \frac{\Gamma_j}{4 \pi} \left( \Phi(r,z-z_{j_+}) - \Phi(r,z-z_{j_-}) \right) ,
  \end{split}
\end{equation}
where
\begin{equation}
  \label{eq:Phi}
    \Phi(r,Z)
    = \frac{1}{r} \left( - \frac{Z}{\sqrt{r^2+Z^2}} \erf{\left(\frac{\sqrt{r^2+Z^2}}{\varepsilon}\right)} + \exp{\left(-\frac{r^2}{\varepsilon^2}\right)} \erf{\left(\frac{Z}{\varepsilon}\right)} \right) .
\end{equation}

Analogously, $\Phi^{i}(r,Z)$ can be defined as
\begin{equation}
  \label{eq:bigPhiideal}
  \begin{split}
    \Phi^{i}(r,Z) \defeq \lim_{\varepsilon \to 0} \Phi(r,Z) = \frac{1}{r} \left( - \frac{Z}{\sqrt{r^2+Z^2}} \right) ,
  \end{split}
\end{equation}
and the velocity induced by a singular vortex segment can be calculated as
\begin{equation}
  \label{eq:velocitythetasegmideal}
  \begin{split}
    u_{\theta}^{vi}(r,z) = \frac{\Gamma_j}{4 \pi} \left( \Phi^i(r,z-z_{j_+}) - \Phi^i(r,z-z_{j_-}) \right) ,
  \end{split}
\end{equation}
which agrees with the classical results for a vortex segment (see \emph{e.g.}~\citep{katz1991low}).

Therefore, in the present work, for each vortex segment, the missing velocity is then calculated as
\begin{equation}
  \begin{split}
  \label{eq:velocitythetamissingtab}
    u_{\theta}^{m}(r,z) = & u_{\theta}^{vi}(r,z)-u_{\theta}^{v}(r,z) \\
    = & \frac{\Gamma_j}{4 \pi} \Big( \left( \Phi^i(r,z-z_{j_+}) - \Phi^i(r,z-z_{j_-}) \right) - \left( \Phi(r,z-z_{j_+}) - \Phi(r,z-z_{j_-}) \right) \Big) .
  \end{split}
\end{equation}
The same correction is applicable to the bound vortices and wake vortices, taking into account the different orientations of the vortices.

For $z=0$, $z_{j_-}=0$ and $z_{j_+} \to +\infty$, the induced velocity is given by
\begin{equation}
    u_{\theta}^{v}(r,0) = \frac{\Gamma_j}{4 \pi r} \left( 1 - \exp{\left(-\frac{r^2}{\varepsilon^2}\right)} \right)
    \label{eq:uthetasemiinf}
\end{equation}
which is the formula for a semi-infinite vortex, used by~\citep{martinez2019filtered} (and consistent with the relationship mentioned by~\citep{dag2020new,meyer2019vortex}). 

However, equation~\eqref{eq:velocitythetamissingtab} is clearly different from the approximations employed in~\citep{dag2020new,meyer2019vortex}. The approximation for $\Phi$ used in these works for a vortex segment introduces errors. For the formula of~\citep{meyer2019vortex}, negative and positive errors compensate each other for a semi-infinite vortex, but the same can not be concluded for other vortex geometries. Such errors are not expected to be present in the application of the correction of \citep{martinez2019filtered} to straight wings in uniform flow. However, the application of the formula of \citep{martinez2019filtered} to the case of rotating blades without adaptations, such as performed in \citep{stanly2022large}, may incur other errors, since the helical vortex structure formed by the blades is being modeled by a vortex sheet formed by straight horseshoe vortices.

Nevertheless, the good results of \citep{dag2020new} and \citep{meyer2019vortex} show that their corrections improve the results when compared to ALM without correction or with other heuristic tip corrections, even using the approximate value of $\Phi$. A possible explanation for that is based on the observation that the missing velocity is directly related to the difference $\Phi^i-\Phi$. For the vortices near the blades which tend to dominate the correction, the value of $\Phi^i$ is much larger than $\Phi$, then the exact value of $\Phi$ is not as relevant, as long as the order of magnitude is the same. The dominance of the vortices near the blades is also a possible explanation for the improved results observed by \citep{stanly2022large}, even considering straight horseshoe vortices instead of helical vortices. The estimate of the order of magnitude of the errors caused by these different approximations is left for further studies. In the present work, the analytical integration, represented by equation~\ref{eq:Phi}, is used to avoid the approximation errors.

\subsection{Forming the vortex sheet} \label{sec:formvortexsheet}

Forming the vortex sheet is an intrinsic part of both the ALM with smearing correction and the lifting line method. It should be noted, however, that the methods are not going to be mathematically identical even if the vortex sheet is formed with the same procedure in both methods. In the actuator line, the induced velocity is formed partly by the smeared vortices created in the CFD simulation and partly by the vortex sheet of the smearing correction, which is formed by one of the strategies detailed below (that may be more or less dependent on the results from the simulation). Hence, even if the vortex sheet from the smearing correction is identical to the vortex sheet of the lifting line method, the induced velocities can differ slightly, because the vortex sheet created in the CFD simulation may be different from the vortex sheet of the lifting line method. This phenomenon can be observed in the results of Section \ref{sec:res_wing}, where its effect is noticeable but negligible.

The formation of the vortex sheet for the smearing correction is different among past works. The method of \citep{martinez2019filtered} does not require an explicit vortex sheet, since the correction is based on analytical semi-infinite vortices for a straight wing under uniform flow. However, the same correction was applied by \citep{stanly2022large}, which would mean that the helical vortex structure is implicitly modeled by horseshoe vortices. As discussed in Section~\ref{sec:lamboseenvelocity}, errors are expected to be introduced by this approximation.

In \citep{dag2020new}, the helical vortices were imposed considering the pitch of each of the helical vortices based on the local relative flow angle at the position where the vortex is released. In \citep{meyer2019vortex} a fixed helical wake was assumed, based on the near-wake model of \citep{pirrung2016coupled,pirrung2017trailed}, and the bookkeeping of the position of released vortices is only necessary for one of the factors involved in the smearing correction. The need for bookkeeping was later removed by \citet{meyer2020brief}, reducing computational cost with negligible effect on the forces.

In the present work, we implement a free-vortex wake method within our simulation, a strategy already suggested by \citet{dag2020new}, assuming that the vortices are advected by the flow velocity. Tracing particles are used to follow the position of the boundaries of the vortex segments. Also, the past values of circulation, that are present in the wake, are directly used, which is a more realistic situation than considering the whole wake as having the value of circulation that the wing currently assumes. This method has the drawback of requiring bookkeeping of the circulation and position of the vortices emitted in previous time steps. Nevertheless, it has the advantage of requiring no \emph{ad hoc} assumption and therefore the method maintains its generality.

There is an ambiguity regarding using the corrected velocity or the uncorrected velocity sampled directly from the numerical simulation, to estimate the position of vortices. However, this difference is of second order for the velocities along the actuator line. Using the corrected velocity might even introduce errors or numerical instability if used throughout the vortex sheet, because of the singularity of ideal vortices: vortices recently released might suffer the effect of the singularity of the bound vortex and curved vortices might induce numerical instabilities.

Hence, in the current implementation, the tracing particles that define the boundaries of the vortex filaments are advected by the velocities sampled from the simulation, without correction. This reduces the number of operations needed so that the correction velocity is only calculated on the control points of the actuator line, not on other points of the vortex sheet. More relevant for our non-iterative method, the vortex sheet can be advected before the computation of the correction velocity at a certain time step. For its lower computational cost, a first-order Euler time-integration method is used to advect the tracing particles. The information related to timesteps older than $n-n_w$ is not stored, where $n$ is the current timestep and $n_w$ is the number of tracing particles.

In order to save computational resources, not all tracing particles are kept. The vortices released in the last $n_{nw}$ time steps are always maintained in the memory. However, a group of vortices is fused if the distance between the tracing particles is below $d_{w}$, with the circulation taken as the average of the circulation of the vortices fused together. The strategy of fusing vortices guarantees a minimum wake length of $(n_w-n_{nw}-1)d_{w}$ near the tip of the blades, the region of interest and where the advection velocity is larger. For the current simulations, the choice of parameters is $n_w=50$, $n_{nw}=10$ and $d_w=\varepsilon/2$, guaranteeing a wake length of approximately $20\varepsilon$ near the tip. Analysis of the correction of equation \eqref{eq:velocitythetamissingtab} indicates that errors involved in discarding vortices farther than this length are negligible. The wake length is lower near the hub but this region is less relevant for the performance of the turbine. A parameter study for the simulation of the NREL 5-MW turbine of Section \ref{sec:res_rotor} showed that doubling $n_w$ and $n_{nw}$ had negligible effects on the circulation.

For unsteady simulations, a vorticity wake in the spanwise direction would be shed when the circulation changes, similar to the starting vortex of Kelvin's circulation theorem. Following a quasi-steady approach, the correction for this spanwise shed vorticity is neglected, hence it is not modeled in our method. For further discussion about this shed vorticity and unsteady effects, the reader is referred to~\citep{kleine2022onstability}. The effect of the Gaussian smearing on unsteady effects is left for future studies.
\section{Iterative lifting line consistent with the actuator line method} \label{sec:NLLL}

The formulation of the actuator line presented in Sections \ref{sec:basicACL} and \ref{sec:itesmearcorr} is very similar to a non-linear lifting line method \citep{anderson1991fundamentals} (see also \cite{phillips2000modern}). Hence, we can develop a numerical non-linear lifting line method consistent with the actuator line method with vortex-based smearing correction. While \citet{anderson1991fundamentals} uses the undisturbed uniform inflow velocity ($\mathbf{U}_{\infty}$) to calculate the circulation, we use the local velocity, basing our method on equation \eqref{eq:GammaNLL}, in an approach similar to~\cite{phillips2000modern}. Also, in this formulation, the undisturbed velocity $\mathbf{U}=(U_x,U_y,U_z)$ (velocity as if the wing was not present in the flow) can change along the span, a feature paramount for the application of the actuator line to rotating blades. The local velocity $\mathbf{u}$, which is previously unknown, can be calculated from the local undisturbed velocity $\mathbf{U}$ and the velocity induced by the vortex system, $\mathbf{u}^{vi}$, at control point $\mathbf{x}_j$, given by
\begin{equation}
  \mathbf{u}(\mathbf{x}_j) = \mathbf{U}(\mathbf{x}_j) + \mathbf{u}^{vi}(\mathbf{x}_j) ,
  \label{eq:uNLL}
\end{equation}

The steps of the non-linear lifting line are:
\begin{enumerate}
  \item Start from a guess of the circulation distribution, for example, applying equation \eqref{eq:GammaNLL} with $\mathbf{u}=\mathbf{U}$;
  \item \label{item:formvortex} Form the vortex sheet by:
  \begin{itemize}
      \item prescribing the vortex sheet, for example, by assuming helical or horseshoe vortices; or
      \item employing a free-vortex wake method.
  \end{itemize}
  \item Calculate the velocity induced by the bound vortex and the vortex sheet at every control point, $\mathbf{u}^{vi}(\mathbf{x}_j)$. Find the local velocity according to equation \eqref{eq:uNLL}; \label{item:velocity}
  \item Calculate the effective angle of attack using equation \eqref{eq:effalpha};
  \item Find the local lift coefficient $C_l$, interpolating from the airfoil data table;
  \item Calculate the new value of circulation $\Gamma_j^{new}$ using equation \eqref{eq:GammaNLL};
  \item Update the current value of circulation using a relaxation factor $r$:
  \begin{equation}
    \Gamma_j=r\Gamma_j^{new} + (1-r) \Gamma_j^{old}
  \end{equation}
  \item Restart from step (\ref{item:formvortex}) if the local velocity affects the formation of the vortex sheet or from step (\ref{item:velocity}) if a prescribed vortex sheet is considered, using the value of circulation calculated in the previous step. Iterate until a chosen convergence criterion is reached.
\end{enumerate}

This iterative lifting line method was used as the reference to compare the results of the ALM in section~\ref{sec:res_wing}, with the parameters detailed in section~\ref{sec:metLL}. This formulation is able to deal with configurations in which the induced velocity can no longer be considered small compared to the undisturbed velocity, as observed in~\citep{kleine2022simulating}. Based on the same equations, a linearized lifting line method is developed in Appendix~\ref{sec:LLL}, which is used to derive the non-iterative smearing correction.

\section{Non-iterative smearing correction based on the linearized lifting line method} \label{sec:nonitesmearcorr}

The steps to apply the smearing correction (Section \ref{sec:itesmearcorr}) are basically equal to the steps of the non-linear lifting line described in Section \ref{sec:NLLL}, except for step (\ref{item:velocity}), which becomes the calculation of the missing velocities, and step (\ref{item:formvortex}), which allows the formation of the vortex sheet using data from the CFD simulation. Therefore, a non-iterative smearing correction based on the formulas of the linearized lifting line described in Appendix \ref{sec:LLL} can be devised.

A minor difference in the process of the iterative lifting line and the smearing correction is the possibility to start from a previous time step, which provides a better guess for the circulation. This minor difference, however, is what allows the linearized, non-iterative, formulation of the smearing correction to have good results even if the airfoil lift coefficient is not linear \emph{w.r.t.} the angle of attack.

The first phase of the method is to find an approximate solution to linearize around. This is achieved by applying the first three iterative steps of Section \ref{sec:itesmearcorr} only once. The corrected velocity calculated from the circulation distribution $\boldsymbol{\Gamma}^{n-1}$ from the previous time-step $n-1$ is
\begin{equation}
  \mathbf{u}^{c}(\mathbf{x}_j,\boldsymbol{\Gamma}^{n-1}) = \mathbf{u}^{s}(\mathbf{x}_j) + \mathbf{u}^{m}(\mathbf{x}_j,\boldsymbol{\Gamma}^{n-1}) .
\end{equation}
Representing the variables around which the functions are linearized by $^\dagger$:
\begin{equation}
  u_y^{\dagger} \defeq u_{y}^{c}(\mathbf{x}_j,\boldsymbol{\Gamma}^{n-1}) = u_{y}^{s}(\mathbf{x}_j) + u_{y}^{m}(\mathbf{x}_j,\boldsymbol{\Gamma}^{n-1})
  \label{eq:UyLsmear}
\end{equation}
\begin{equation}
  u_z^{\dagger} \defeq u_{z}^{c}(\mathbf{x}_j,\boldsymbol{\Gamma}^{n-1}) = u_{z}^{s}(\mathbf{x}_j) + u_{z}^{m}(\mathbf{x}_j,\boldsymbol{\Gamma}^{n-1}) .
  \label{eq:UzLsmear}
\end{equation}
Analogously to Appendix \ref{sec:LLL}:
\begin{equation}
  u_r^\dagger \defeq \sqrt{(u_z^{\dagger2}+u_y^{\dagger2})}, \qquad
  \alpha_{j}^\dagger \defeq \alpha_{gj} + \arctan{\left(\frac{u_y^\dagger}{u_z^\dagger}\right)} \qquad \text{and} \qquad
  \Gamma_{j}^\dagger \defeq \frac{1}{2} u_r^\dagger \, c_j \, C_l(\alpha_{j}^\dagger) .
\end{equation}

Phase two of the linearization process involves finding another linear relation between the corrected velocity and the circulation. It is assumed that the missing velocity is not relevant for the formation of the vortex sheet, as discussed in Section \ref{sec:formvortexsheet}, and the CFD velocity is used directly to advect the vortices. This implies that the current value of circulation does not affect the geometry of the vortex sheet, what simplifies the method and is consistent with a first-order approximation.

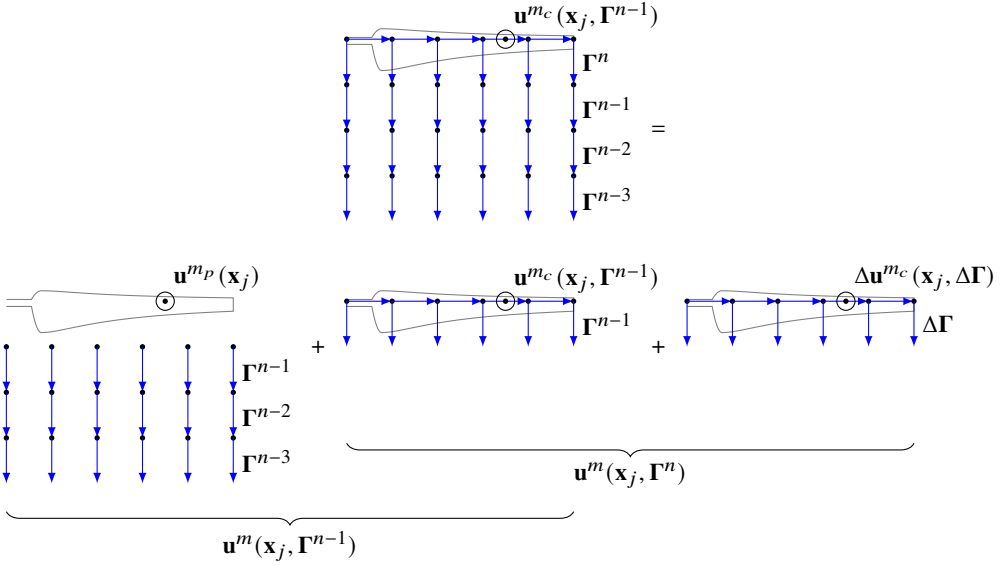
\begin{figure}
  \centering
  \begin{tikzpicture}
  \def\R{3.0}
  \def\dn{0.5}
  \def\ln{-0.2}
  \def\hn{-0.35}
  \def\maxi{5}
  \def\maxj{3}
  \def\d{\dn*\R}
  \def\l{\ln*\R}
  \def\h{\hn*\R}
  \def\D{\R+\d}
  \def\y{\maxj*\l+\l+\h}
  \def\x{-2*\D-3*\d}
  \draw[scale=\R,gray] plot file{Figures/blade.dat};
  \foreach \i in {0,...,\maxi}
  {
    \ifthenelse{\i>0}{\draw[-latex,blue] (\i/\maxi*\R - \R/\maxi,0)--(\i/\maxi*\R,0);}{}
    \foreach \j in {0,...,\maxj}
    {
      \fill (\i/\maxi*\R,\j*\l) circle[radius=1pt];
      \draw[-latex, blue] (\i/\maxi*\R,\j*\l)--(\i/\maxi*\R,\j*\l+\l);
    }
  } 
  \node[right] at (\R,0*\l+\l/2) {$\boldsymbol{\Gamma}^{n}$};
  \foreach \j in {1,...,\maxj}
  {
    \node[right] at (\R,\j*\l+\l/2) {$\boldsymbol{\Gamma}^{n-\j}$};
  }
  \node[right] at (\R-3*\R/\maxi/2,-\l/2) {$\mathbf{u}^{m_{c}}(\mathbf{x}_j,\boldsymbol{\Gamma}^{n-1})$};
  \fill (\R-3*\R/\maxi/2,0) circle[radius=1pt];
  \draw (\R-3*\R/\maxi/2,0) circle[radius=\R/\maxi/5];
  \node[black] at (\R+0.75*\d,2.0*\l) {$=$};
  
  \draw[scale=\R,gray,shift={(-1-\dn,\maxj*\ln+\ln+\hn)}] plot file{Figures/blade.dat};
  \foreach \i in {0,...,\maxi}
  {
    \foreach \j in {1,...,\maxj}
    {
      \fill (\i/\maxi*\R+\x+\D,\j*\l+\y) circle[radius=1pt];
      \draw[-latex, blue] (\i/\maxi*\R+\x+\D,\j*\l+\y)--(\i/\maxi*\R+\x+\D,\j*\l+\l+\y);
    }
  } 
  \foreach \j in {1,...,\maxj}
  {
    \node[right] at (\R+\x+\D,\y+\j*\l+\l/2) {$\boldsymbol{\Gamma}^{n-\j}$};
  }
  \node[right] at (\R-3*\R/\maxi/2+\x+\D,\y-\l/2) {$\mathbf{u}^{m_{p}}(\mathbf{x}_j)$};
  \fill (\R-3*\R/\maxi/2+\x+\D,\y) circle[radius=1pt];
  \draw (\R-3*\R/\maxi/2+\x+\D,\y) circle[radius=\R/\maxi/5];
  \node[black] at (\R+0.75*\d+\x+\D,1.0*\l+\y) {$+$};
  
  \draw[scale=\R,gray,shift={(0,\maxj*\ln+\ln+\hn)}] plot file{Figures/blade.dat};
  \foreach \i in {0,...,\maxi}
  {
    \ifthenelse{\i>0}{\draw[-latex,blue] (\i/\maxi*\R - \R/\maxi+\x+2*\D+\d,\y)--(\i/\maxi*\R+\x+2*\D+\d,\y);}{}
    \foreach \j in {0,...,0}
    {
      \fill (\i/\maxi*\R+\x+2*\D+\d,\j*\l+\y) circle[radius=1pt];
      \draw[-latex, blue] (\i/\maxi*\R+\x+2*\D+\d,\j*\l+\y)--(\i/\maxi*\R+\x+2*\D+\d,\j*\l+\l+\y);
    }
  }
  \node[right] at (\R+\x+2*\D+\d,\y+\l/2) {$\boldsymbol{\Gamma}^{n-1}$};
  \node[right] at (\R-3*\R/\maxi/2+\x+2*\D+\d,\y-\l/2) {$\mathbf{u}^{m_{c}}(\mathbf{x}_j,\boldsymbol{\Gamma}^{n-1})$};
  \fill (\R-3*\R/\maxi/2+\x+2*\D+\d,\y) circle[radius=1pt];
  \draw (\R-3*\R/\maxi/2+\x+2*\D+\d,\y) circle[radius=\R/\maxi/5];
  \node[black] at (\R+0.75*\d+\x+2*\D+\d,1.0*\l+\y) {$+$};
  
  \draw[scale=\R,gray,shift={(1+1*\dn,\maxj*\ln+\ln+\hn)}] plot file{Figures/blade.dat};
  \foreach \i in {0,...,\maxi}
  {
    \ifthenelse{\i>0}{\draw[-latex,blue] (\i/\maxi*\R - \R/\maxi+\x+3*\D+2*\d,\y)--(\i/\maxi*\R+\x+3*\D+2*\d,\y);}{}
    \foreach \j in {0,...,0}
    {
      \fill (\i/\maxi*\R+\x+3*\D+2*\d,\j*\l+\y) circle[radius=1pt];
      \draw[-latex, blue] (\i/\maxi*\R+\x+3*\D+2*\d,\j*\l+\y)--(\i/\maxi*\R+\x+3*\D+2*\d,\j*\l+\l+\y);
    }
  }
  \node[right] at (\R+\x+3*\D+2*\d,\y+\l/2) {$\Delta\boldsymbol{\Gamma}$};
  \node[right] at (\R-3*\R/\maxi/2+\x+3*\D+2*\d,\y-\l/2) {$\Delta \mathbf{u}^{m_{c}}(\mathbf{x}_j,\Delta\boldsymbol{\Gamma})$};
  \fill (\R-3*\R/\maxi/2+\x+3*\D+2*\d,\y) circle[radius=1pt];
  \draw (\R-3*\R/\maxi/2+\x+3*\D+2*\d,\y) circle[radius=\R/\maxi/5];
  
  \draw[decoration={brace,mirror,raise=2pt,amplitude=5pt},decorate]  (\x+\D,2*\y+1.5*\l) -- (\R+\x+2*\D+\d,2*\y+1.5*\l) node[midway,below=6pt,black]{$\mathbf{u}^{m}(\mathbf{x}_j,\boldsymbol{\Gamma}^{n-1})$};
  \draw[decoration={brace,mirror,raise=2pt,amplitude=5pt},decorate]  (\x+2*\D+\d,2*\y) -- (\R+\x+3*\D+2*\d,2*\y) node[midway,below=6pt,black]{$\mathbf{u}^{m}(\mathbf{x}_j,\boldsymbol{\Gamma}^{n})$};
  
\end{tikzpicture}

  \caption{Schematic representation of equation \eqref{eq:umLsmearndelta}. The missing velocity is the sum of the missing velocity due to the vortex sheet emitted in the previous time steps (leftmost figure of the second row) and the missing velocity due to the bound vortex and the vortex sheet emitted in the current time step, which can also be split into two terms (central and rightmost figures of the second row).}
  \label{fig:vortexsheet}
\end{figure}

The unknown missing velocity for the value of circulation at the current time step, $\boldsymbol{\Gamma}^{n}$,  can be divided in two components:
\begin{equation}
  \mathbf{u}^{m}(\mathbf{x}_j,\boldsymbol{\Gamma}^{n}) = \mathbf{u}^{m_{p}}(\mathbf{x}_j,\boldsymbol{\Gamma}^{n}) + \mathbf{u}^{m_{c}}(\mathbf{x}_j,\boldsymbol{\Gamma}^{n})
  \label{eq:umLsmearn}
\end{equation}
where $\mathbf{u}^{m_{p}}$ is the velocity induced by the vortex sheet emitted in the previous time steps and $\mathbf{u}^{m_{c}}$ is the velocity induced by the vortex sheet emitted in the current time step and by the bound vortices. The value of circulation at the current time step does not affect the vortex wake emitted in the previous time steps, hence, the velocity induced by its vortices, $\mathbf{u}^{m_{p}}$, is already computed. Considering the linearity of the relationship between velocity and circulation, equation~\eqref{eq:umLsmearn} can be rewritten as
\begin{equation}
  \mathbf{u}^{m}(\mathbf{x}_j,\boldsymbol{\Gamma}^{n}) = \mathbf{u}^{m_{p}}(\mathbf{x}_j) + \mathbf{u}^{m_{c}}(\mathbf{x}_j,\boldsymbol{\Gamma}^{n-1}) + \Delta \mathbf{u}^{m_{c}}(\mathbf{x}_j,\Delta \boldsymbol{\Gamma}) = \mathbf{u}^{m}(\mathbf{x}_j,\boldsymbol{\Gamma}^{n-1}) + \Delta \mathbf{u}^{m_{c}}(\mathbf{x}_j,\Delta \boldsymbol{\Gamma}),
  \label{eq:umLsmearndelta}
\end{equation}
where $\Delta \mathbf{u}^{m_{c}} = \mathbf{u}^{m_{c}}(\mathbf{x}_j,\boldsymbol{\Gamma}^{n})-\mathbf{u}^{m_{c}}(\mathbf{x}_j,\boldsymbol{\Gamma}^{n-1})$ and $\Delta \boldsymbol{\Gamma}=\boldsymbol{\Gamma}^{n}-\boldsymbol{\Gamma}^{n-1}$. A schematic representation is shown in figure \ref{fig:vortexsheet}.

We write the velocities induced at control point $j$ by the vortex system $k$ which includes the bound vortex and the vortex sheet created by the segment $k$ at the current time step, as:
\begin{equation}
  u_{y}^{m_{ck}}(\mathbf{x}_j,\Gamma_k^{n}) = a_{y,jk}^{m_c} \Gamma_k^{n}
  \label{eq:u_yLsmearn_jk}
\end{equation}
\begin{equation}
  u_{z}^{m_{ck}}(\mathbf{x}_j,\Gamma_k^{n}) = a_{z,jk}^{m_c} \Gamma_k^{n}
  \label{eq:u_zLsmearn_jk}
\end{equation}
where $a_{y,jk}^{m_c}$ and $a_{y,jk}^{m_c}$ are the influence coefficients obtained considering only the bound vortex and the vortex sheet emitted at the current time step (see figure \ref{fig:vortexsheet}). It should be noted that the coefficients $a_{y,jk}^{m_c}$ and $a_{y,jk}^{m_c}$ should already be known to perform step (\ref{item:velocitysm}) of the usual iterative version of the smearing correction (Section \ref{sec:itesmearcorr}). Equations \eqref{eq:u_yLsmearn_jk} and \eqref{eq:u_zLsmearn_jk} written in matrix form are:
\begin{equation}
  \mathsfbi{u_y}^{m_{c}}(\boldsymbol{\Gamma}^{n}) = \mathsfbi{A_y}^{m_c} \boldsymbol{\Gamma}^{n}
  \label{eq:u_yLsmearn_1}
\end{equation}
\begin{equation}
  \mathsfbi{u_z}^{m_{c}}(\boldsymbol{\Gamma}^{n}) = \mathsfbi{A_z}^{m_c} \boldsymbol{\Gamma}^{n} .
  \label{eq:u_zLsmearn_1}
\end{equation}

The missing velocity $\mathsfbi{u_y}^{m}(\boldsymbol{\Gamma}^{n})$ is then written as
\begin{equation}
  \begin{aligned}
    \mathsfbi{u_y}^{m}(\boldsymbol{\Gamma}^{n}) =  \mathsfbi{u_y}^{m}(\boldsymbol{\Gamma}^{n-1}) + \mathsfbi{A_y}^{m_c} \Delta\boldsymbol{\Gamma} .
  \end{aligned}
  \label{eq:u_ymLsmear}
\end{equation}
Combining equations~\eqref{eq:UyLsmear} and \eqref{eq:u_ymLsmear}, the corrected velocity $\mathsfbi{u_y}^{c}$ can be written as
\begin{equation}
  \begin{aligned}
    \mathsfbi{u_y}^{c}(\boldsymbol{\Gamma}^{n}) = \mathsfbi{u_y}^\dagger + \mathsfbi{A_y}^{m_c} \Delta\boldsymbol{\Gamma}
  \end{aligned} .
  \label{eq:u_ycLsmear}
\end{equation}
Analogously for $\mathsfbi{u_z}^{c}$
\begin{equation}
  \begin{aligned}
    \mathsfbi{u_z}^{c}(\boldsymbol{\Gamma}^{n}) = \mathsfbi{u_z}^\dagger + \mathsfbi{A_z}^{m_c} \Delta\boldsymbol{\Gamma} .
  \end{aligned}
  \label{eq:u_zcLsmear}
\end{equation}

Finally, the linear equation can be found following the steps of Appendix \ref{sec:LLL}. Similar to equation \eqref{eq:Gamma_jLLL2}, the linearization of equation \eqref{eq:GammaNLL} becomes
\begin{equation}
    \Gamma_j^{n} = \Gamma_{j}^\dagger + \frac{1}{2} \, c_j \left[ \left( C_l(\alpha_{j}^\dagger) \frac{u_y^\dagger}{u_r^\dagger} + \frac{\partial C_l}{\partial \alpha}(\alpha_{j}^\dagger) \frac{u_z^\dagger}{u_r^\dagger} \right) \Delta u_{y}^{m_{c}} + \left( C_l(\alpha_{j}^\dagger) \frac{u_z^\dagger}{u_r^\dagger} - \frac{\partial C_l}{\partial \alpha}(\alpha_{j}^\dagger) \frac{u_y^\dagger}{u_r^\dagger} \right) \Delta u_{z}^{m_{c}} \right] ,
  \label{eq:Gamma_jLsmear2}
\end{equation}
which can be written in matrix form as
\begin{equation}
  \left(\mathsfbi{I} - (diag(\mathsfbi{b_y}^\dagger) \mathsfbi{A_y}^{m_c}) - (diag(\mathsfbi{b_z}^\dagger) \mathsfbi{A_z}^{m_c} ) \right) \Delta\boldsymbol{\Gamma} = \boldsymbol{\Gamma}^\dagger - \boldsymbol{\Gamma}^{n-1} .
  \label{eq:circLsmear}
\end{equation}
where $\mathsfbi{b_y}^\dagger$ and $\mathsfbi{b_z}^\dagger$ can be found using the corresponding version of equations \eqref{eq:byLLL} and \eqref{eq:bzLLL}. $\Delta\boldsymbol{\Gamma}$ is found by solving the linear system and the corrected velocities are calculated using equations \eqref{eq:u_ycLsmear} and \eqref{eq:u_zcLsmear}.

If the lift coefficient has a linear relation with the angle of attack, this method is expected to give good results, even if the differences in the circulation are high, just like a conventional lifting line method has good results despite starting from a completely wrong circulation distribution. However, the strength of the method is in the fact that the functions are linearized around the velocities calculated at the first iteration of a conventional smearing correction. The differences in the velocities and circulation are expected to be small in most simulations, justifying this linearization, even if the lift coefficient is not linear. The method is only expected to perform poorly if the lift coefficient slope changes with the angle of attack and the flow conditions vary greatly between time steps. For these extreme cases, the linear method presented here can be used to accelerate the convergence of an iterative method. The linear system is relatively small, of size $N$, usually several orders of magnitude smaller than the CFD solver, which makes its solution computationally inexpensive.

For multiple turbines, each turbine can be solved in isolation, since the distance between turbines is much greater than $\varepsilon$. In fact, if the distance between the blades is much larger than $\varepsilon$ or if forces at the hub are not relevant, the system of equations can be divided and each blade solved independently. This last simplification was not implemented for the results of Section \ref{sec:results}.

All the aspects regarding the formation of the vortex sheet and calculation of the induced velocity are accounted for in matrices $\mathsfbi{A_y}^{m_c}$ and $\mathsfbi{A_z}^{m_c}$. This means that the method of previous works \citep{dag2020new,martinez2019filtered,meyer2019vortex,meyer2020brief,stanly2022large} could benefit from the procedure outlined in this section, just by forming matrices $\mathsfbi{A_y}^{m_c}$ and $\mathsfbi{A_z}^{m_c}$ compatible with their respective methods. In particular, if desirable, all of the methods for increasing the speed of calculations detailed in \citep{meyer2020brief} could be easily implemented and would only affect the coefficients of these matrices.

For the methods that prescribe the shape of the vortex sheet, the separation of the missing velocity in the components of equation \eqref{eq:umLsmearn} could be skipped, however, it would reduce errors relative to the linearization if implemented. This means that the non-iterative strategy should also work even without bookkeeping of previous values of circulation. A mixed strategy that could be used to reduce the cost is to use a prescribed wake to calculate $\mathsfbi{u_y}^{m}(\boldsymbol{\Gamma}^{n-1})$ considering the circulation along the whole wake as $\boldsymbol{\Gamma}^{n-1}$ (substituting older values of $\boldsymbol{\Gamma}$ in figure \ref{fig:vortexsheet} by $\boldsymbol{\Gamma}^{n-1}$) and another prescribed near wake to define $\Delta \mathsfbi{u_y}^{m_{c}}$, reducing the magnitude of the influence coefficients (compared to the case without any bookkeeping).

\section{Numerical method} \label{sec:numerical}

\subsection{Numerical solver of Navier-Stokes equations with the actuator line} 
The incompressible three-dimensional Navier-Stokes equations are solved in the spectral-element code \texttt{Nek5000} \citep{fischer2008nek5000}. The system of equations is expressed in weak form and the solution is expanded in terms of basis and test functions on each of these elements \citep{offermans2020adaptive}. In this work, the basis for the velocity space are seventh-order Lagrange polynomial interpolants on Gauss-Lobatto-Legendre quadrature points in each element. The $\mathbb{P}_N-\mathbb{P}_{N-2}$ formulation is used \citep{maday1989spectral}. For temporal discretization, a third-order implicit/explicit scheme (BDF3/EXT3) \citep{fischer2003implementation} is employed. Filtering of the higher modes is applied to stabilize the simulation \citep{fischer2001filter}.
 
In the current simulations, an adaptive mesh refinement (AMR) strategy with a spectral error indicator \citep{offermans2019aspects,offermans2020adaptive,tanarro2020enabling} is used to reduce the computational cost. We choose to force maximum discretization of the grid around the actuator lines, independently of the error indicator, to guarantee adequate discretization for the chosen value of the smearing parameter.

The actuator line was previously implemented in \texttt{Nek5000} \citep{kleusberg2016actuator,kleusberg2017high,kleusberg2019wind} with Prandtl's tip correction. The code was extensively validated, showing good agreement between numerical and experimental results \citep{kleusberg2017high,muhle2018blind,kleusberg2019wind,kleusberg2020parametric}. Since the focus of those studies was on the wake and on total forces, the errors caused by the approximation of the tip correction were not considered to greatly affect the results. The low dissipation and dispersion of the spectral element method make it well-suited for stability analysis, hence, this code was previously used in several vortex stability studies \citep{kleusberg2019wind,kleusberg2019tip,kleine2019tip,kleine2022stability}.

The iterative and direct smearing corrections of Sections~\ref{sec:ACL} and~\ref{sec:nonitesmearcorr}, with the correction velocity detailed in section~\ref{sec:lamboseenvelocity}, were implemented in the code. For the iterative smearing correction, the circulation is considered converged when $\| \boldsymbol{\Gamma}^{new} - \boldsymbol{\Gamma}^{old} \|/\| \boldsymbol{\Gamma}^{new} \| < 10^{-5}$ (where $\| \boldsymbol{\Gamma} \|$ represents the Euclidean norm of the circulation vector). Also implemented is the analytical form of the convolution of the Gaussian function with a force distribution formed by segments of constant force presented in Section~\ref{sec:vorticityshed}.

For the NREL 5-MW reference turbine \citep{jonkman2009definition}, the airfoil is defined only for a few spanwise positions in the original reference. In \citep{martinez2018comparison} two different approaches have been employed by different codes, one considering abrupt changes in the airfoil and the other by considering interpolation of coefficients between sections. In the present implementation, the airfoil data from \citep{jonkman2009definition} is interpolated between sections, as performed by the DTU code EllipSys3D in \citep{martinez2018comparison}, making the force curves smoother and avoiding problems with the discontinuities.

The non-iterative method requires the value of the lift coefficient slope. A third-order polynomial interpolation of $C_l$ is used to avoid discontinuities in the lift coefficient slope. The tabulated airfoil lift coefficient is modeled using a shape-preserving piecewise cubic polynomial \citep{mathworks2022interp1}.

\begin{figure}
  \centering
  \begin{tikzpicture}
  \def\L{2.0}
  \def\lx{-1.5}
  \def\ly{1.0}
  \def\lz{-2.5}
  \def\lzm{-1.0}
  \def\rn{0.2}
  \def\Lx{\lx*\L}
  \def\Ly{\ly*\L}
  \def\Lz{\lz*\L}
  \def\Lzm{\lzm*\L}
  \def\R{\rn*\L}
  \def\Vd{0.1*\Lzm}
  \def\Vu{0.5*\Lzm}
  \def\xV{0.0*\Lx}
  \def\xVl{1.0*\Lx}
  
  \draw[-] (0,0,0)--(\Lx,0,0);
  \draw[-] (\Lx,0,0)--(\Lx,\Ly,0);
  \draw[-] (\Lx,\Ly,0)--(0,\Ly,0);
  \draw[-] (0,\Ly,0)--(0,0,0);
  \draw[dashed] (0,0,\Lz)--(\Lx,0,\Lz);
  \draw[dashed] (\Lx,0,\Lz)--(\Lx,\Ly,\Lz);
  \draw[-] (\Lx,\Ly,\Lz)--(0,\Ly,\Lz);
  \draw[-] (0,\Ly,\Lz)--(0,0,\Lz);
  \draw[-] (0,0,0)--(0,0,\Lz);
  \draw[dashed] (\Lx,0,0)--(\Lx,0,\Lz);
  \draw[-] (\Lx,\Ly,0)--(\Lx,\Ly,\Lz);
  \draw[-] (0,\Ly,0)--(0,\Ly,\Lz);
  \fill[fill=gray,fill opacity=0.2] (0,0,\Lzm) rectangle (\Lx,\Ly,\Lzm);
  \draw[scale=\R,black!90!white,shift={(\lx/\rn/2,\ly/\rn/2,\lzm/\rn)},xscale=-1,fill=black!90!white] plot file{Figures/blade.dat};
  \draw[scale=\R,black!90!white,shift={(\lx/\rn/2,\ly/\rn/2,\lzm/\rn)},xscale=-1,fill=black!90!white,rotate=120] plot file{Figures/blade.dat};
  \draw[scale=\R,black!90!white,shift={(\lx/\rn/2,\ly/\rn/2,\lzm/\rn)},xscale=-1,fill=black!90!white,rotate=240] plot file{Figures/blade.dat};
  
  \draw[->] (0.5*\L,0,\Lzm)--++(-\L/5,0,0) node[below]{\scriptsize{$x$}};
  \draw[->] (0.5*\L,0,\Lzm)--++(0,\L/5,0) node[left]{\scriptsize{$y$}};
  \draw[->] (0.5*\L,0,\Lzm)--++(0,0,-\L/3) node[below]{\scriptsize{$z$}};
  
  \draw[latex-latex] (0,-\L/10,0)--(\Lx,-\L/10,0) node[midway,below]{\scriptsize{$L_x$}};
  \draw[latex-latex] (\Lx-\L/10,0,0)--(\Lx-\L/10,\Ly,0) node[midway,left]{\scriptsize{$L_y$}};
  \draw[latex-latex] (0.95*\Lx,\Ly,0)--(0.95*\Lx,\Ly,\Lz) node[midway,right]{\scriptsize{$L_z$}};
  \draw[latex-latex] (0.7*\Lx,\Ly,0)--(0.7*\Lx,\Ly,\Lzm) node[midway,right]{\scriptsize{$L_{zin}$}};
  \draw[latex-latex] (0.7*\Lx,\Ly,\Lzm)--(0.7*\Lx,\Ly,\Lz) node[midway,right]{\scriptsize{$L_{zout}$}};
  
  \draw[->] (\xV,0,0)--(\xV,0,\Vd);
  \draw[->] (\xV,0.125*\Ly,0)--(\xV,0.125*\Ly,0.125*\Vu+0.875*\Vd);
  \draw[->] (\xV,0.25*\Ly,0)--(\xV,0.25*\Ly,0.25*\Vu+0.75*\Vd);
  \draw[->] (\xV,0.375*\Ly,0)--(\xV,0.375*\Ly,0.375*\Vu+0.625*\Vd);
  \draw[->] (\xV,0.5*\Ly,0)--(\xV,0.5*\Ly,0.5*\Vu+0.5*\Vd);
  \draw[->] (\xV,0.625*\Ly,0)--(\xV,0.625*\Ly,0.625*\Vu+0.375*\Vd);
  \draw[->] (\xV,0.75*\Ly,0)--(\xV,0.75*\Ly,0.75*\Vu+0.25*\Vd);
  \draw[->] (\xV,0.875*\Ly,0)--(\xV,0.875*\Ly,0.875*\Vu+0.125*\Vd);
  \draw[->] (\xV,\Ly,0)--(\xV,\Ly,\Vu);
  \draw[dotted] (\xV,0,\Vd)--(\xV,\Ly,\Vu) node[midway,right]{\scriptsize{$\mathbf{U_{\infty}}$}};
  
  \draw[->] (\xVl,0,0)--(\xVl,0,\Vd);
  \draw[->] (\xVl,0.125*\Ly,0)--(\xVl,0.125*\Ly,0.125*\Vu+0.875*\Vd);
  \draw[->] (\xVl,0.25*\Ly,0)--(\xVl,0.25*\Ly,0.25*\Vu+0.75*\Vd);
  \draw[->] (\xVl,0.375*\Ly,0)--(\xVl,0.375*\Ly,0.375*\Vu+0.625*\Vd);
  \draw[->] (\xVl,0.5*\Ly,0)--(\xVl,0.5*\Ly,0.5*\Vu+0.5*\Vd);
  \draw[->] (\xVl,0.625*\Ly,0)--(\xVl,0.625*\Ly,0.625*\Vu+0.375*\Vd);
  \draw[->] (\xVl,0.75*\Ly,0)--(\xVl,0.75*\Ly,0.75*\Vu+0.25*\Vd);
  \draw[->] (\xVl,0.875*\Ly,0)--(\xVl,0.875*\Ly,0.875*\Vu+0.125*\Vd);
  \draw[->] (\xVl,\Ly,0)--(\xVl,\Ly,\Vu);
  \draw[dotted] (\xVl,0,\Vd)--(\xVl,\Ly,\Vu);
  
\end{tikzpicture}
  \caption{Reference system and schematic view of the computational domain. The center of the turbine/wing is equidistant to the upper, lower and lateral boundary conditions. Distances to the inflow ($L_{zin}$) and to the outflow ($L_{zout}$) may be different. The center of the turbine/wing is located at the origin of the coordinate system.}
  \label{fig:domain}
\end{figure}
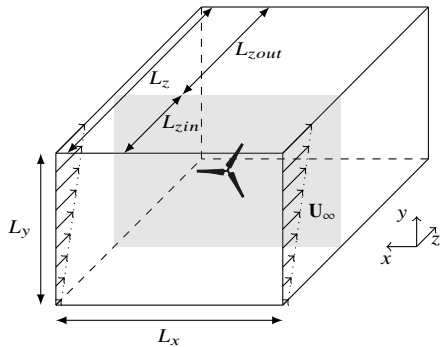

A schematic representation of the computational domain can be seen in figure \ref{fig:domain}. Dirichlet boundary conditions are used for the inflow, upper, lower and lateral boundary conditions. The natural outflow boundary condition is imposed at the outlet.

\subsection{Lifting line method} \label{sec:metLL}

The non-linear iterative lifting line method of Section \ref{sec:NLLL} was implemented considering horseshoe vortices aligned with the $z$-direction. The same discretization of the actuator line method is used. Since the lifting line method is much faster to run, a stricter convergence criterion is used. The iteration is considered converged when the difference of circulation at every control point is lower than $10^{-8}$ of the mean absolute value of circulation:
\begin{equation}
    \frac{\max\limits_{1 \leq j \leq N} \left(|\Gamma_j^{new} - \Gamma_j^{old}| \right) }{\frac{1}{N}\sum_{j=1}^N|\Gamma_j^{new}|} < 10^{-8} .
\end{equation}

\section{Results} \label{sec:results}

\subsection{Comparison with the lifting line method} \label{sec:res_wing}

The results of the simulation of the actuator line method (ALM) with smearing correction were compared with the results of a non-linear iterative lifting line for a planar straight wing with constant chord in uniform flow. Because the flow has reached the steady state, no difference was expected between the iterative and the direct method for this case, which is confirmed by the results. Nevertheless, we present the results for both strategies for completeness.

The parameters of the geometry and simulation are presented in table \ref{tab:param_LLACLstraight}. Except where explicitly stated, all values are non-dimensionalized by the span ($R$), the velocity at infinity ($U_{ref}=U_z$) and the density ($\rho$). An ideal airfoil without drag and with lift coefficient $C_l = 2 \pi \alpha$ is chosen (considering the standard non-dimensionalization of $C_l$ using the chord). The angle of attack $\alpha_g=1/(2\pi)$ is such that a lift coefficient of $C_l(\alpha_g)=1$ and a circulation of $\Gamma_0/(R U_z)=0.05$ would be expected for a two-dimensional simulation (case with zero induced velocity).

\begin{table}
    \centering
    \begin{tabular}{c c c c c c}
        Aspect ratio, $R/c$ & $\alpha_g$ (rad) & $\alpha_g$ (deg) & $(L_x/R,L_y/R,L_z/R)$ & $L_{zin}/R$ \\
        $10$ & $1/(2\pi)$ & $9.189$ & $(12,12,12)$ & $6$
    \end{tabular}
    \caption{Parameters of the simulation of the wing with constant chord in uniform flow.}
    \label{tab:param_LLACLstraight}
\end{table}

For the ALM simulations, the Reynolds number based on the chord is $Re_c = c U_z/\nu = 10^4$ (where $\nu$ is the kinematic viscosity). The Reynolds number is not applicable to the lifting line method.

The average grid spacing in the region of the actuator line is $\Delta x=R/56$. The results for two values of smearing parameter are compared: $\varepsilon = 3.5 \Delta x = R/16$ and $\varepsilon = 7 \Delta x = R/8$. The lower chosen value is higher than the usual minimum ($2 \Delta x \leq \varepsilon \leq 3 \Delta x$), but it provides a better reference for validation because it allows better resolution of the vortex core, a choice also adopted by~\citet{meyer2019vortex}. It is also of practical interest: $\varepsilon = 3.5 \Delta x$ has been used in vortex stability studies~\citep{kleine2022stability}, based on parametric studies that showed that a larger smearing parameter creates lower numerical oscillations~\citep{kleusberg2019parametric}. The last value, $\varepsilon = 7 \Delta x$, is only included as a reference, in order to evaluate the correction for a very large kernel, but large kernels are not usually employed in practice (as far as we are aware). The case of $\varepsilon = 2 \Delta x = R/28$ is discussed in Appendix~\ref{app:grid}, where the effect of grid resolution is analyzed.

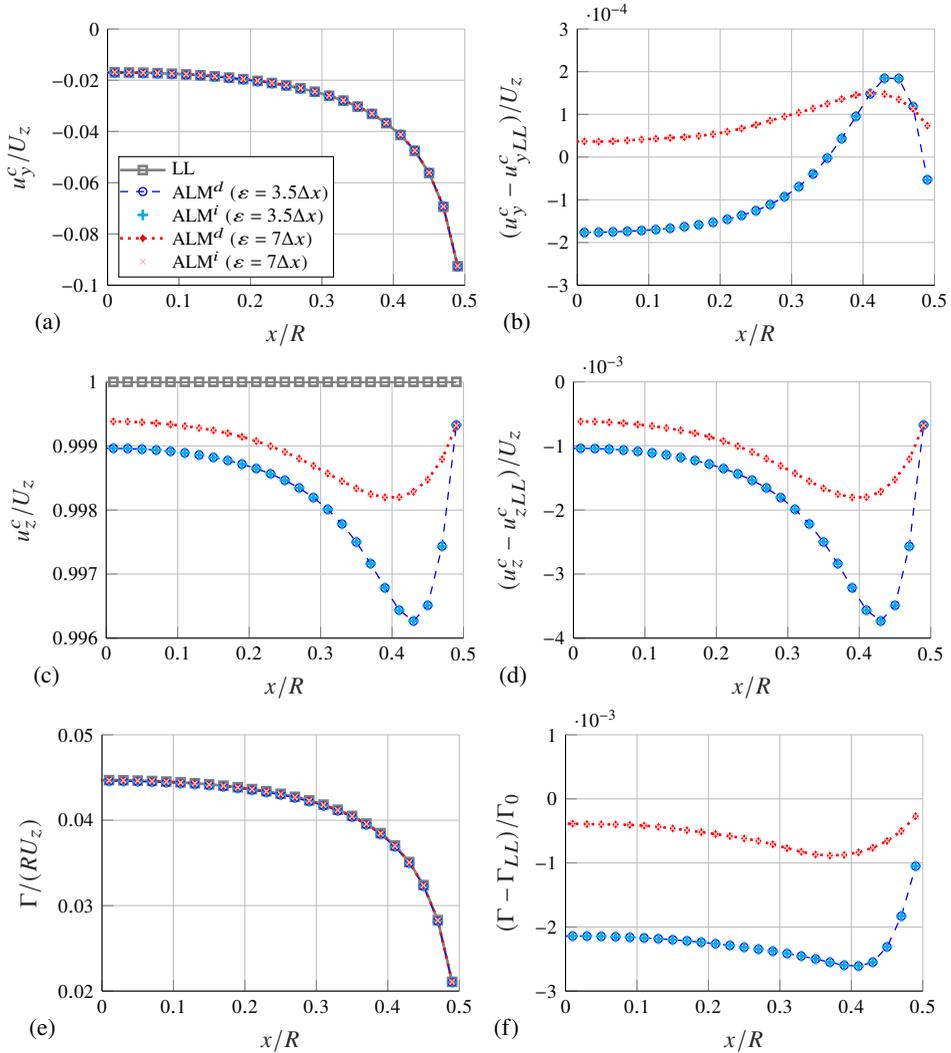
\begin{figure}
  \centering
%
%
\definecolor{mycolor1}{rgb}{1.00000,1.00000,0.00000}%
\definecolor{mycolor2}{rgb}{0.00000,1.00000,1.00000}%
\begin{tikzpicture}

\begin{axis}[%
width=0.35\textwidth,
height=0.25\textwidth,
scale only axis,
xmin=0.0,
xmax=0.5,
xtick={0.0, 0.1, 0.2, 0.3, 0.4, 0.5},
xlabel style={font=\color{white!15!black}},
xlabel={$x/R$},
ymin=-0.1,
ymax=0,
ylabel style={font=\color{white!15!black}},
ylabel={$u_y^c/U_z$},
yticklabel style={
        /pgf/number format/fixed,
        /pgf/number format/precision=3
},
scaled y ticks=false,
xmajorgrids,
ymajorgrids,
tick label style={font=\scriptsize},
axis background/.style={fill=white},
axis x line*=bottom,
axis y line*=left,
legend style={legend cell align=left, align=left, font=\scriptsize, row sep=-0.1cm, inner sep=0pt},
legend pos=south west
]
\addplot [color=gray, line width=1.0pt, mark size=1.5pt, mark=square, mark options={solid, gray}]
  table[row sep=crcr]{%
-0.49	-0.0925870209333405\\
-0.47	-0.0693908787624665\\
-0.45	-0.056255599662689\\
-0.43	-0.0475888881957751\\
-0.41	-0.0414029726521644\\
-0.39	-0.0367652904276586\\
-0.37	-0.0331680223644067\\
-0.35	-0.0303071796442326\\
-0.33	-0.0279883520990819\\
-0.31	-0.0260809577524658\\
-0.29	-0.0244938967178062\\
-0.27	-0.0231616672961077\\
-0.25	-0.0220360080591876\\
-0.23	-0.021080642539291\\
-0.21	-0.0202678544196067\\
-0.19	-0.019576188115161\\
-0.17	-0.0189888658666816\\
-0.15	-0.0184926750658085\\
-0.13	-0.018077172599072\\
-0.11	-0.0177341082410082\\
-0.09	-0.0174570029865848\\
-0.07	-0.0172408395803871\\
-0.05	-0.0170818363614657\\
-0.03	-0.0169772847904103\\
-0.01	-0.0169254373905926\\
0.01	-0.0169254373905931\\
0.03	-0.0169772847904105\\
0.05	-0.0170818363614658\\
0.0700000000000001	-0.0172408395803864\\
0.09	-0.0174570029865845\\
0.11	-0.0177341082410081\\
0.13	-0.0180771725990724\\
0.15	-0.0184926750658089\\
0.17	-0.0189888658666816\\
0.19	-0.0195761881151601\\
0.21	-0.0202678544196065\\
0.23	-0.0210806425392912\\
0.25	-0.0220360080591879\\
0.27	-0.0231616672961081\\
0.29	-0.0244938967178065\\
0.31	-0.0260809577524667\\
0.33	-0.0279883520990814\\
0.35	-0.0303071796442318\\
0.37	-0.0331680223644066\\
0.39	-0.0367652904276586\\
0.41	-0.0414029726521647\\
0.43	-0.0475888881957757\\
0.45	-0.0562555996626887\\
0.47	-0.0693908787624662\\
0.49	-0.0925870209333404\\
};
\addlegendentry{LL}

\addplot [color=blue, dashed, line width=0.5pt, mark size=1.5pt, mark=o, mark options={solid, blue}]
  table[row sep=crcr]{%
-0.49	-0.0926399628496373\\
-0.47	-0.0692729210901141\\
-0.45	-0.05607157282118\\
-0.43	-0.0474043128489364\\
-0.41	-0.0412548028826863\\
-0.39	-0.0366701943484689\\
-0.37	-0.0331249413268098\\
-0.35	-0.0303094519903163\\
-0.33	-0.0280279532631833\\
-0.31	-0.0261502225763328\\
-0.29	-0.0245866494209166\\
-0.27	-0.0232729470595629\\
-0.25	-0.022161552828071\\
-0.23	-0.0212172770767773\\
-0.21	-0.0204135212720414\\
-0.19	-0.0197290127988029\\
-0.17	-0.0191474179765746\\
-0.15	-0.0186557978876427\\
-0.13	-0.0182439563427135\\
-0.11	-0.0179038317528572\\
-0.09	-0.0176290588600361\\
-0.07	-0.0174146709411361\\
-0.05	-0.0172569660440704\\
-0.03	-0.0171532788953324\\
-0.01	-0.0171018473138927\\
0.01	-0.0171018462403649\\
0.03	-0.0171532768523142\\
0.05	-0.0172569619900255\\
0.07	-0.0174146658304909\\
0.09	-0.0176290560051333\\
0.11	-0.0179038302301311\\
0.13	-0.0182439499829858\\
0.15	-0.0186557986316945\\
0.17	-0.0191474160046604\\
0.19	-0.019729009991626\\
0.21	-0.020413521049607\\
0.23	-0.0212172787149726\\
0.25	-0.0221615556916193\\
0.27	-0.0232729529319505\\
0.29	-0.0245866560285746\\
0.31	-0.0261502302451681\\
0.33	-0.0280279598589182\\
0.35	-0.0303094563137114\\
0.37	-0.0331249487869997\\
0.39	-0.0366701977210746\\
0.41	-0.0412548055284462\\
0.43	-0.0474043159192413\\
0.45	-0.0560715750204841\\
0.47	-0.0692729218148137\\
0.49	-0.0926399625462617\\
};
\addlegendentry{ALM$^d$ ($\varepsilon = 3.5 \Delta x$)}

\addplot [color=cyan, line width=1.0pt, only marks, mark size=2.0pt, mark=+, mark options={solid, cyan}]
  table[row sep=crcr]{%
-0.49	-0.0926401164680461\\
-0.47	-0.0692731198987469\\
-0.45	-0.0560717119005029\\
-0.43	-0.0474043677975882\\
-0.41	-0.0412547860939118\\
-0.39	-0.0366701431604364\\
-0.37	-0.0331248881401518\\
-0.35	-0.0303094123254846\\
-0.33	-0.0280279283902998\\
-0.31	-0.0261502089362342\\
-0.29	-0.0245866423887171\\
-0.27	-0.023272946004795\\
-0.25	-0.0221615562260162\\
-0.23	-0.0212172755379488\\
-0.21	-0.0204135151994997\\
-0.19	-0.0197290031887127\\
-0.17	-0.0191474085461174\\
-0.15	-0.0186557908352192\\
-0.13	-0.0182439462613996\\
-0.11	-0.0179038241930319\\
-0.09	-0.0176290501580599\\
-0.07	-0.0174146614545484\\
-0.05	-0.0172569580965725\\
-0.03	-0.0171532735747702\\
-0.01	-0.0171018433932162\\
0.01	-0.017101843793039\\
0.03	-0.0171532763797481\\
0.05	-0.0172569632237636\\
0.07	-0.0174146689342202\\
0.09	-0.0176290601281641\\
0.11	-0.0179038355665048\\
0.13	-0.0182439585686839\\
0.15	-0.0186558041517276\\
0.17	-0.0191474216271923\\
0.19	-0.0197290165279252\\
0.21	-0.0204135279176566\\
0.23	-0.0212172855436136\\
0.25	-0.0221615614845993\\
0.27	-0.0232729507279545\\
0.29	-0.0245866459631545\\
0.31	-0.0261502120198691\\
0.33	-0.0280279321317996\\
0.35	-0.0303094163624458\\
0.37	-0.0331248919766133\\
0.39	-0.0366701468126204\\
0.41	-0.0412547896959533\\
0.43	-0.047404371304487\\
0.45	-0.0560717144544435\\
0.47	-0.0692731219984206\\
0.49	-0.0926401183659905\\
};
\addlegendentry{ALM$^i$ ($\varepsilon = 3.5 \Delta x$)}

\addplot [color=red, dotted, line width=1.0pt, mark size=0.75pt, mark=diamond, mark options={solid, red}]
  table[row sep=crcr]{%
-0.49	-0.0925134466726856\\
-0.47	-0.0692775606396011\\
-0.45	-0.0561208027253486\\
-0.43	-0.0474414136354333\\
-0.41	-0.0412526360446312\\
-0.39	-0.036620050638865\\
-0.37	-0.0330321888893847\\
-0.35	-0.030182468384102\\
-0.33	-0.0278744722247284\\
-0.31	-0.0259769864747114\\
-0.29	-0.024399122773907\\
-0.27	-0.0230762407567972\\
-0.25	-0.0219607742014483\\
-0.23	-0.0210138235978915\\
-0.21	-0.0202084591862245\\
-0.19	-0.0195227372613571\\
-0.17	-0.0189395259231292\\
-0.15	-0.0184459280524193\\
-0.13	-0.0180323028835572\\
-0.11	-0.0176910766021929\\
-0.09	-0.0174159757954401\\
-0.07	-0.0172019245330358\\
-0.05	-0.0170447731044712\\
-0.03	-0.0169412179498227\\
-0.01	-0.0168886607025865\\
0.01	-0.0168886600541682\\
0.03	-0.0169412136327629\\
0.05	-0.0170447711819101\\
0.07	-0.0172019248472423\\
0.09	-0.0174159767679247\\
0.11	-0.0176910816871676\\
0.13	-0.018032313636208\\
0.15	-0.0184459383984052\\
0.17	-0.0189395394143951\\
0.19	-0.0195227508153105\\
0.21	-0.0202084707462373\\
0.23	-0.0210138372508306\\
0.25	-0.0219607946424383\\
0.27	-0.0230762634478468\\
0.29	-0.0243991477059922\\
0.31	-0.0259770131202511\\
0.33	-0.0278745015488875\\
0.35	-0.0301825003212058\\
0.37	-0.0330322210006926\\
0.39	-0.0366200840099151\\
0.41	-0.0412526656520037\\
0.43	-0.0474414364661295\\
0.45	-0.0561208176509505\\
0.47	-0.0692775677605057\\
0.49	-0.0925134463127865\\
};
\addlegendentry{ALM$^d$ ($\varepsilon = 7 \Delta x$)}

\addplot [color=pink, line width=0.5pt, only marks, mark size=1.5pt, mark=x, mark options={solid, pink}]
  table[row sep=crcr]{%
-0.49	-0.092513465960044\\
-0.47	-0.0692775846508255\\
-0.45	-0.0561208357551724\\
-0.43	-0.0474414542200039\\
-0.41	-0.0412526811824748\\
-0.39	-0.0366200964468988\\
-0.37	-0.0330322164663706\\
-0.35	-0.0301824869107151\\
-0.33	-0.0278744785439993\\
-0.31	-0.0259769874578648\\
-0.29	-0.0243991278417173\\
-0.27	-0.0230762504553147\\
-0.25	-0.0219607705680457\\
-0.23	-0.0210138038856765\\
-0.21	-0.0202084335247809\\
-0.19	-0.0195227155589873\\
-0.17	-0.018939512184015\\
-0.15	-0.0184459248648592\\
-0.13	-0.0180323067901936\\
-0.11	-0.0176910693419412\\
-0.09	-0.0174159524973329\\
-0.07	-0.0172018838834044\\
-0.05	-0.0170447180877034\\
-0.03	-0.0169411769553421\\
-0.01	-0.0168886723662887\\
0.01	-0.0168886772481796\\
0.03	-0.0169412008868476\\
0.05	-0.0170447567125644\\
0.07	-0.0172019348136526\\
0.09	-0.0174160129685424\\
0.11	-0.0176911301134202\\
0.13	-0.0180323572812357\\
0.15	-0.0184459695061038\\
0.17	-0.0189395443723982\\
0.19	-0.0195227313777858\\
0.21	-0.020208438125329\\
0.23	-0.0210138039586305\\
0.25	-0.0219607697640517\\
0.27	-0.0230762496552227\\
0.29	-0.0243991286572209\\
0.31	-0.0259769860402495\\
0.33	-0.0278744706224887\\
0.35	-0.0301824801874584\\
0.37	-0.0330322180651905\\
0.39	-0.0366200877356405\\
0.41	-0.0412526755233387\\
0.43	-0.0474414513453692\\
0.45	-0.056120832567887\\
0.47	-0.0692775813092914\\
0.49	-0.0925134640908617\\
};
\addlegendentry{ALM$^i$ ($\varepsilon = 7 \Delta x$)}

\end{axis}
\node [left] at (-0.5,-0.5) {(a)};
\end{tikzpicture}%
%
%
\definecolor{mycolor1}{rgb}{1.00000,1.00000,0.00000}%
\definecolor{mycolor2}{rgb}{0.00000,1.00000,1.00000}%
\begin{tikzpicture}

\begin{axis}[%
width=0.35\textwidth,
height=0.25\textwidth,
scale only axis,
xmin=0.0,
xmax=0.5,
xtick={0.0, 0.1, 0.2, 0.3, 0.4, 0.5},
xlabel style={font=\color{white!15!black}},
xlabel={$x/R$},
ymin=-0.0003,
ymax=0.0003,
ytick={-0.0003, -0.0002,..., 0.0003},
ylabel style={font=\color{white!15!black}},
ylabel={$(u^c_y - u^c_{yLL})/U_z$},
xmajorgrids,
ymajorgrids,
tick label style={font=\scriptsize},
axis background/.style={fill=white},
axis x line*=bottom,
axis y line*=left,
]

\addplot [color=blue, dashed, line width=0.5pt, mark size=1.5pt, mark=o, mark options={solid, blue}]
  table[row sep=crcr]{%
-0.49	-5.29419162967754e-05\\
-0.47	0.00011795767235237\\
-0.45	0.000184026841509033\\
-0.43	0.000184575346838738\\
-0.41	0.000148169769478131\\
-0.39	9.50960791896924e-05\\
-0.37	4.30810375968843e-05\\
-0.35	-2.27234608367669e-06\\
-0.33	-3.96011641013681e-05\\
-0.31	-6.92648238669949e-05\\
-0.29	-9.27527031104186e-05\\
-0.27	-0.000111279763455183\\
-0.25	-0.000125544768883396\\
-0.23	-0.000136634537486309\\
-0.21	-0.000145666852434695\\
-0.19	-0.000152824683641927\\
-0.17	-0.000158552109892969\\
-0.15	-0.000163122821834165\\
-0.13	-0.000166783743641478\\
-0.11	-0.00016972351184898\\
-0.09	-0.000172055873451333\\
-0.07	-0.000173831360749006\\
-0.05	-0.000175129682604747\\
-0.03	-0.000175994104922108\\
-0.01	-0.000176409923300119\\
0.01	-0.000176408849771825\\
0.03	-0.000175992061903663\\
0.05	-0.000175125628559696\\
0.07	-0.00017382625010455\\
0.09	-0.000172053018548751\\
0.11	-0.000169721989123049\\
0.13	-0.000166777383913366\\
0.15	-0.000163123565885638\\
0.17	-0.000158550137978837\\
0.19	-0.000152821876465932\\
0.21	-0.00014566663000054\\
0.23	-0.000136636175681451\\
0.25	-0.000125547632431389\\
0.27	-0.000111285635842364\\
0.29	-9.27593107680792e-05\\
0.31	-6.9272492701427e-05\\
0.33	-3.96077598368086e-05\\
0.35	-2.2766694795541e-06\\
0.37	4.30735774069393e-05\\
0.39	9.50927065839699e-05\\
0.41	0.00014816712371845\\
0.43	0.000184572276534394\\
0.45	0.000184024642204565\\
0.47	0.000117956947652542\\
0.49	-5.29416129213117e-05\\
};
\addlegendentry{ALM$^d$ ($\varepsilon = 3.5 \Delta x$)}

\addplot [color=cyan, line width=1.0pt, only marks, mark size=2.0pt, mark=+, mark options={solid, cyan}]
  table[row sep=crcr]{%
-0.49	-5.30955347055745e-05\\
-0.47	0.000117758863719564\\
-0.45	0.000183887762186134\\
-0.43	0.000184520398186935\\
-0.41	0.000148186558252633\\
-0.39	9.51472672221926e-05\\
-0.37	4.31342242548857e-05\\
-0.35	-2.23268125197629e-06\\
-0.33	-3.95762912178688e-05\\
-0.31	-6.92511837683922e-05\\
-0.29	-9.27456709109188e-05\\
-0.27	-0.000111278708687285\\
-0.25	-0.000125548166828596\\
-0.23	-0.00013663299865781\\
-0.21	-0.000145660779892996\\
-0.19	-0.000152815073551728\\
-0.17	-0.000158542679435769\\
-0.15	-0.000163115769410666\\
-0.13	-0.00016677366232758\\
-0.11	-0.000169715952023683\\
-0.09	-0.000172047171475131\\
-0.07	-0.000173821874161307\\
-0.05	-0.000175121735106847\\
-0.03	-0.000175988784359909\\
-0.01	-0.000176406002623621\\
0.01	-0.000176406402445928\\
0.03	-0.000175991589337562\\
0.05	-0.000175126862297797\\
0.07	-0.000173829353833848\\
0.09	-0.00017205714157955\\
0.11	-0.000169727325496749\\
0.13	-0.000166785969611467\\
0.15	-0.000163129085918739\\
0.17	-0.000158555760510735\\
0.19	-0.000152828412765132\\
0.21	-0.000145673498050142\\
0.23	-0.000136643004322451\\
0.25	-0.000125553425411391\\
0.27	-0.000111283431846364\\
0.29	-9.27492453479811e-05\\
0.31	-6.92542674024244e-05\\
0.33	-3.9580032718211e-05\\
0.35	-2.23671821395627e-06\\
0.37	4.31303877933414e-05\\
0.39	9.51436150381652e-05\\
0.41	0.00014818295621135\\
0.43	0.000184516891288691\\
0.45	0.000183885208245166\\
0.47	0.000117756764045643\\
0.49	-5.30974326501121e-05\\
};
\addlegendentry{ALM$^i$ ($\varepsilon = 3.5 \Delta x$)}

\addplot [color=red, dotted, line width=1.0pt, mark size=0.75pt, mark=diamond, mark options={solid, red}]
  table[row sep=crcr]{%
-0.49	7.35742606549211e-05\\
-0.47	0.000113318122865372\\
-0.45	0.000134796937340434\\
-0.43	0.000147474560341837\\
-0.41	0.000150336607533233\\
-0.39	0.000145239788793598\\
-0.37	0.000135833475021982\\
-0.35	0.000124711260130621\\
-0.33	0.000113879874353531\\
-0.31	0.000103971277754408\\
-0.29	9.47739438991804e-05\\
-0.27	8.54265393105159e-05\\
-0.25	7.52338577393041e-05\\
-0.23	6.68189413994894e-05\\
-0.21	5.93952333822062e-05\\
-0.19	5.34508538038714e-05\\
-0.17	4.93399435524296e-05\\
-0.15	4.67470133892374e-05\\
-0.13	4.48697155148205e-05\\
-0.11	4.3031638815319e-05\\
-0.09	4.10271911446686e-05\\
-0.07	3.89150473512935e-05\\
-0.05	3.70632569944504e-05\\
-0.03	3.60668405875904e-05\\
-0.01	3.67766880060814e-05\\
0.01	3.67773364248723e-05\\
0.03	3.60711576476366e-05\\
0.05	3.70651795557064e-05\\
0.07	3.89147331440499e-05\\
0.09	4.10262186598502e-05\\
0.11	4.30265538404508e-05\\
0.13	4.48589628644323e-05\\
0.15	4.67366674036622e-05\\
0.17	4.93264522864641e-05\\
0.19	5.34372998495652e-05\\
0.21	5.93836733691587e-05\\
0.23	6.68052884605502e-05\\
0.25	7.52134167496116e-05\\
0.27	8.5403848261334e-05\\
0.29	9.47490118143185e-05\\
0.31	0.000103944632215573\\
0.33	0.000113850550193889\\
0.35	0.000124679323026046\\
0.37	0.000135801363714043\\
0.39	0.000145206417743467\\
0.41	0.00015030700016095\\
0.43	0.000147451729646191\\
0.45	0.000134782011738166\\
0.47	0.000113311001960548\\
0.49	7.3574620553879e-05\\
};
\addlegendentry{ALM$^d$ ($\varepsilon = 7 \Delta x$)}

\addplot [color=pink, line width=0.5pt, only marks, mark size=1.5pt, mark=x, mark options={solid, pink}]
  table[row sep=crcr]{%
-0.49	7.35549732965163e-05\\
-0.47	0.000113294111640974\\
-0.45	0.000134763907516634\\
-0.43	0.000147433975771237\\
-0.41	0.000150291469689631\\
-0.39	0.000145193980759795\\
-0.37	0.000135805898036086\\
-0.35	0.000124692733517524\\
-0.33	0.00011387355508263\\
-0.31	0.000103970294601005\\
-0.29	9.47688760888805e-05\\
-0.27	8.54168407930164e-05\\
-0.25	7.52374911419072e-05\\
-0.23	6.6838653614492e-05\\
-0.21	5.94208948258047e-05\\
-0.19	5.34725561736719e-05\\
-0.17	4.93536826666301e-05\\
-0.15	4.67502009493352e-05\\
-0.13	4.48658088784207e-05\\
-0.11	4.30388990670175e-05\\
-0.09	4.1050489251869e-05\\
-0.07	3.8955696982694e-05\\
-0.05	3.71182737622516e-05\\
-0.03	3.61078350681922e-05\\
-0.01	3.67650243038795e-05\\
0.01	3.67601424134736e-05\\
0.03	3.6083903562939e-05\\
0.05	3.70796489014037e-05\\
0.07	3.89047667337522e-05\\
0.09	4.09900180421488e-05\\
0.11	4.29781275878525e-05\\
0.13	4.48153178367328e-05\\
0.15	4.67055597050611e-05\\
0.17	4.93214942833622e-05\\
0.19	5.34567373742662e-05\\
0.21	5.94162942774597e-05\\
0.23	6.68385806606511e-05\\
0.25	7.52382951362124e-05\\
0.27	8.54176408854358e-05\\
0.29	9.47680605856219e-05\\
0.31	0.000103971712217175\\
0.33	0.000113881476592689\\
0.35	0.000124699456773444\\
0.37	0.000135804299216141\\
0.39	0.000145202692018065\\
0.41	0.000150297128825951\\
0.43	0.000147436850406489\\
0.45	0.000134767094801667\\
0.47	0.000113297453174843\\
0.49	7.35568424786864e-05\\
};
\addlegendentry{ALM$^i$ ($\varepsilon = 7 \Delta x$)}

\legend{}
\end{axis}
\node [left] at (-0.5,-0.5) {(b)};
\end{tikzpicture}%
%
%
\definecolor{mycolor1}{rgb}{1.00000,1.00000,0.00000}%
\definecolor{mycolor2}{rgb}{0.00000,1.00000,1.00000}%
\begin{tikzpicture}

\begin{axis}[%
width=0.35\textwidth,
height=0.25\textwidth,
scale only axis,
xmin=0.0,
xmax=0.5,
xtick={0.0, 0.1, 0.2, 0.3, 0.4, 0.5},
xlabel style={font=\color{white!15!black}},
xlabel={$x/R$},
ymin=0.996,
ymax=1,
ylabel style={font=\color{white!15!black}},
ylabel={$u_z^c/U_z$},
yticklabel style={
        /pgf/number format/fixed,
        /pgf/number format/precision=3
},
scaled y ticks=false,
xmajorgrids,
ymajorgrids,
tick label style={font=\scriptsize},
axis background/.style={fill=white},
axis x line*=bottom,
axis y line*=left,
legend style={legend cell align=left, align=left, font=\scriptsize, row sep=-0.1cm},
legend pos=south west
]
\addplot [color=gray, line width=1.0pt, mark size=1.5pt, mark=square, mark options={solid, gray}]
  table[row sep=crcr]{%
-0.49	1\\
-0.47	1\\
-0.45	1\\
-0.43	1\\
-0.41	1\\
-0.39	1\\
-0.37	1\\
-0.35	1\\
-0.33	1\\
-0.31	1\\
-0.29	1\\
-0.27	1\\
-0.25	1\\
-0.23	1\\
-0.21	1\\
-0.19	1\\
-0.17	1\\
-0.15	1\\
-0.13	1\\
-0.11	1\\
-0.09	1\\
-0.07	1\\
-0.05	1\\
-0.03	1\\
-0.01	1\\
0.01	1\\
0.03	1\\
0.05	1\\
0.0700000000000001	1\\
0.09	1\\
0.11	1\\
0.13	1\\
0.15	1\\
0.17	1\\
0.19	1\\
0.21	1\\
0.23	1\\
0.25	1\\
0.27	1\\
0.29	1\\
0.31	1\\
0.33	1\\
0.35	1\\
0.37	1\\
0.39	1\\
0.41	1\\
0.43	1\\
0.45	1\\
0.47	1\\
0.49	1\\
};
\addlegendentry{LL}

\addplot [color=blue, dashed, line width=0.5pt, mark size=1.5pt, mark=o, mark options={solid, blue}]
  table[row sep=crcr]{%
-0.49	0.999326742747289\\
-0.47	0.997434378497135\\
-0.45	0.996511663035622\\
-0.43	0.996266260682468\\
-0.41	0.996436071651367\\
-0.39	0.996785395413477\\
-0.37	0.997162368363097\\
-0.35	0.997500429980152\\
-0.33	0.997782133897739\\
-0.31	0.998010515200853\\
-0.29	0.998194921844666\\
-0.27	0.998344411655561\\
-0.25	0.99846482405222\\
-0.23	0.998564261956546\\
-0.21	0.998647359157389\\
-0.19	0.998716148870283\\
-0.17	0.99877325173349\\
-0.15	0.99882041462281\\
-0.13	0.998859172078037\\
-0.11	0.998890717381133\\
-0.09	0.998915930805182\\
-0.07	0.998935290492729\\
-0.05	0.998949401092711\\
-0.03	0.998958680045278\\
-0.01	0.998963282522259\\
0.01	0.998963278604793\\
0.03	0.99895868174179\\
0.05	0.998949390618958\\
0.07	0.998935268270086\\
0.09	0.998915921242693\\
0.11	0.998890717851614\\
0.13	0.998859142376726\\
0.15	0.998820424526969\\
0.17	0.998773237143175\\
0.19	0.998716128529259\\
0.21	0.998647353596756\\
0.23	0.998564255913057\\
0.25	0.998464815418356\\
0.27	0.998344413846537\\
0.29	0.998194925798657\\
0.31	0.998010530270584\\
0.33	0.99778214523782\\
0.35	0.997500424028467\\
0.37	0.99716238211855\\
0.39	0.996785390393375\\
0.41	0.996436073408576\\
0.43	0.996266271275172\\
0.45	0.996511669536913\\
0.47	0.997434378717617\\
0.49	0.999326741943547\\
};
\addlegendentry{ALM$^d$ ($\varepsilon = 3.5 \Delta x$)}

\addplot [color=cyan, line width=1.0pt, only marks, mark size=2.0pt, mark=+, mark options={solid, cyan}]
  table[row sep=crcr]{%
-0.49	0.999326744136666\\
-0.47	0.997434373669498\\
-0.45	0.996511664176326\\
-0.43	0.99626627161504\\
-0.41	0.996436082546511\\
-0.39	0.996785399173766\\
-0.37	0.997162383925163\\
-0.35	0.997500427614807\\
-0.33	0.997782140044989\\
-0.31	0.998010522067997\\
-0.29	0.998194918028459\\
-0.27	0.99834440862978\\
-0.25	0.998464821868617\\
-0.23	0.998564262453947\\
-0.21	0.998647356385842\\
-0.19	0.998716135484074\\
-0.17	0.998773244400023\\
-0.15	0.998820423645003\\
-0.13	0.998859156166231\\
-0.11	0.998890719618741\\
-0.09	0.99891592753042\\
-0.07	0.998935280550935\\
-0.05	0.998949395919194\\
-0.03	0.998958682271858\\
-0.01	0.998963283163309\\
0.01	0.998963273153252\\
0.03	0.998958670633283\\
0.05	0.998949383999175\\
0.07	0.998935270097898\\
0.09	0.998915924581152\\
0.11	0.998890717658421\\
0.13	0.998859148609703\\
0.15	0.998820420852711\\
0.17	0.998773237058864\\
0.19	0.998716131646572\\
0.21	0.998647357993011\\
0.23	0.998564262582424\\
0.25	0.998464821805356\\
0.27	0.998344406765033\\
0.29	0.998194913281068\\
0.31	0.998010517163762\\
0.33	0.997782136601098\\
0.35	0.997500420564296\\
0.37	0.997162376965232\\
0.39	0.996785393626498\\
0.41	0.996436078321765\\
0.43	0.996266271955494\\
0.45	0.996511666241073\\
0.47	0.997434377891451\\
0.49	0.999326751018131\\
};
\addlegendentry{ALM$^i$ ($\varepsilon = 3.5 \Delta x$)}

\addplot [color=red, dotted, line width=1.0pt, mark size=0.75pt, mark=diamond, mark options={solid, red}]
  table[row sep=crcr]{%
-0.49	0.999320815240245\\
-0.47	0.998792655380176\\
-0.45	0.998474025202664\\
-0.43	0.998285734212428\\
-0.41	0.998200671005867\\
-0.39	0.998196440511222\\
-0.37	0.998248519902661\\
-0.35	0.998337820075167\\
-0.33	0.998449773022972\\
-0.31	0.998570861544983\\
-0.29	0.998690044887821\\
-0.27	0.998801872190123\\
-0.25	0.998906173476686\\
-0.23	0.998996762778848\\
-0.21	0.999075430055988\\
-0.19	0.999142789649552\\
-0.17	0.999198572373675\\
-0.15	0.999243418053371\\
-0.13	0.999280064926216\\
-0.11	0.999310250723702\\
-0.09	0.999335144415434\\
-0.07	0.999354771971945\\
-0.05	0.999369315616205\\
-0.03	0.999379025746172\\
-0.01	0.999381181449309\\
0.01	0.999381188242814\\
0.03	0.999379030934095\\
0.05	0.999369345604124\\
0.07	0.999354816839802\\
0.09	0.999335193038642\\
0.11	0.999310309511599\\
0.13	0.999280135307597\\
0.15	0.999243458019798\\
0.17	0.999198595706248\\
0.19	0.999142790619503\\
0.21	0.999075411294635\\
0.23	0.998996744819676\\
0.25	0.998906174004517\\
0.27	0.99880186040771\\
0.29	0.998690012278773\\
0.31	0.998570825281577\\
0.33	0.998449746324036\\
0.35	0.998337800783032\\
0.37	0.998248494556048\\
0.39	0.998196433730376\\
0.41	0.998200675525368\\
0.43	0.998285741520466\\
0.45	0.998474030258709\\
0.47	0.998792665345125\\
0.49	0.999320827468525\\
};
\addlegendentry{ALM$^d$ ($\varepsilon = 7 \Delta x$)}

\addplot [color=pink, line width=0.5pt, only marks, mark size=1.5pt, mark=x, mark options={solid, pink}]
  table[row sep=crcr]{%
-0.49	0.99932083245318\\
-0.47	0.998792671768662\\
-0.45	0.998474037900359\\
-0.43	0.998285747969454\\
-0.41	0.998200694383393\\
-0.39	0.998196472192554\\
-0.37	0.998248484323566\\
-0.35	0.998337808116093\\
-0.33	0.998449761017967\\
-0.31	0.998570830931451\\
-0.29	0.99869001560245\\
-0.27	0.998801867643004\\
-0.25	0.998906148872384\\
-0.23	0.998996749505106\\
-0.21	0.999075424817092\\
-0.19	0.999142789947503\\
-0.17	0.999198573965159\\
-0.15	0.999243429394914\\
-0.13	0.999280117226453\\
-0.11	0.999310278574635\\
-0.09	0.999335185538045\\
-0.07	0.999354829339418\\
-0.05	0.999369359846515\\
-0.03	0.999379038740888\\
-0.01	0.999381182917825\\
0.01	0.999381152616413\\
0.03	0.999379009160077\\
0.05	0.999369320205906\\
0.07	0.999354769028365\\
0.09	0.999335128867358\\
0.11	0.999310244594321\\
0.13	0.999280062508372\\
0.15	0.99924338111638\\
0.17	0.999198527245624\\
0.19	0.999142749266051\\
0.21	0.999075400895118\\
0.23	0.998996743460322\\
0.25	0.998906148887891\\
0.27	0.998801877201492\\
0.29	0.998690046776674\\
0.31	0.99857086590933\\
0.33	0.998449776983\\
0.35	0.998337825436332\\
0.37	0.998248540770747\\
0.39	0.99819645088732\\
0.41	0.998200702904851\\
0.43	0.998285775459846\\
0.45	0.998474054477219\\
0.47	0.998792687639055\\
0.49	0.999320849217706\\
};
\addlegendentry{ALM$^i$ ($\varepsilon = 7 \Delta x$)}

\legend{}
\end{axis}
\node [left] at (-0.5,-0.5) {(c)};
\end{tikzpicture}%
%
%
\definecolor{mycolor1}{rgb}{1.00000,1.00000,0.00000}%
\definecolor{mycolor2}{rgb}{0.00000,1.00000,1.00000}%
\begin{tikzpicture}

\begin{axis}[%
width=0.35\textwidth,
height=0.25\textwidth,
scale only axis,
xmin=0.0,
xmax=0.5,
xtick={0.0, 0.1, 0.2, 0.3, 0.4, 0.5},
xlabel style={font=\color{white!15!black}},
xlabel={$x/R$},
ymin=-0.004,
ymax=0.000,
ylabel style={font=\color{white!15!black}},
ylabel={$(u^c_z - u^c_{zLL})/U_z$},
xmajorgrids,
ymajorgrids,
tick label style={font=\scriptsize},
axis background/.style={fill=white},
axis x line*=bottom,
axis y line*=left,
]

\addplot [color=blue, dashed, line width=0.5pt, mark size=1.5pt, mark=o, mark options={solid, blue}]
  table[row sep=crcr]{%
-0.49	-0.00067325725271139\\
-0.47	-0.00256562150286466\\
-0.45	-0.00348833696437834\\
-0.43	-0.00373373931753196\\
-0.41	-0.00356392834863284\\
-0.39	-0.00321460458652334\\
-0.37	-0.0028376316369032\\
-0.35	-0.00249957001984791\\
-0.33	-0.00221786610226105\\
-0.31	-0.00198948479914685\\
-0.29	-0.00180507815533415\\
-0.27	-0.00165558834443891\\
-0.25	-0.00153517594778008\\
-0.23	-0.00143573804345365\\
-0.21	-0.001352640842611\\
-0.19	-0.00128385112971663\\
-0.17	-0.00122674826651004\\
-0.15	-0.00117958537718987\\
-0.13	-0.00114082792196302\\
-0.11	-0.00110928261886667\\
-0.09	-0.00108406919481763\\
-0.07	-0.00106470950727056\\
-0.05	-0.00105059890728942\\
-0.03	-0.00104131995472156\\
-0.01	-0.00103671747774137\\
0.01	-0.00103672139520661\\
0.03	-0.00104131825821019\\
0.05	-0.00105060938104229\\
0.07	-0.00106473172991417\\
0.09	-0.00108407875730721\\
0.11	-0.00110928214838568\\
0.13	-0.00114085762327432\\
0.15	-0.00117957547303127\\
0.17	-0.00122676285682455\\
0.19	-0.00128387147074105\\
0.21	-0.0013526464032444\\
0.23	-0.00143574408694302\\
0.25	-0.00153518458164437\\
0.27	-0.00165558615346251\\
0.29	-0.00180507420134346\\
0.31	-0.0019894697294156\\
0.33	-0.00221785476217995\\
0.35	-0.00249957597153316\\
0.37	-0.00283761788145043\\
0.39	-0.00321460960662501\\
0.41	-0.0035639265914239\\
0.43	-0.0037337287248278\\
0.45	-0.00348833046308694\\
0.47	-0.00256562128238291\\
0.49	-0.00067325805645313\\
};
\addlegendentry{ALM$^d$ ($\varepsilon = 3.5 \Delta x$)}

\addplot [color=cyan, line width=1.0pt, only marks, mark size=2.0pt, mark=+, mark options={solid, cyan}]
  table[row sep=crcr]{%
-0.49	-0.000673255863334221\\
-0.47	-0.00256562633050184\\
-0.45	-0.00348833582367369\\
-0.43	-0.00373372838495967\\
-0.41	-0.00356391745348916\\
-0.39	-0.00321460082623437\\
-0.37	-0.00283761607483657\\
-0.35	-0.00249957238519272\\
-0.33	-0.00221785995501123\\
-0.31	-0.00198947793200299\\
-0.29	-0.00180508197154083\\
-0.27	-0.00165559137021953\\
-0.25	-0.00153517813138282\\
-0.23	-0.0014357375460532\\
-0.21	-0.00135264361415777\\
-0.19	-0.00128386451592566\\
-0.17	-0.0012267555999772\\
-0.15	-0.00117957635499688\\
-0.13	-0.00114084383376878\\
-0.11	-0.00110928038125935\\
-0.09	-0.00108407246958009\\
-0.07	-0.00106471944906494\\
-0.05	-0.00105060408080582\\
-0.03	-0.00104131772814209\\
-0.01	-0.00103671683669115\\
0.01	-0.00103672684674816\\
0.03	-0.00104132936671719\\
0.05	-0.0010506160008249\\
0.07	-0.0010647299021016\\
0.09	-0.00108407541884814\\
0.11	-0.00110928234157859\\
0.13	-0.0011408513902974\\
0.15	-0.00117957914728917\\
0.17	-0.00122676294113588\\
0.19	-0.00128386835342842\\
0.21	-0.00135264200698859\\
0.23	-0.00143573741757574\\
0.25	-0.00153517819464355\\
0.27	-0.00165559323496656\\
0.29	-0.00180508671893198\\
0.31	-0.00198948283623757\\
0.33	-0.00221786339890151\\
0.35	-0.00249957943570411\\
0.37	-0.00283762303476787\\
0.39	-0.00321460637350179\\
0.41	-0.00356392167823505\\
0.43	-0.00373372804450556\\
0.45	-0.0034883337589271\\
0.47	-0.00256562210854927\\
0.49	-0.000673248981868824\\
};
\addlegendentry{ALM$^i$ ($\varepsilon = 3.5 \Delta x$)}

\addplot [color=red, dotted, line width=1.0pt, mark size=0.75pt, mark=diamond, mark options={solid, red}]
  table[row sep=crcr]{%
-0.49	-0.00067918475975548\\
-0.47	-0.00120734461982441\\
-0.45	-0.00152597479733574\\
-0.43	-0.00171426578757239\\
-0.41	-0.00179932899413271\\
-0.39	-0.00180355948877797\\
-0.37	-0.00175148009733894\\
-0.35	-0.00166217992483286\\
-0.33	-0.00155022697702756\\
-0.31	-0.00142913845501702\\
-0.29	-0.00130995511217913\\
-0.27	-0.00119812780987749\\
-0.25	-0.00109382652331425\\
-0.23	-0.0010032372211517\\
-0.21	-0.000924569944012466\\
-0.19	-0.000857210350447835\\
-0.17	-0.00080142762632518\\
-0.15	-0.00075658194662942\\
-0.13	-0.000719935073784206\\
-0.11	-0.000689749276298279\\
-0.09	-0.00066485558456586\\
-0.07	-0.000645228028054778\\
-0.05	-0.000630684383794655\\
-0.03	-0.000620974253828011\\
-0.01	-0.000618818550690792\\
0.01	-0.000618811757186255\\
0.03	-0.000620969065905141\\
0.05	-0.000630654395876395\\
0.07	-0.000645183160198015\\
0.09	-0.000664806961357955\\
0.11	-0.000689690488401395\\
0.13	-0.000719864692402705\\
0.15	-0.000756541980202474\\
0.17	-0.000801404293751973\\
0.19	-0.000857209380497048\\
0.21	-0.000924588705365159\\
0.23	-0.00100325518032396\\
0.25	-0.00109382599548336\\
0.27	-0.00119813959228954\\
0.29	-0.00130998772122737\\
0.31	-0.00142917471842263\\
0.33	-0.00155025367596395\\
0.35	-0.00166219921696764\\
0.37	-0.00175150544395153\\
0.39	-0.00180356626962408\\
0.41	-0.00179932447463205\\
0.43	-0.00171425847953444\\
0.45	-0.00152596974129127\\
0.47	-0.00120733465487466\\
0.49	-0.000679172531475092\\
};
\addlegendentry{ALM$^d$ ($\varepsilon = 7 \Delta x$)}

\addplot [color=pink, line width=0.5pt, only marks, mark size=1.5pt, mark=x, mark options={solid, pink}]
  table[row sep=crcr]{%
-0.49	-0.00067916754682007\\
-0.47	-0.0012073282313384\\
-0.45	-0.00152596209964051\\
-0.43	-0.00171425203054554\\
-0.41	-0.00179930561660679\\
-0.39	-0.00180352780744641\\
-0.37	-0.00175151567643417\\
-0.35	-0.00166219188390737\\
-0.33	-0.0015502389820331\\
-0.31	-0.00142916906854929\\
-0.29	-0.00130998439754959\\
-0.27	-0.00119813235699562\\
-0.25	-0.00109385112761551\\
-0.23	-0.00100325049489447\\
-0.21	-0.000924575182907783\\
-0.19	-0.000857210052496837\\
-0.17	-0.000801426034841013\\
-0.15	-0.000756570605085827\\
-0.13	-0.000719882773547398\\
-0.11	-0.000689721425364609\\
-0.09	-0.00066481446195521\\
-0.07	-0.000645170660581562\\
-0.05	-0.000630640153485462\\
-0.03	-0.000620961259112263\\
-0.01	-0.00061881708217526\\
0.01	-0.000618847383586862\\
0.03	-0.000620990839922841\\
0.05	-0.00063067979409448\\
0.07	-0.000645230971635269\\
0.09	-0.000664871132642242\\
0.11	-0.000689755405679171\\
0.13	-0.00071993749162802\\
0.15	-0.000756618883620108\\
0.17	-0.000801472754375809\\
0.19	-0.000857250733948933\\
0.21	-0.000924599104881518\\
0.23	-0.00100325653967792\\
0.25	-0.00109385111210925\\
0.27	-0.00119812279850784\\
0.29	-0.00130995322332617\\
0.31	-0.00142913409067047\\
0.33	-0.0015502230169997\\
0.35	-0.00166217456366835\\
0.37	-0.00175145922925268\\
0.39	-0.00180354911267977\\
0.41	-0.00179929709514903\\
0.43	-0.00171422454015369\\
0.45	-0.00152594552278096\\
0.47	-0.00120731236094507\\
0.49	-0.000679150782293525\\
};
\addlegendentry{ALM$^i$ ($\varepsilon = 7 \Delta x$)}

\legend{}
\end{axis}
\node [left] at (-0.5,-0.5) {(d)};
\end{tikzpicture}%
%
%
\definecolor{mycolor1}{rgb}{1.00000,1.00000,0.00000}%
\definecolor{mycolor2}{rgb}{0.00000,1.00000,1.00000}%
\begin{tikzpicture}

\begin{axis}[%
width=0.35\textwidth,
height=0.25\textwidth,
scale only axis,
xmin=0.0,
xmax=0.5,
xtick={0.0, 0.1, 0.2, 0.3, 0.4, 0.5},
xlabel style={font=\color{white!15!black}},
xlabel={$x/R$},
ymin=0.02,
ymax=0.05,
ylabel style={font=\color{white!15!black}},
ylabel={$\Gamma/(R U_z)$},
yticklabel style={
        /pgf/number format/fixed,
        /pgf/number format/precision=3
},
scaled y ticks=false,
xmajorgrids,
ymajorgrids,
tick label style={font=\scriptsize},
axis background/.style={fill=white},
axis x line*=bottom,
axis y line*=left,
legend image post style={scale=0.8},
legend style={legend cell align=left, align=left, font=\scriptsize, row sep=-0.1cm},
legend pos=south west
]
\addplot [color=gray, line width=1.0pt, mark size=1.5pt, mark=square, mark options={solid, gray}]
  table[row sep=crcr]{%
-0.49	0.021085411690138\\
-0.47	0.0283029863148513\\
-0.45	0.0323965179766576\\
-0.43	0.0351004433104566\\
-0.41	0.0370319783395984\\
-0.39	0.0384810035500162\\
-0.37	0.039605522004944\\
-0.35	0.0405001954223245\\
-0.33	0.0412256062423345\\
-0.31	0.0418224718196333\\
-0.29	0.0423192136272753\\
-0.27	0.0427362764776541\\
-0.25	0.043088730136613\\
-0.23	0.0433879060358545\\
-0.21	0.0436424644066526\\
-0.19	0.0438591110659145\\
-0.17	0.0440430910415767\\
-0.15	0.0441985356806217\\
-0.13	0.0443287109395114\\
-0.11	0.0444361973697492\\
-0.09	0.0445230217727025\\
-0.07	0.0445907538449183\\
-0.05	0.0446405768176967\\
-0.03	0.0446733382132682\\
-0.01	0.0446895848558835\\
0.01	0.0446895848558834\\
0.03	0.0446733382132681\\
0.05	0.0446405768176968\\
0.0700000000000001	0.0445907538449184\\
0.09	0.0445230217727027\\
0.11	0.0444361973697492\\
0.13	0.0443287109395113\\
0.15	0.0441985356806216\\
0.17	0.0440430910415767\\
0.19	0.0438591110659148\\
0.21	0.0436424644066528\\
0.23	0.0433879060358545\\
0.25	0.0430887301366129\\
0.27	0.042736276477654\\
0.29	0.0423192136272751\\
0.31	0.0418224718196331\\
0.33	0.0412256062423346\\
0.35	0.0405001954223247\\
0.37	0.039605522004944\\
0.39	0.0384810035500162\\
0.41	0.0370319783395983\\
0.43	0.0351004433104565\\
0.45	0.0323965179766576\\
0.47	0.0283029863148514\\
0.49	0.021085411690138\\
};
\addlegendentry{LL}

\addplot [color=blue, dashed, line width=0.5pt, mark size=1.5pt, mark=o, mark options={solid, blue}]
  table[row sep=crcr]{%
-0.49	0.0210328457881224\\
-0.47	0.0282114690403853\\
-0.45	0.0322809457083041\\
-0.43	0.0349729447985345\\
-0.41	0.0369014311100247\\
-0.39	0.0383511099460064\\
-0.37	0.0394780040716356\\
-0.35	0.0403752182546051\\
-0.33	0.0411028912339072\\
-0.31	0.0417017770134672\\
-0.29	0.0422002946013307\\
-0.27	0.042618957718507\\
-0.25	0.0429729061895931\\
-0.23	0.0432735336902305\\
-0.21	0.0435293799934778\\
-0.19	0.0437471935651482\\
-0.17	0.0439322102451383\\
-0.15	0.0440885602449923\\
-0.13	0.0442195089149021\\
-0.11	0.0443276370777054\\
-0.09	0.0444149796721557\\
-0.07	0.0444831142818337\\
-0.05	0.044533229251861\\
-0.03	0.0445661793153355\\
-0.01	0.0445825235945866\\
0.01	0.0445825237350938\\
0.03	0.0445661800402505\\
0.05	0.0445332299984141\\
0.07	0.0444831147720378\\
0.09	0.0444149800885205\\
0.11	0.044327637578263\\
0.13	0.0442195094223529\\
0.15	0.0440885605070441\\
0.17	0.0439322101334042\\
0.19	0.0437471934274907\\
0.21	0.0435293797851615\\
0.23	0.0432735328751605\\
0.25	0.042972904861477\\
0.27	0.0426189559897867\\
0.29	0.0422002927309618\\
0.31	0.0417017753671087\\
0.33	0.0411028897374118\\
0.35	0.0403752166051065\\
0.37	0.0394780024269057\\
0.39	0.038351108641278\\
0.41	0.0369014303715008\\
0.43	0.0349729443693773\\
0.45	0.032280945347035\\
0.47	0.028211468825583\\
0.49	0.0210328458423599\\
};
\addlegendentry{ALM$^d$ ($\varepsilon = 3.5 \Delta x$)}

\addplot [color=cyan, line width=1.0pt, only marks, mark size=2.0pt, mark=+, mark options={solid, cyan}]
  table[row sep=crcr]{%
-0.49	0.0210328784991675\\
-0.47	0.0282115145870338\\
-0.45	0.0322810027410848\\
-0.43	0.0349730081386325\\
-0.41	0.0369014910282024\\
-0.39	0.0383511570363602\\
-0.37	0.0394780362117121\\
-0.35	0.0403752382224836\\
-0.33	0.0411029035493914\\
-0.31	0.0417017844215429\\
-0.29	0.0422002991206733\\
-0.27	0.0426189608409355\\
-0.25	0.0429729086507709\\
-0.23	0.0432735370252788\\
-0.21	0.0435293840583714\\
-0.19	0.0437471978661197\\
-0.17	0.0439322145819221\\
-0.15	0.0440885645761589\\
-0.13	0.0442195129640048\\
-0.11	0.044327640970738\\
-0.09	0.0444149833881189\\
-0.07	0.0444831177010605\\
-0.05	0.0445332322869906\\
-0.03	0.0445661818138281\\
-0.01	0.0445825253968827\\
0.01	0.0445825248826025\\
0.03	0.0445661806067594\\
0.05	0.0445332304872127\\
0.07	0.0444831152983043\\
0.09	0.0444149806295666\\
0.11	0.0443276379883057\\
0.13	0.0442195095378471\\
0.15	0.0440885608839275\\
0.17	0.0439322106675923\\
0.19	0.0437471940635114\\
0.21	0.0435293806623874\\
0.23	0.0432735342359324\\
0.25	0.0429729069995335\\
0.27	0.0426189594601642\\
0.29	0.0422002980224954\\
0.31	0.0417017835132942\\
0.33	0.0411029026071255\\
0.35	0.0403752371093025\\
0.37	0.0394780352477241\\
0.39	0.0383511561700961\\
0.41	0.0369014901598715\\
0.43	0.0349730073993845\\
0.45	0.0322810022306526\\
0.47	0.0282115142429638\\
0.49	0.0210328782787403\\
};
\addlegendentry{ALM$^i$ ($\varepsilon = 3.5 \Delta x$)}

\addplot [color=red, dotted, line width=1.0pt, mark size=0.75pt, mark=diamond, mark options={solid, red}]
  table[row sep=crcr]{%
-0.49	0.0210718838134021\\
-0.47	0.0282778345975889\\
-0.45	0.0323635991378535\\
-0.43	0.0350622487085209\\
-0.41	0.0369902746785553\\
-0.39	0.0384372895892114\\
-0.37	0.0395612806868689\\
-0.35	0.0404567815736327\\
-0.33	0.0411842736817134\\
-0.31	0.0417839920068168\\
-0.29	0.0422837349994479\\
-0.27	0.0427034009246884\\
-0.25	0.0430578296341934\\
-0.23	0.0433588616700446\\
-0.21	0.0436149983699601\\
-0.19	0.0438331274975935\\
-0.17	0.044018590123297\\
-0.15	0.0441754485219511\\
-0.13	0.0443068544144626\\
-0.11	0.0444152630735451\\
-0.09	0.0445026951138672\\
-0.07	0.0445707397450254\\
-0.05	0.0446207045025995\\
-0.03	0.0446536357822048\\
-0.01	0.0446702109430398\\
0.01	0.0446702114858309\\
0.03	0.0446536373941919\\
0.05	0.0446207066041466\\
0.07	0.0445707418896392\\
0.09	0.0445026972399765\\
0.11	0.0444152644194155\\
0.13	0.0443068545644437\\
0.15	0.044175447279005\\
0.17	0.0440185870638098\\
0.19	0.0438331233008858\\
0.21	0.0436149938117798\\
0.23	0.043358856497049\\
0.25	0.0430578232606681\\
0.27	0.0427033932324996\\
0.29	0.0422837255661998\\
0.31	0.0417839818565506\\
0.33	0.0411842631738049\\
0.35	0.0404567706216498\\
0.37	0.0395612693815561\\
0.39	0.0384372788224611\\
0.41	0.0369902656573465\\
0.43	0.035062241947506\\
0.45	0.0323635947350109\\
0.47	0.0282778328762601\\
0.49	0.0210718845350963\\
};
\addlegendentry{ALM$^d$ ($\varepsilon = 7 \Delta x$)}

\addplot [color=pink, line width=0.5pt, only marks, mark size=1.5pt, mark=x, mark options={solid, pink}]
  table[row sep=crcr]{%
-0.49	0.0210718928498637\\
-0.47	0.0282778491411136\\
-0.45	0.032363617448971\\
-0.43	0.0350622694107451\\
-0.41	0.0369902965225824\\
-0.39	0.0384373116985536\\
-0.37	0.0395613002747579\\
-0.35	0.0404567936054828\\
-0.33	0.0411842773366397\\
-0.31	0.041783988116467\\
-0.29	0.0422837258811665\\
-0.27	0.042703389119616\\
-0.25	0.0430578158769259\\
-0.23	0.0433588486282709\\
-0.21	0.0436149892803889\\
-0.19	0.0438331245758784\\
-0.17	0.0440185944457088\\
-0.15	0.044175459136462\\
-0.13	0.0443068686606771\\
-0.11	0.0444152784347684\\
-0.09	0.0445027081945401\\
-0.07	0.0445707476247772\\
-0.05	0.0446207072209177\\
-0.03	0.0446536351825724\\
-0.01	0.0446702085696756\\
0.01	0.0446702055758309\\
0.03	0.0446536274952469\\
0.05	0.0446206949506571\\
0.07	0.0445707306683655\\
0.09	0.0445026886245463\\
0.11	0.04441526010491\\
0.13	0.044306852679338\\
0.15	0.0441754445452031\\
0.17	0.044018583537835\\
0.19	0.0438331186769788\\
0.21	0.0436149874211178\\
0.23	0.043358848546951\\
0.25	0.0430578160493581\\
0.27	0.0427033898724826\\
0.29	0.042283727010196\\
0.31	0.041783990071433\\
0.33	0.041184280173364\\
0.35	0.040456796033411\\
0.37	0.0395613016134249\\
0.39	0.0384373125279339\\
0.41	0.0369902979170839\\
0.43	0.0350622706269899\\
0.45	0.0323636179426466\\
0.47	0.0282778495910279\\
0.49	0.0210718935942173\\
};
\addlegendentry{ALM$^i$ ($\varepsilon = 7 \Delta x$)}

\legend{}
\end{axis}
\node [left] at (-0.5,-0.5) {(e)};
\end{tikzpicture}%
%
%
\definecolor{mycolor1}{rgb}{1.00000,1.00000,0.00000}%
\definecolor{mycolor2}{rgb}{0.00000,1.00000,1.00000}%
\begin{tikzpicture}

\begin{axis}[%
width=0.35\textwidth,
height=0.25\textwidth,
scale only axis,
xmin=0.0,
xmax=0.5,
xtick={0.0, 0.1, 0.2, 0.3, 0.4, 0.5},
xlabel style={font=\color{white!15!black}},
xlabel={$x/R$},
ymin=-0.003,
ymax=0.001,
ylabel style={font=\color{white!15!black}},
ylabel={$(\Gamma-\Gamma_{LL})/\Gamma_{0}$},
xmajorgrids,
ymajorgrids,
tick label style={font=\scriptsize},
axis background/.style={fill=white},
axis x line*=bottom,
axis y line*=left,
legend image post style={scale=0.8},
legend style={legend cell align=left, align=left, font=\scriptsize},
legend pos=outer north east
]

\addplot [color=blue, dashed, line width=0.5pt, mark size=1.5pt, mark=o, mark options={solid, blue}]
  table[row sep=crcr]{%
-0.49	-0.00105131804018723\\
-0.47	-0.00183034548910408\\
-0.45	-0.00231144536679596\\
-0.43	-0.00254997023813998\\
-0.41	-0.0026109445911658\\
-0.39	-0.00259787207988958\\
-0.37	-0.00255035866586561\\
-0.35	-0.00249954335409196\\
-0.33	-0.00245430016825608\\
-0.31	-0.00241389612303618\\
-0.29	-0.0023783805186105\\
-0.27	-0.00234637518266377\\
-0.25	-0.00231647894012324\\
-0.23	-0.00228744691220969\\
-0.21	-0.00226168826322852\\
-0.19	-0.00223835001506156\\
-0.17	-0.00221761592850523\\
-0.15	-0.0021995087123269\\
-0.13	-0.00218404049192751\\
-0.11	-0.00217120584061992\\
-0.09	-0.00216084201068107\\
-0.07	-0.00215279126143699\\
-0.05	-0.00214695131646091\\
-0.03	-0.00214317795840076\\
-0.01	-0.00214122522568505\\
0.01	-0.0021412224155396\\
0.03	-0.00214316346009848\\
0.05	-0.00214693638539897\\
0.07	-0.00215278145735694\\
0.09	-0.0021608336833875\\
0.11	-0.00217119582946788\\
0.13	-0.0021840303429095\\
0.15	-0.00219950347128915\\
0.17	-0.00221761816318823\\
0.19	-0.00223835276821621\\
0.21	-0.00226169242955761\\
0.23	-0.00228746321360867\\
0.25	-0.00231650550244474\\
0.27	-0.00234640975706863\\
0.29	-0.00237841792598506\\
0.31	-0.00241392905020203\\
0.33	-0.00245433009816477\\
0.35	-0.00249957634406743\\
0.37	-0.00255039156046446\\
0.39	-0.00259789817445719\\
0.41	-0.00261095936164176\\
0.43	-0.00254997882128237\\
0.45	-0.00231145259217861\\
0.47	-0.0018303497851517\\
0.49	-0.00105131695543779\\
};
\addlegendentry{ALM$^d$ ($\varepsilon = 3.5 \Delta x$)}

\addplot [color=cyan, line width=1.0pt, only marks, mark size=2.0pt, mark=+, mark options={solid, cyan}]
  table[row sep=crcr]{%
-0.49	-0.00105066381928527\\
-0.47	-0.00182943455613416\\
-0.45	-0.00231030471118211\\
-0.43	-0.00254870343618013\\
-0.41	-0.002609746227612\\
-0.39	-0.00259693027281361\\
-0.37	-0.00254971586433563\\
-0.35	-0.00249914399652194\\
-0.33	-0.00245405385857221\\
-0.31	-0.00241374796152217\\
-0.29	-0.00237829013175848\\
-0.27	-0.00234631273409385\\
-0.25	-0.00231642971656732\\
-0.23	-0.00228738021124369\\
-0.21	-0.00226160696535648\\
-0.19	-0.00223826399563166\\
-0.17	-0.00221752919282924\\
-0.15	-0.00219942208899493\\
-0.13	-0.00218395950987354\\
-0.11	-0.00217112797996798\\
-0.09	-0.00216076769141701\\
-0.07	-0.00215272287690105\\
-0.05	-0.00214689061386894\\
-0.03	-0.00214312798854876\\
-0.01	-0.00214118917976311\\
0.01	-0.00214119946536548\\
0.03	-0.00214315212992058\\
0.05	-0.00214692660942699\\
0.07	-0.00215277093202689\\
0.09	-0.00216082286246556\\
0.11	-0.00217118762861399\\
0.13	-0.00218402803302555\\
0.15	-0.00219949593362117\\
0.17	-0.00221760747942616\\
0.19	-0.00223834004780219\\
0.21	-0.00226167488503955\\
0.23	-0.00228743599817071\\
0.25	-0.0023164627413147\\
0.27	-0.00234634034951852\\
0.29	-0.00237831209531307\\
0.31	-0.00241376612649211\\
0.33	-0.00245407270389085\\
0.35	-0.00249916626014746\\
0.37	-0.00254973514409661\\
0.39	-0.00259694759809524\\
0.41	-0.00260976359422792\\
0.43	-0.00254871822113853\\
0.45	-0.00231031491982687\\
0.47	-0.00182944143753586\\
0.49	-0.00105066822782984\\
};
\addlegendentry{ALM$^i$ ($\varepsilon = 3.5 \Delta x$)}

\addplot [color=red, dotted, line width=1.0pt, mark size=0.75pt, mark=diamond, mark options={solid, red}]
  table[row sep=crcr]{%
-0.49	-0.000270557534685609\\
-0.47	-0.000503034345189101\\
-0.45	-0.000658376776003593\\
-0.43	-0.000763892038623358\\
-0.41	-0.00083407322076413\\
-0.39	-0.000874279215993478\\
-0.37	-0.000884826361396641\\
-0.35	-0.000868276973732917\\
-0.33	-0.00082665121232467\\
-0.31	-0.000769596256238715\\
-0.29	-0.000709572556463898\\
-0.27	-0.000657511059235701\\
-0.25	-0.000618010048318308\\
-0.23	-0.000580887316129589\\
-0.21	-0.000549320733785143\\
-0.19	-0.000519671366358961\\
-0.17	-0.000490018365535585\\
-0.15	-0.000461743173356583\\
-0.13	-0.000437130500924213\\
-0.11	-0.000418685924033356\\
-0.09	-0.00040653317665859\\
-0.07	-0.000400281997810326\\
-0.05	-0.000397446301897855\\
-0.03	-0.000394048621221726\\
-0.01	-0.00038747825682856\\
0.01	-0.000387467401005099\\
0.03	-0.000394016381477522\\
0.05	-0.000397404270956078\\
0.07	-0.000400239105536293\\
0.09	-0.000406490654475182\\
0.11	-0.000418659006625335\\
0.13	-0.000437127501300192\\
0.15	-0.00046176803227678\\
0.17	-0.000490079555280515\\
0.19	-0.000519755300517602\\
0.21	-0.000549411897394155\\
0.23	-0.000580990776040631\\
0.25	-0.000618137518823694\\
0.27	-0.000657664903010346\\
0.29	-0.000709761221422531\\
0.31	-0.000769799261558606\\
0.33	-0.000826861370495372\\
0.35	-0.000868496013394438\\
0.37	-0.000885052467653526\\
0.39	-0.000874494550999113\\
0.41	-0.000834253644938044\\
0.43	-0.000764027258919691\\
0.45	-0.000658464832856226\\
0.47	-0.000503068771766797\\
0.49	-0.000270543100802148\\
};
\addlegendentry{ALM$^d$ ($\varepsilon = 7 \Delta x$)}

\addplot [color=pink, line width=0.5pt, only marks, mark size=1.5pt, mark=x, mark options={solid, pink}]
  table[row sep=crcr]{%
-0.49	-0.000270376805453634\\
-0.47	-0.000502743474695125\\
-0.45	-0.000658010553653691\\
-0.43	-0.000763477994139431\\
-0.41	-0.000833636340222161\\
-0.39	-0.000873837029149584\\
-0.37	-0.000884434603616721\\
-0.35	-0.000868036336730954\\
-0.33	-0.000826578113798696\\
-0.31	-0.00076967406323483\\
-0.29	-0.000709754922091891\\
-0.27	-0.000657747160683649\\
-0.25	-0.000618285193668252\\
-0.23	-0.000581148151603611\\
-0.21	-0.000549502525209119\\
-0.19	-0.000519729800660934\\
-0.17	-0.000489931917299664\\
-0.15	-0.000461530883138564\\
-0.13	-0.000436845576634252\\
-0.11	-0.000418378699567386\\
-0.09	-0.000406271563200641\\
-0.07	-0.000400124402774296\\
-0.05	-0.000397391935533812\\
-0.03	-0.00039406061386971\\
-0.01	-0.000387525724112547\\
0.01	-0.000387585601005012\\
0.03	-0.000394214360377439\\
0.05	-0.000397637340746046\\
0.07	-0.000400463531010304\\
0.09	-0.000406662963079074\\
0.11	-0.000418745296735243\\
0.13	-0.000437165203414119\\
0.15	-0.000461822708314776\\
0.17	-0.000490150074776583\\
0.19	-0.000519847778657598\\
0.21	-0.000549539710634132\\
0.23	-0.000581149778000683\\
0.25	-0.000618281745023688\\
0.27	-0.000657732103350353\\
0.29	-0.000709732341498527\\
0.31	-0.000769634963910639\\
0.33	-0.000826521379313399\\
0.35	-0.000867987778170454\\
0.37	-0.000884407830277623\\
0.39	-0.000873820441543159\\
0.41	-0.000833608450190075\\
0.43	-0.000763453669241782\\
0.45	-0.000658000680142378\\
0.47	-0.000502734476410787\\
0.49	-0.00027036191838213\\
};
\addlegendentry{ALM$^i$ ($\varepsilon = 7 \Delta x$)}

\legend{}
\end{axis}
\node [left] at (-0.5,-0.5) {(f)};
\end{tikzpicture}%
  \caption{Comparison of the results for the straight wing in uniform flow. LL - Non-linear iterative lifting line method of Section \ref{sec:NLLL}; ALM$^d$ - ALM with non-iterative correction of Section \ref{sec:nonitesmearcorr}; ALM$^i$ - ALM with iterative correction of Section \ref{sec:itesmearcorr}. (a) Corrected velocity in $y$-direction. (b) Difference between $y$-velocities obtained by the ALM and LL. (c) Corrected velocity in $z$-direction. (d) Difference between $z$-velocities obtained by the ALM and LL. (d) Circulation. (e) Relative difference between circulation obtained by the ALM and LL.}
  \label{fig:res_LLACLstraight}
\end{figure}

Because of symmetry, just the results for half of the wing are shown. As can be seen in figure \ref{fig:res_LLACLstraight}, the agreement is excellent for all cases. The difference in the induced velocity $u_y$ is in the order of $10^{-4}$ ($0.01 \%$ of the undisturbed velocity), an agreement much better than the ones reported by \citet{dag2020new} and \citet{meyer2019vortex}. The difference is in the order of the square of $u_y$, which is consistent with a first-order method and, therefore, a better agreement was not expected. The other works are mentioned not as a criticism, but to bring context to the current values, showing the strength of the method. For most practical applications of the actuator line, the errors of the other methods can also be considered negligible.

We believe that the remarkable agreement can be mainly explained by three factors. First, the convolution of the force with the Gaussian kernel was performed analytically using equation \eqref{eq:fcode}, considering constant force in a segment, which is the case that is mathematically equivalent to the lifting line method. Second, the lifting line method was carefully constructed to be consistent with the actuator line method. Third, the use of the velocity induced by a smeared vortex segment of Section \ref{sec:lamboseenvelocity} has lower associated errors when compared to other strategies where the vortices are also discretized.


The difference between the results for the two values of the smearing parameter is negligible. This is exactly the aim of the smearing correction: to make the forces on the blades insensitive to variations of the smearing correction and approach the lifting line method, which is the limit $\varepsilon \to 0$.

The velocity in the $z$-direction is also shown in figure \ref{fig:res_LLACLstraight}, because it influences the circulation. The effect of the smearing correction on $u^c_z$ is minimal. For most points, the missing velocity $u^m_z$ is at least one order of magnitude lower than the difference observed in figure \ref{fig:res_LLACLstraight}(d). Hence, the differences observed are mainly due to the velocity sampled from the numerical simulation. The reason for the differences is probably the discrepancies in the vortex sheet created in the CFD domain and the vortex sheet of the lifting line method, as mentioned in the first paragraph of Section \ref{sec:formvortexsheet}. The prescribed horseshoe vortices of the lifting line method do not induce velocity in the $z$-direction. In the CFD simulation, the vortices are free. As a consequence, the downwash from the wing inclines the vortex sheet, which induces velocity in the negative $z$-direction. A more advanced lifting line method might be able to capture those effects. Hence, the differences in $u_z$ (and its consequence in $\Gamma$) can be, at least partially, attributed to a limitation of the current implementation of the lifting line method.

The relative differences in the values of circulation are in the order of $10^{-3}$ (differences in the order of $0.1 \%$ of $\Gamma_0$), which is a consequence of the errors in $u_y$ and $u_z$. This difference is negligible for practical purposes. The circulation for the higher value of the smearing parameter agrees better with the lifting line. This apparently contradictory result has two simple explanations. First, the vortex core is larger, so the representation of the vorticity is better, reducing numerical errors (see Appendix~\ref{app:grid}). Second, for higher values of $\varepsilon$ the results depend less on the data from the simulation and more on the smearing correction. In the limit $\varepsilon \to \infty$, no information from the numerical simulation is used and the force is determined only by the correction, that regress identically to the lifting line method. Nevertheless, increasing the smearing parameter is not usually beneficial, because the value of $\varepsilon$ has other effects on the simulation beyond its influence on the forces, as discussed in Section \ref{sec:res_rotor}. It should be noted, however, that the lifting line method has its own limitations on reproducing the results of wings and blades. Notwithstanding, it is a relevant tool for middle or low-fidelity aerodynamic calculations.

Before the development of the smearing correction, the use of the ALM was almost exclusively restricted to simulations of rotating blades, probably due to its inability to accurately reproduce the induced velocities, which affected the lift of translating wings. The remarkable agreement achieved here, which further confirms the good agreement of previous works~\citep{dag2020new,martinez2019filtered,meyer2019vortex}, might encourage the adoption of the ALM for other applications beyond rotating blades. For example, the aeronautics community could use it to simulate wings or the lifting surfaces of aircrafts. The ALM enables the integration of lifting lines with a CFD solver, allowing more complex configurations and flow conditions than a classical lifting line method. The low grid requirement is a clear advantage over traditional CFD methods that employ wall boundary conditions for the wing. Also, the free-vortex method described here maintains the generality of the method, so diverse configurations are allowed, and it does not suffer from the instability problems of traditional vortex filament methods~\citep{leishman2000principles}.

\subsection{Unsteady rotor simulations} \label{sec:res_rotor}

The NREL 5-MW wind turbine subjected to a sheared inflow was simulated using the iterative and direct smearing corrections. The undisturbed wind speed varies in the vertical direction ($y$), in order to simulate a condition in which the circulation of the blade changes every time step. Hence, for every time step of this simulation, the initial guess of circulation is different from the final value of circulation, both for the iterative and the non-iterative corrections.

The domain was reduced to $L_x/R=L_y/R=8$ to make the inflow velocity $U_{z}=y/5+1$ always positive and avoid problems with the outflow boundary condition, which is not stable if the velocity is negative. The distance from the inflow to the blades, $L_{zin}/R=4$, was also reduced, maintaining the distance from the blades to the outflow $L_{zout}/R=6$. Since this is a comparative study, the reduction of the domain does not affect any analysis.

The parameters of the simulation are presented in table \ref{tab:param_rotor}. Except where explicitly stated, all values are non-dimensionalized by the radius ($R$), the velocity at the center of the turbine ($U_{ref}=U(0,0,0)$) and the density ($\rho$). Hence, the geometry, detailed in \citep{jonkman2009definition}, was non-dimensionalized by the radius. For this choice of reference values, the tip speed ratio (which is the ratio of $\Omega R$ and the velocity at the center of the turbine) is $\Omega R/U_{ref}=7.55$. The average grid spacing in the region of the actuator line is the same as the simulation of the straight wing, $\Delta x=R/56$. The Reynolds number based on the radius is $Re_R = R U_{ref}/\nu = 5 \cdot 10^4$. A time step of $\Delta t = T/400$ was used, where $T$ is the period of rotation.

\begin{table}
    \centering
    \begin{tabular}{c c c c}
        $\mathbf{U}(x,y,z)/U_{ref}$ & $\Omega R/U_{ref}$ & $(L_x/R,L_y/R,L_z/R)$ & $L_{zin}/R$ \\
        $(0,0,y/(5R)+1)$ & $7.55$ & $(8,8,10)$ & $4$ 
    \end{tabular}
    \caption{Parameters of the simulation of the NREL 5-MW wind turbine in sheared flow.}
    \label{tab:param_rotor}
\end{table}

\begin{figure}
  \centering
%
%
\definecolor{mycolor1}{rgb}{1.00000,1.00000,0.00000}%
\definecolor{mycolor2}{rgb}{0.00000,1.00000,1.00000}%
\begin{tikzpicture}

\begin{axis}[%
width=0.35\textwidth,
height=0.25\textwidth,
scale only axis,
xmin=0,
xmax=1,
xlabel style={font=\color{white!15!black}},
xlabel={$r/R$},
ymin=0.0,
ymax=1.0,
ylabel style={font=\color{white!15!black}},
ylabel={$f_l$},
yticklabel style={
        /pgf/number format/fixed,
        /pgf/number format/precision=3
},
scaled y ticks=false,
xmajorgrids,
ymajorgrids,
tick label style={font=\scriptsize},
axis background/.style={fill=white},
axis x line*=bottom,
axis y line*=left,
legend image post style={scale=0.8},
legend style={legend cell align=left, align=left, font=\scriptsize},
]

\addplot [color=blue, dashed, line width=0.5pt, mark size=1.5pt, mark=o, mark options={solid, blue}]
  table[row sep=crcr]{%
0.0360119047619048	1.21430643318376e-17\\
0.0604166666666667	2.86229373536173e-17\\
0.0848214285714286	3.12250225675825e-17\\
0.109226190476191	4.16333634234434e-17\\
0.133630952380952	0.00346274003062088\\
0.158035714285714	0.0707748114551115\\
0.182440476190476	0.136890956356438\\
0.206845238095238	0.177704163633965\\
0.23125	0.207566914502804\\
0.255654761904762	0.23280309824081\\
0.280059523809524	0.252594577179059\\
0.304464285714286	0.270531719823948\\
0.328869047619048	0.288533602819923\\
0.353273809523809	0.307777627844749\\
0.377678571428571	0.328614086994959\\
0.402083333333333	0.352476887246613\\
0.426488095238095	0.379541242132527\\
0.450892857142857	0.408223383048729\\
0.475297619047619	0.433809575671262\\
0.499702380952381	0.459836653320025\\
0.524107142857143	0.4879555083079\\
0.548511904761905	0.518554930465666\\
0.572916666666667	0.550684444967719\\
0.597321428571429	0.581207418547207\\
0.62172619047619	0.610618737519636\\
0.646130952380952	0.637213497575796\\
0.670535714285714	0.656019608163549\\
0.694940476190476	0.673807158914787\\
0.719345238095238	0.694366781185579\\
0.74375	0.718062410295873\\
0.768154761904762	0.742282709398468\\
0.792559523809524	0.766088670590398\\
0.816964285714286	0.788859180792483\\
0.841369047619048	0.809710994595058\\
0.86577380952381	0.827135929440342\\
0.890178571428571	0.838095969157006\\
0.914583333333333	0.832834143698174\\
0.938988095238095	0.799538100854106\\
0.963392857142857	0.699536741996754\\
0.987797619047619	0.536259438822617\\
};
\addlegendentry{ALM$^d$ ($\varepsilon = 3.5 \Delta x$)}

\addplot [color=cyan, line width=1.0pt, only marks, mark size=2.0pt, mark=+, mark options={solid, cyan}]
  table[row sep=crcr]{%
0.0360119047619048	-1.47451495458029e-17\\
0.0604166666666667	-2.16840434497101e-17\\
0.0848214285714286	1.73472347597681e-17\\
0.109226190476191	2.42861286636753e-17\\
0.133630952380952	0.00346273990798104\\
0.158035714285714	0.0707747943268461\\
0.182440476190476	0.13689092202341\\
0.206845238095238	0.177704139768152\\
0.23125	0.207566909578581\\
0.255654761904762	0.232803113809745\\
0.280059523809524	0.252594601140549\\
0.304464285714286	0.270531743654806\\
0.328869047619048	0.28853362037636\\
0.353273809523809	0.307777639200918\\
0.377678571428571	0.328614099886711\\
0.402083333333333	0.352476890837832\\
0.426488095238095	0.379541240620582\\
0.450892857142857	0.408223375298091\\
0.475297619047619	0.43380956299146\\
0.499702380952381	0.459836644048995\\
0.524107142857143	0.487955488551181\\
0.548511904761905	0.518554904663923\\
0.572916666666667	0.550684407651775\\
0.597321428571429	0.581207368599957\\
0.62172619047619	0.61061868009685\\
0.646130952380952	0.63721345030713\\
0.670535714285714	0.656019578105716\\
0.694940476190476	0.673807140039663\\
0.719345238095238	0.694366774530488\\
0.74375	0.718062399676784\\
0.768154761904762	0.742282716993027\\
0.792559523809524	0.766088692480819\\
0.816964285714286	0.788859220837927\\
0.841369047619048	0.809711057418791\\
0.86577380952381	0.827136011107561\\
0.890178571428571	0.838096100831328\\
0.914583333333333	0.832834341050552\\
0.938988095238095	0.799538357614827\\
0.963392857142857	0.699537018242013\\
0.987797619047619	0.53625961271515\\
};
\addlegendentry{ALM$^i$ ($\varepsilon = 3.5 \Delta x$)}

\addplot [color=red, dotted, line width=1.0pt, mark size=0.75pt, mark=diamond, mark options={solid, red}]
  table[row sep=crcr]{%
0.0360119047619048	-2.51534904016637e-17\\
0.0604166666666667	-1.38777878078145e-17\\
0.0848214285714286	-2.60208521396521e-17\\
0.109226190476191	3.46944695195361e-17\\
0.133630952380952	0.00349118883707165\\
0.158035714285714	0.0735830069723706\\
0.182440476190476	0.141984076252307\\
0.206845238095238	0.180238489974645\\
0.23125	0.209910851342959\\
0.255654761904762	0.234576844949031\\
0.280059523809524	0.254089020525102\\
0.304464285714286	0.27204951051484\\
0.328869047619048	0.290209041330016\\
0.353273809523809	0.30968158007497\\
0.377678571428571	0.330799041695996\\
0.402083333333333	0.354890454283797\\
0.426488095238095	0.38216467073909\\
0.450892857142857	0.411188287824978\\
0.475297619047619	0.437111289199217\\
0.499702380952381	0.463402864970993\\
0.524107142857143	0.491647271991456\\
0.548511904761905	0.522367930242002\\
0.572916666666667	0.554610289287496\\
0.597321428571429	0.585399394535642\\
0.62172619047619	0.615162039183028\\
0.646130952380952	0.642285586131705\\
0.670535714285714	0.661417874772546\\
0.694940476190476	0.679543915624362\\
0.719345238095238	0.700447246958358\\
0.74375	0.724546651912552\\
0.768154761904762	0.749211932913228\\
0.792559523809524	0.773470322238319\\
0.816964285714286	0.796705658814421\\
0.841369047619048	0.818033805361769\\
0.86577380952381	0.835994989077727\\
0.890178571428571	0.847547340681586\\
0.914583333333333	0.842506632867716\\
0.938988095238095	0.809108056899455\\
0.963392857142857	0.707378979873861\\
0.987797619047619	0.543329324867795\\
};
\addlegendentry{ALM$^d$ ($\varepsilon = 7 \Delta x$)}

\addplot [color=pink, line width=0.5pt, only marks, mark size=1.5pt, mark=x, mark options={solid, pink}]
  table[row sep=crcr]{%
0.0360119047619048	4.77048955893622e-17\\
0.0604166666666667	-2.25514051876985e-17\\
0.0848214285714286	0\\
0.109226190476191	1.73472347597681e-18\\
0.133630952380952	0.00349118828922267\\
0.158035714285714	0.0735829918308487\\
0.182440476190476	0.141984056530995\\
0.206845238095238	0.180238476056226\\
0.23125	0.209910841607636\\
0.255654761904762	0.234576842122568\\
0.280059523809524	0.254089027077111\\
0.304464285714286	0.272049524192737\\
0.328869047619048	0.290209059607628\\
0.353273809523809	0.309681599551255\\
0.377678571428571	0.330799060693146\\
0.402083333333333	0.354890470622053\\
0.426488095238095	0.382164685307951\\
0.450892857142857	0.411188301074512\\
0.475297619047619	0.437111301667844\\
0.499702380952381	0.463402875703982\\
0.524107142857143	0.491647277606891\\
0.548511904761905	0.52236792879556\\
0.572916666666667	0.554610282066296\\
0.597321428571429	0.58539938443936\\
0.62172619047619	0.615162033982116\\
0.646130952380952	0.642285589233515\\
0.670535714285714	0.661417894939963\\
0.694940476190476	0.679543954633186\\
0.719345238095238	0.700447302070552\\
0.74375	0.724546721275428\\
0.768154761904762	0.749212016112223\\
0.792559523809524	0.773470419115938\\
0.816964285714286	0.796705771979849\\
0.841369047619048	0.818033944140416\\
0.86577380952381	0.835995159481253\\
0.890178571428571	0.84754755271442\\
0.914583333333333	0.84250689787714\\
0.938988095238095	0.809108360720193\\
0.963392857142857	0.707379279814551\\
0.987797619047619	0.543329471534289\\
};
\addlegendentry{ALM$^i$ ($\varepsilon = 7 \Delta x$)}

\legend{}
\end{axis}
\node [left] at (-0.5,-0.5) {(a)};
\end{tikzpicture}%
%
%
\definecolor{mycolor1}{rgb}{1.00000,1.00000,0.00000}%
\definecolor{mycolor2}{rgb}{0.00000,1.00000,1.00000}%
\begin{tikzpicture}

\begin{axis}[%
width=0.35\textwidth,
height=0.25\textwidth,
scale only axis,
xmin=0,
xmax=1,
xlabel style={font=\color{white!15!black}},
xlabel={$r/R$},
ymin=0,
ymax=0.04,
ylabel style={font=\color{white!15!black}},
ylabel={$f_d$},
yticklabel style={
        /pgf/number format/fixed,
        /pgf/number format/precision=3
},
scaled y ticks=false,
xmajorgrids,
ymajorgrids,
tick label style={font=\scriptsize},
axis background/.style={fill=white},
axis x line*=bottom,
axis y line*=left,
legend image post style={scale=0.8},
legend style={legend cell align=left, align=left, font=\scriptsize, row sep=-0.1cm, inner sep=0pt},
]

\addplot [color=blue, dashed, line width=0.5pt, mark size=1.5pt, mark=o, mark options={solid, blue}]
  table[row sep=crcr]{%
0.0360119047619048	0.0142333204001958\\
0.0604166666666667	0.0166273424314403\\
0.0848214285714286	0.0205163050614673\\
0.109226190476191	0.0225402656714163\\
0.133630952380952	0.0274584836025464\\
0.158035714285714	0.0333641111703193\\
0.182440476190476	0.00862223571767378\\
0.206845238095238	0.0040186386324047\\
0.23125	0.00285491553854925\\
0.255654761904762	0.00222220111028808\\
0.280059523809524	0.00242589937167052\\
0.304464285714286	0.00268035123264153\\
0.328869047619048	0.00290419871367291\\
0.353273809523809	0.00304383639055429\\
0.377678571428571	0.00318744626837146\\
0.402083333333333	0.00326114338291376\\
0.426488095238095	0.00315723242106103\\
0.450892857142857	0.00307116894180379\\
0.475297619047619	0.00327528912435072\\
0.499702380952381	0.00348236993015681\\
0.524107142857143	0.00368005461383573\\
0.548511904761905	0.00382517879825897\\
0.572916666666667	0.00393393670642413\\
0.597321428571429	0.00412979742718829\\
0.62172619047619	0.0043409796502473\\
0.646130952380952	0.00449867647329699\\
0.670535714285714	0.00442588520951394\\
0.694940476190476	0.00418150883287033\\
0.719345238095238	0.00406320894050181\\
0.74375	0.00419980866203511\\
0.768154761904762	0.00433468145670331\\
0.792559523809524	0.00446435081850156\\
0.816964285714286	0.00458639436789841\\
0.841369047619048	0.00469737352022842\\
0.86577380952381	0.00479160706089325\\
0.890178571428571	0.00485765651121763\\
0.914583333333333	0.00483231174886062\\
0.938988095238095	0.00466040539142513\\
0.963392857142857	0.00404391324662546\\
0.987797619047619	0.00317129378475198\\
};
\addlegendentry{ALM$^d$ ($\varepsilon = 3.5 \Delta x$)}

\addplot [color=cyan, line width=1.0pt, only marks, mark size=2.0pt, mark=+, mark options={solid, cyan}]
  table[row sep=crcr]{%
0.0360119047619048	0.0142333202985477\\
0.0604166666666667	0.0166273420476201\\
0.0848214285714286	0.0205163044907806\\
0.109226190476191	0.0225402637800395\\
0.133630952380952	0.0274584815556599\\
0.158035714285714	0.0333640842160324\\
0.182440476190476	0.00862222190493741\\
0.206845238095238	0.00401863642397746\\
0.23125	0.00285491554136484\\
0.255654761904762	0.00222220127974852\\
0.280059523809524	0.00242589946207712\\
0.304464285714286	0.00268035129031675\\
0.328869047619048	0.00290419876940583\\
0.353273809523809	0.00304383641778905\\
0.377678571428571	0.00318744629103242\\
0.402083333333333	0.00326114340726202\\
0.426488095238095	0.00315723240445741\\
0.450892857142857	0.00307116889362233\\
0.475297619047619	0.00327528905256848\\
0.499702380952381	0.00348236987374946\\
0.524107142857143	0.00368005448418771\\
0.548511904761905	0.00382517852850915\\
0.572916666666667	0.0039339363418225\\
0.597321428571429	0.00412979691138751\\
0.62172619047619	0.00434097906529148\\
0.646130952380952	0.00449867602499161\\
0.670535714285714	0.00442588491690793\\
0.694940476190476	0.0041815087284811\\
0.719345238095238	0.00406320892110482\\
0.74375	0.00419980861359415\\
0.768154761904762	0.00433468148042782\\
0.792559523809524	0.00446435090570446\\
0.816964285714286	0.00458639452753944\\
0.841369047619048	0.00469737377159005\\
0.86577380952381	0.00479160735660016\\
0.890178571428571	0.00485765699924783\\
0.914583333333333	0.00483231248352577\\
0.938988095238095	0.00466040631393333\\
0.963392857142857	0.00404391426464995\\
0.987797619047619	0.00317129437384623\\
};
\addlegendentry{ALM$^i$ ($\varepsilon = 3.5 \Delta x$)}

\addplot [color=red, dotted, line width=1.0pt, mark size=0.75pt, mark=diamond, mark options={solid, red}]
  table[row sep=crcr]{%
0.0360119047619048	0.0147220972654284\\
0.0604166666666667	0.0173389888576299\\
0.0848214285714286	0.021522794939227\\
0.109226190476191	0.0235345836721894\\
0.133630952380952	0.0277140085881492\\
0.158035714285714	0.0342032908242565\\
0.182440476190476	0.00853117095145061\\
0.206845238095238	0.00406836081426278\\
0.23125	0.00288327176583879\\
0.255654761904762	0.00223954628317845\\
0.280059523809524	0.00243591761861556\\
0.304464285714286	0.00268638150023048\\
0.328869047619048	0.0029108046491395\\
0.353273809523809	0.00305132027632732\\
0.377678571428571	0.00319493841672161\\
0.402083333333333	0.00327981841121038\\
0.426488095238095	0.00317041374979159\\
0.450892857142857	0.00308787788744659\\
0.475297619047619	0.00329317507610624\\
0.499702380952381	0.00350120779099528\\
0.524107142857143	0.00370453117540075\\
0.548511904761905	0.00386321932343979\\
0.572916666666667	0.00397046732042843\\
0.597321428571429	0.00417064754325197\\
0.62172619047619	0.00438524568465276\\
0.646130952380952	0.00455995460542769\\
0.670535714285714	0.00447712580553053\\
0.694940476190476	0.00421363028009974\\
0.719345238095238	0.00408567114795479\\
0.74375	0.00422381089438324\\
0.768154761904762	0.00436058678629628\\
0.792559523809524	0.00449225062396454\\
0.816964285714286	0.00461648474112603\\
0.841369047619048	0.00472988385100226\\
0.86577380952381	0.00482685214939051\\
0.890178571428571	0.00489587667844038\\
0.914583333333333	0.00487157972033159\\
0.938988095238095	0.00469895510784263\\
0.963392857142857	0.00407546924206653\\
0.987797619047619	0.00320104155891748\\
};
\addlegendentry{ALM$^d$ ($\varepsilon = 7 \Delta x$)}

\addplot [color=pink, line width=0.5pt, only marks, mark size=1.5pt, mark=x, mark options={solid, pink}]
  table[row sep=crcr]{%
0.0360119047619048	0.0147220967686178\\
0.0604166666666667	0.0173389877836608\\
0.0848214285714286	0.0215227927564112\\
0.109226190476191	0.0235345801345635\\
0.133630952380952	0.0277139978182094\\
0.158035714285714	0.0342032697048128\\
0.182440476190476	0.0085311668511651\\
0.206845238095238	0.00406835946736699\\
0.23125	0.00288327167007366\\
0.255654761904762	0.00223954626676671\\
0.280059523809524	0.00243591760484403\\
0.304464285714286	0.0026863814901229\\
0.328869047619048	0.00291080468233809\\
0.353273809523809	0.00305132031729327\\
0.377678571428571	0.00319493844266545\\
0.402083333333333	0.00327981852805265\\
0.426488095238095	0.00317041380391436\\
0.450892857142857	0.00308787795246494\\
0.475297619047619	0.00329317513075734\\
0.499702380952381	0.00350120783493885\\
0.524107142857143	0.00370453120891917\\
0.548511904761905	0.00386321931641129\\
0.572916666666667	0.0039704672568137\\
0.597321428571429	0.00417064744669295\\
0.62172619047619	0.00438524561511881\\
0.646130952380952	0.0045599546716731\\
0.670535714285714	0.0044771260128241\\
0.694940476190476	0.00421363049504045\\
0.719345238095238	0.00408567133613853\\
0.74375	0.00422381113189674\\
0.768154761904762	0.00436058707587676\\
0.792559523809524	0.0044922509661639\\
0.816964285714286	0.00461648514702354\\
0.841369047619048	0.00472988435577769\\
0.86577380952381	0.00482685277464777\\
0.890178571428571	0.0048958774527772\\
0.914583333333333	0.0048715806842132\\
0.938988095238095	0.00469895618659623\\
0.963392857142857	0.00407547036795197\\
0.987797619047619	0.0032010420597411\\
};
\addlegendentry{ALM$^i$ ($\varepsilon = 7 \Delta x$)}

\end{axis}
\node [left] at (-0.5,-0.5) {(b)};
\end{tikzpicture}%
%
%
\definecolor{mycolor1}{rgb}{1.00000,1.00000,0.00000}%
\definecolor{mycolor2}{rgb}{0.00000,1.00000,1.00000}%
\begin{tikzpicture}

\begin{axis}[%
width=0.35\textwidth,
height=0.25\textwidth,
scale only axis,
xmin=0,
xmax=1,
xlabel style={font=\color{white!15!black}},
xlabel={$r/R$},
ymin=0,
ymax=0.15,
ylabel style={font=\color{white!15!black}},
ylabel={$\Gamma/(R U_{ref})$},
yticklabel style={
        /pgf/number format/fixed,
        /pgf/number format/precision=3
},
scaled y ticks=false,
xmajorgrids,
ymajorgrids,
tick label style={font=\scriptsize},
axis background/.style={fill=white},
axis x line*=bottom,
axis y line*=left,
legend image post style={scale=0.8},
legend style={legend cell align=left, align=left, font=\scriptsize},
]

\addplot [color=blue, dashed, line width=0.5pt, mark size=1.5pt, mark=o, mark options={solid, blue}]
  table[row sep=crcr]{%
0.0360119047619048	0\\
0.0604166666666667	0\\
0.0848214285714286	0\\
0.109226190476191	0\\
0.133630952380952	0.002328778312043\\
0.158035714285714	0.046046978244746\\
0.182440476190476	0.0807641898751912\\
0.206845238095238	0.0951527632751932\\
0.23125	0.102071171033331\\
0.255654761904762	0.105702599993723\\
0.280059523809524	0.106483687157724\\
0.304464285714286	0.106390382704511\\
0.328869047619048	0.106298358075058\\
0.353273809523809	0.106615163039457\\
0.377678571428571	0.107394689110902\\
0.402083333333333	0.108996540695008\\
0.426488095238095	0.111352575371956\\
0.450892857142857	0.11390334139504\\
0.475297619047619	0.115363679527267\\
0.499702380952381	0.116785953150963\\
0.524107142857143	0.11856620281508\\
0.548511904761905	0.120793601952271\\
0.572916666666667	0.123172724896125\\
0.597321428571429	0.125006030808482\\
0.62172619047619	0.126458654522691\\
0.646130952380952	0.127233497230584\\
0.670535714285714	0.12643343695699\\
0.694940476190476	0.125495801815428\\
0.719345238095238	0.125116573697704\\
0.74375	0.125313108204466\\
0.768154761904762	0.125587844320833\\
0.792559523809524	0.125778180201326\\
0.816964285714286	0.125792555564518\\
0.841369047619048	0.125510830898947\\
0.86577380952381	0.124728063904113\\
0.890178571428571	0.123040198433567\\
0.914583333333333	0.11911467323933\\
0.938988095238095	0.111469901277527\\
0.963392857142857	0.0951211854343262\\
0.987797619047619	0.071154105906242\\
};
\addlegendentry{ALM$^d$ ($\varepsilon = 3.5 \Delta x$)}

\addplot [color=cyan, line width=1.0pt, only marks, mark size=2.0pt, mark=+, mark options={solid, cyan}]
  table[row sep=crcr]{%
0.0360119047619048	0\\
0.0604166666666667	0\\
0.0848214285714286	0\\
0.109226190476191	0\\
0.133630952380952	0.0023287641420235\\
0.158035714285714	0.046046909245355\\
0.182440476190476	0.0807641023905299\\
0.206845238095238	0.0951526481307926\\
0.23125	0.102071026854926\\
0.255654761904762	0.105702427167243\\
0.280059523809524	0.106483479962924\\
0.304464285714286	0.106390139271242\\
0.328869047619048	0.106298080735665\\
0.353273809523809	0.106614854528847\\
0.377678571428571	0.107394353017552\\
0.402083333333333	0.108996176539469\\
0.426488095238095	0.111352186848967\\
0.450892857142857	0.113902930535747\\
0.475297619047619	0.1153632490383\\
0.499702380952381	0.116785507682572\\
0.524107142857143	0.118565740123672\\
0.548511904761905	0.120793125519798\\
0.572916666666667	0.123172234705345\\
0.597321428571429	0.125005526089948\\
0.62172619047619	0.126458135090445\\
0.646130952380952	0.127232964326028\\
0.670535714285714	0.126432890905276\\
0.694940476190476	0.125495241253052\\
0.719345238095238	0.125116000563866\\
0.74375	0.125312523014023\\
0.768154761904762	0.125587253587805\\
0.792559523809524	0.125777586687772\\
0.816964285714286	0.12579196372912\\
0.841369047619048	0.125510245802154\\
0.86577380952381	0.124727492596697\\
0.890178571428571	0.123039654214078\\
0.914583333333333	0.119114174043339\\
0.938988095238095	0.111469470681147\\
0.963392857142857	0.0951208494484571\\
0.987797619047619	0.0711539009432058\\
};
\addlegendentry{ALM$^i$ ($\varepsilon = 3.5 \Delta x$)}

\addplot [color=red, dotted, line width=1.0pt, mark size=0.75pt, mark=diamond, mark options={solid, red}]
  table[row sep=crcr]{%
0.0360119047619048	0\\
0.0604166666666667	0\\
0.0848214285714286	0\\
0.109226190476191	0\\
0.133630952380952	0.0023373855855386\\
0.158035714285714	0.0468581390213077\\
0.182440476190476	0.0819490016277182\\
0.206845238095238	0.0958108553568072\\
0.23125	0.102611896104017\\
0.255654761904762	0.106192655484138\\
0.280059523809524	0.106991146438743\\
0.304464285714286	0.106949406397867\\
0.328869047619048	0.106908113085295\\
0.353273809523809	0.107270014568641\\
0.377678571428571	0.108086634789902\\
0.402083333333333	0.109703995659368\\
0.426488095238095	0.112068207200997\\
0.450892857142857	0.114661030316517\\
0.475297619047619	0.116165791236136\\
0.499702380952381	0.117618148153727\\
0.524107142857143	0.119402717141658\\
0.548511904761905	0.121607773740838\\
0.572916666666667	0.123962878416957\\
0.597321428571429	0.125813106675996\\
0.62172619047619	0.127306553605302\\
0.646130952380952	0.128139700188837\\
0.670535714285714	0.127394428063174\\
0.694940476190476	0.126504115557683\\
0.719345238095238	0.126161898566219\\
0.74375	0.126392062636634\\
0.768154761904762	0.126700188794182\\
0.792559523809524	0.126923541761104\\
0.816964285714286	0.12696856762689\\
0.841369047619048	0.126713029503714\\
0.86577380952381	0.12595893936505\\
0.890178571428571	0.124294301113412\\
0.914583333333333	0.120354318022301\\
0.938988095238095	0.112657679131526\\
0.963392857142857	0.0961041697475125\\
0.987797619047619	0.0719430829732067\\
};
\addlegendentry{ALM$^d$ ($\varepsilon = 7 \Delta x$)}

\addplot [color=pink, line width=0.5pt, only marks, mark size=1.5pt, mark=x, mark options={solid, pink}]
  table[row sep=crcr]{%
0.0360119047619048	0\\
0.0604166666666667	0\\
0.0848214285714286	0\\
0.109226190476191	0\\
0.133630952380952	0.0023373808567263\\
0.158035714285714	0.0468580783266627\\
0.182440476190476	0.0819488911830198\\
0.206845238095238	0.0958106904315728\\
0.23125	0.102611686021063\\
0.255654761904762	0.106192405575725\\
0.280059523809524	0.106990862037655\\
0.304464285714286	0.106949091391143\\
0.328869047619048	0.106907770681586\\
0.353273809523809	0.107269647792345\\
0.377678571428571	0.10808624591006\\
0.402083333333333	0.109703586202428\\
0.426488095238095	0.112067779229452\\
0.450892857142857	0.114660585778254\\
0.475297619047619	0.116165331896959\\
0.499702380952381	0.117617675326897\\
0.524107142857143	0.11940223162285\\
0.548511904761905	0.121607276410257\\
0.572916666666667	0.123962370274497\\
0.597321428571429	0.125812588006021\\
0.62172619047619	0.127306024944321\\
0.646130952380952	0.128139161128711\\
0.670535714285714	0.127393880083673\\
0.694940476190476	0.126503560121627\\
0.719345238095238	0.126161337431728\\
0.74375	0.126391498691279\\
0.768154761904762	0.126699625730892\\
0.792559523809524	0.126922983863243\\
0.816964285714286	0.12696801998743\\
0.841369047619048	0.126712498956358\\
0.86577380952381	0.125958433713429\\
0.890178571428571	0.124293831717086\\
0.914583333333333	0.120353900249808\\
0.938988095238095	0.112657329523597\\
0.963392857142857	0.0961039048536268\\
0.987797619047619	0.0719429200312423\\
};
\addlegendentry{ALM$^i$ ($\varepsilon = 7 \Delta x$)}

\legend{}
\end{axis}
\node [left] at (-0.5,-0.5) {(c)};
\end{tikzpicture}%
%
%
\begin{tikzpicture}

\begin{axis}[%
width=0.35\textwidth,
height=0.25\textwidth,
scale only axis,
xmin=0,
xmax=1,
xlabel style={font=\color{white!15!black}},
xlabel={$r/R$},
ymin=-0.00001,
ymax=0.0015,
ylabel style={font=\color{white!15!black}},
ylabel={$\frac{\Gamma - \Gamma_{ref}}{\max(\Gamma_{ref})}$},
tick label style={font=\scriptsize},
axis background/.style={fill=white},
axis x line*=bottom,
axis y line*=left,
xmajorgrids,
ymajorgrids,
legend image post style={scale=0.8},
legend style={legend cell align=left, align=left, font=\scriptsize},
]
\addplot [color=cyan, line width=1.0pt, only marks, mark size=2.0pt, mark=+, mark options={solid, cyan}]
  table[row sep=crcr]{%
0.0360119047619048	0\\
0.0604166666666667	0\\
0.0848214285714286	0\\
0.109226190476191	0\\
0.133630952380952	-1.41700195002183e-08\\
0.158035714285714	-6.89993909994135e-08\\
0.182440476190476	-8.74846612930469e-08\\
0.206845238095238	-1.15144400597567e-07\\
0.23125	-1.44178405603324e-07\\
0.255654761904762	-1.72826480612787e-07\\
0.280059523809524	-2.07194800108668e-07\\
0.304464285714286	-2.43433269098148e-07\\
0.328869047619048	-2.77339392409925e-07\\
0.353273809523809	-3.08510609392521e-07\\
0.377678571428571	-3.36093349401212e-07\\
0.402083333333333	-3.64155539001043e-07\\
0.426488095238095	-3.88522988600815e-07\\
0.450892857142857	-4.10859292393484e-07\\
0.475297619047619	-4.30488966904341e-07\\
0.499702380952381	-4.45468391105375e-07\\
0.524107142857143	-4.62691408395499e-07\\
0.548511904761905	-4.76432473611732e-07\\
0.572916666666667	-4.90190779800503e-07\\
0.597321428571429	-5.04718533494986e-07\\
0.62172619047619	-5.19432246076645e-07\\
0.646130952380952	-5.32904555400338e-07\\
0.670535714285714	-5.46051714095386e-07\\
0.694940476190476	-5.60562376100693e-07\\
0.719345238095238	-5.73133838388529e-07\\
0.74375	-5.85190442298211e-07\\
0.768154761904762	-5.90733028493773e-07\\
0.792559523809524	-5.9351355330417e-07\\
0.816964285714286	-5.91835397883989e-07\\
0.841369047619048	-5.85096792682371e-07\\
0.86577380952381	-5.71307415403455e-07\\
0.890178571428571	-5.44219488904374e-07\\
0.914583333333333	-4.99195990405488e-07\\
0.938988095238095	-4.30596380204817e-07\\
0.963392857142857	-3.35985869098776e-07\\
0.987797619047619	-2.04963036196104e-07\\
};
\addlegendentry{$\Gamma$(ALM$^i$ ($\varepsilon = 3.5 \Delta x$)) - $\Gamma$(ALM$^d$ ($\varepsilon = 3.5 \Delta x$))}

\addplot [color=red, dotted, line width=1.0pt, mark size=0.75pt, mark=diamond, mark options={solid, red}]
  table[row sep=crcr]{%
0.0360119047619048	0\\
0.0604166666666667	0\\
0.0848214285714286	0\\
0.109226190476191	0\\
0.133630952380952	8.60727349559975e-06\\
0.158035714285714	0.000811160776561702\\
0.182440476190476	0.00118481175252701\\
0.206845238095238	0.000658092081613992\\
0.23125	0.000540725070685394\\
0.255654761904762	0.000490055490414698\\
0.280059523809524	0.00050745928101889\\
0.304464285714286	0.0005590236933561\\
0.328869047619048	0.000609755010237495\\
0.353273809523809	0.000654851529184305\\
0.377678571428571	0.000691945679000508\\
0.402083333333333	0.000707454964359602\\
0.426488095238095	0.000715631829041496\\
0.450892857142857	0.000757688921477609\\
0.475297619047619	0.000802111708869097\\
0.499702380952381	0.0008321950027632\\
0.524107142857143	0.000836514326577997\\
0.548511904761905	0.0008141717885672\\
0.572916666666667	0.000790153520831907\\
0.597321428571429	0.000807075867513801\\
0.62172619047619	0.000847899082610903\\
0.646130952380952	0.000906202958253388\\
0.670535714285714	0.00096099110618339\\
0.694940476190476	0.0010083137422543\\
0.719345238095238	0.00104532486851519\\
0.74375	0.00107895443216841\\
0.768154761904762	0.00111234447334882\\
0.792559523809524	0.00114536155977799\\
0.816964285714286	0.00117601206237242\\
0.841369047619048	0.00120219860476689\\
0.86577380952381	0.00123087546093671\\
0.890178571428571	0.00125410267984501\\
0.914583333333333	0.0012396447829717\\
0.938988095238095	0.0011877778539984\\
0.963392857142857	0.000982984313186303\\
0.987797619047619	0.000788977066964708\\
};
\addlegendentry{$\Gamma$(ALM$^d$ ($\varepsilon = 7 \Delta x$)) - $\Gamma$(ALM$^d$ ($\varepsilon = 3.5 \Delta x$))}

\addplot [color=pink, line width=0.5pt, only marks, mark size=1.5pt, mark=x, mark options={solid, pink}]
  table[row sep=crcr]{%
0.0360119047619048	0\\
0.0604166666666667	0\\
0.0848214285714286	0\\
0.109226190476191	0\\
0.133630952380952	8.60254468330011e-06\\
0.158035714285714	0.000811100081916705\\
0.182440476190476	0.00118470130782861\\
0.206845238095238	0.000657927156379592\\
0.23125	0.000540514987731805\\
0.255654761904762	0.000489805582002192\\
0.280059523809524	0.000507174879930891\\
0.304464285714286	0.000558708686631798\\
0.328869047619048	0.000609412606527901\\
0.353273809523809	0.000654484752887804\\
0.377678571428571	0.000691556799158305\\
0.402083333333333	0.000707045507419499\\
0.426488095238095	0.000715203857496793\\
0.450892857142857	0.000757244383214409\\
0.475297619047619	0.000801652369691999\\
0.499702380952381	0.000831722175933289\\
0.524107142857143	0.000836028807770298\\
0.548511904761905	0.0008136744579854\\
0.572916666666667	0.000789645378371395\\
0.597321428571429	0.000806557197539415\\
0.62172619047619	0.000847370421630106\\
0.646130952380952	0.000905663898127407\\
0.670535714285714	0.000960443126682309\\
0.694940476190476	0.0010077583061984\\
0.719345238095238	0.00104476373402379\\
0.74375	0.00107839048681382\\
0.768154761904762	0.00111178141005933\\
0.792559523809524	0.00114480366191738\\
0.816964285714286	0.001175464422912\\
0.841369047619048	0.00120166805741082\\
0.86577380952381	0.00123036980931661\\
0.890178571428571	0.00125363328351979\\
0.914583333333333	0.001239227010478\\
0.938988095238095	0.0011874282460699\\
0.963392857142857	0.000982719419300601\\
0.987797619047619	0.000788814125000301\\
};
\addlegendentry{$\Gamma$(ALM$^i$ ($\varepsilon = 7 \Delta x$)) - $\Gamma$(ALM$^d$ ($\varepsilon = 3.5 \Delta x$))}

\legend{}
\end{axis}
\node [left] at (-0.5,-0.5) {(d)};
\end{tikzpicture}%
  \caption{Circulation and forces along the radial direction ($r$) at time $t/T=12$, for the blade positioned at null azimuthal angle (blade aligned with the $x$-axis)}. (a) Lift force. (b) Drag force. (c) Circulation. (d) Difference of circulation, taking as reference the circulation of ALM$^d$ with $\varepsilon = 3.5 \Delta x$, normalized by the maximum of the circulation.
  \label{fig:res_rotor}
\end{figure}
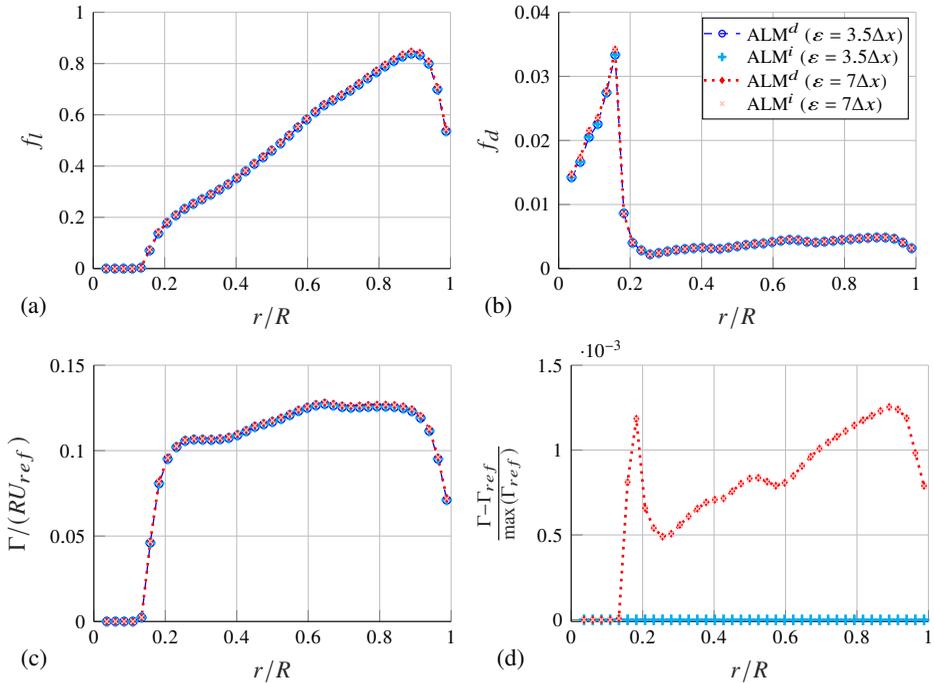

\begin{figure}
  \centering
  \input{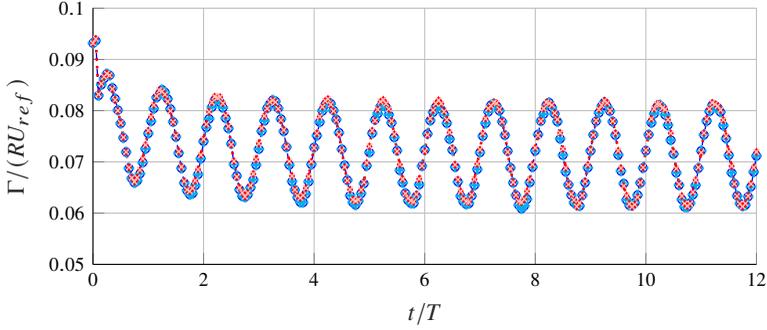}
  \caption{Evolution in time of the circulation at the point closest to the tip. The data is saved with a period of $T/20$, which is higher than the actual time step in the simulation. The correction was turned off for the first 2 points.}
  \label{fig:res_circtime}
\end{figure}

In figure \ref{fig:res_rotor} the circulation and forces at time $t=12T$ are compared for the blade that has null azimuthal angle at this time (blade aligned with the $x$-axis). The evolution in time of the circulation of the control point closest to the tip of this blade is shown in figure \ref{fig:res_circtime}.

As can be seen in figures \ref{fig:res_rotor} and \ref{fig:res_circtime}, the agreement is very good for all cases. The iterative and the direct methods have results that are practically identical. The differences in the value of circulation are lower than $10^{-5}$ if the same smearing parameter $\varepsilon$ is used. This confirms that the non-iterative procedure does not introduce errors for this unsteady flow.

The differences between the circulation for the two values of $\varepsilon$ are in the order of $1 \%$ of the value of the circulation. The differences in the forces are of the same order of magnitude. Similar differences have already been observed in previous studies \citep{meyer2019wake}. These differences are believed to be in the order of magnitude of errors due to the actuator line approximation. For example, in the code comparison of \citep{martinez2018comparison}, larger differences were observed between different implementations of the actuator line. Hence, for practical purposes, such differences can be considered negligible.

The bookkeeping of position and circulation of the free-wake (see Section \ref{sec:formvortexsheet}) begins at the first time step. But the application of the correction is turned on after $t=T/20$. The same orders of magnitude of differences were observed in figure \ref{fig:res_circtime}, even for the points right after the correction is turned on (after the first 2 points).

It should be noted that the effects of the smearing parameter can never be completely eliminated. The smearing correction has the objective of correcting the forces at the blades. Possibly, a better understanding of the role of drag, viscous effects, unsteadiness or other phenomena might reduce even further the errors caused by the Gaussian smearing. However, some errors cannot be eliminated, because the discretization used in actuator line simulations does not usually support vortices with very low vortex core size. Hence, the vorticity that is shed is still smeared. The smeared vortices create a wake that is different depending on the choice of $\varepsilon$, as can be seen in \citep{meyer2019wake}, especially in the near-wake. As a consequence, differences in the wake can have an effect on the forces. Errors in the velocity have a first-order effect in the circulation and forces, hence, if the modified near-wake has a slightly different blockage profile, then the velocities at the turbine and, consequently, the forces would be affected.

\section{Conclusions} \label{sec:conclusions}

The conception of the smearing correction by \citet{dag2020new} (originally in \citep{dag2017combined}) and by \citet{martinez2019filtered} (originally in \citep{martinez2017large}), the formal analysis provided by \citet{martinez2019filtered} and its improvement and analogy to a viscous lifting line by \citet{meyer2019vortex,meyer2020brief} were great advancements to the actuator line method. These steps provided not only a more accurate, general and reliable method but also an understanding of the mathematical and physical reasons for the errors of the uncorrected actuator line. However, these methods resort to iterative procedures with a relaxation factor, which are generally slower, less stable and less deterministic than direct methods. The previous method that did not employ an iterative method~\citep{martinez2019filtered} had to waive the compatibility between circulation and induced velocity at each time step, which may cause errors in unsteady simulations.

In the present work, a non-iterative vortex-based smearing correction for the actuator line method is proposed and validated. Based on the linearization of a lifting line method, the iterative procedures of previous works have been substituted by the direct solution of a small linear system. For the cases tested, no significant difference is observed in the results of the iterative method and the non-iterative method.  All differences are presumed to be orders of magnitude lower than the accuracy of the actuator line method.

The smearing correction reduces the effect of the smearing parameter $\varepsilon$ on the forces. However, differences, which are in the order of the errors of the actuator line approximation, were observed in the forces and circulation for different smearing parameters. This indicates that the smearing parameter still affects the simulation, which agrees with the observation that smearing has an effect on the near wake \citep{meyer2019wake}. Nevertheless, for practical applications, the differences observed in the forces can be considered negligible.

Another contribution of the present work includes the use of a correction function based on the velocity induced by a smeared vortex segment, which eliminates the need for approximations of the correction function. Also, we implement a free-vortex wake model to define the vortex sheet based on the CFD velocities. This keeps the generality of the method and makes it applicable to several configurations without the need for \emph{ad hoc} assumptions. If computational cost is prioritized, one could use approximate functions, prescribe the wakes, or apply one of the methods for increasing the speed from \citet{meyer2020brief}. The focus of this study is on presenting and validating the technique. Many options to reduce the computational cost even further could be implemented. Additionally, the optimal choice of parameters for a desired accuracy is a matter of investigation. For these reasons, the analysis of the computational cost of different choices of options and parameters of the correction is left for future studies.

In order to further understand the limitations of the smearing correction, the comparison to blade-resolved simulations is suggested for future studies. Also, the implementation of three-dimensional corrections for drag and the understanding of unsteady effects are current areas of active research (see~\citep{kleine2022onstability}).

Additionally, by carefully constructing a non-linear lifting line method and an actuator line method that are consistent with each other, we arrive at differences in the induced velocity that are considerably better than differences reported in the literature. The good agreement serves as an \emph{a posteriori} justification for the linearization of the smearing correction, in addition to the proof that the iterative lifting line method and the ALM with vortex-based smearing correction are mathematically identical if the circulation in each segment of the actuator line is assumed constant.

So far, the actuator line method has been used primarily by the rotor aerodynamics community, possibly because of its known limitations to replicate the forces on non-rotating wings. The use of the smearing correction with a general formulation removes this limitation and reproduces the results of a lifting line method, as shown here and in previous works. This could motivate other communities, such as the aeronautical community, to also adopt this technique to simulate wings, taking advantage of the lower grid requirements allowed by the ALM, in flow conditions more complex than the ones allowed by the lifting line method.

\backsection[Acknowledgements]{The computations were performed on resources provided by the Swedish National Infrastructure for Computing (SNIC) at the High Performance Computing Center North (HPC2N). This work was conducted within StandUp for Wind. V.G.K. thanks KTH Engineering Mechanics for partially funding this work. The authors appreciate the contributions of the anonymous reviewers, especially the reviewer who contributed to the development of the analytical formula for the velocity induced by a smeared vortex segment.}

\backsection[Declaration of interests]{The authors report no conflict of interest.}

\backsection[Author ORCID]{V.G. Kleine, https://orcid.org/0000-0001-9360-7300; A. Hanifi, https://orcid.org/0000-0002-5913-5431; D.S. Henningson, https://orcid.org/0000-0001-7864-3071}

\appendix
\section{Linearized lifting line} \label{sec:LLL}

Equation \eqref{eq:GammaNLL} can be linearized around the undisturbed velocities using the Taylor series expansion:
\begin{equation}
  \Gamma_j = \Gamma_{0j} + \frac{1}{2} \, c_j \left[ \left( C_l(\alpha_{0j}) \frac{U_y}{U_0} + \frac{\partial C_l}{\partial \alpha}(\alpha_{0j}) \frac{U_z}{U_0} \right) u_{y}^{vi} + \left( C_l(\alpha_{0j}) \frac{U_z}{U_0} - \frac{\partial C_l}{\partial \alpha}(\alpha_{0j}) \frac{U_y}{U_0} \right) u_{z}^{vi} \right] ,
  \label{eq:Gamma_jLLL2}
\end{equation}
where
\begin{equation}
  U_{0} \defeq \sqrt{(U_z^2+U_y^2)},
\end{equation}
\begin{equation}
  \alpha_{0j} \defeq \alpha_{gj} + \arctan{\left(\frac{U_y}{U_z}\right)}
\end{equation}
\begin{equation}
  \Gamma_{0j} \defeq \frac{1}{2} U_{0} \, c_j \, C_l(\alpha_{0j}) ,
\end{equation}
and $\frac{\partial C_l}{\partial \alpha}(\alpha_{0j})$ is the slope of the lift coefficient evaluated at angle $\alpha_{0j}$.

Defining
\begin{equation}
  b_{yj} \defeq \frac{1}{2} \, c_j \left( C_l(\alpha_{0j}) \frac{U_y}{U_0} + \frac{\partial C_l}{\partial \alpha}(\alpha_{0j}) \frac{U_z}{U_0} \right) 
  \label{eq:byLLL}
\end{equation}
\begin{equation}
  b_{zj} \defeq \frac{1}{2} \, c_j  \left( C_l(\alpha_{0j}) \frac{U_z}{U_0} - \frac{\partial C_l}{\partial \alpha}(\alpha_{0j}) \frac{U_y}{U_0} \right)
  \label{eq:bzLLL}
\end{equation}
we arrive at
\begin{equation}
  \Gamma_j = \Gamma_{0j} + b_{yj} u_{y}^{vi} + b_{zj} u_{z}^{vi} .
\label{eq:LNLL}
\end{equation}
where we see that coefficients $b_{yj}$ and $b_{zj}$ are a measure of the sensitivity of the circulation to a change in the velocities in $y$ and $z$ directions.

To avoid an iterative process, another relationship between the induced velocities and circulation must be used. This relationship is the one described in step (\ref{item:velocity}) of the iterative process of Section \ref{sec:NLLL}. The velocities induced at control point $j$ by the vortex system $k$ (which includes the bound vortex and the vortex sheet created by the segment $k$) can be written as:
\begin{equation}
  u_{y}^{vi_k}(\mathbf{x}_j) = a_{y,jk} \Gamma_k
\end{equation}
\begin{equation}
  u_{z}^{vi_k}(\mathbf{x}_j) = a_{z,jk} \Gamma_k
\end{equation}
where $a_{y,jk}$ and $a_{y,jk}$ are called the influence coefficients. The values of the influence coefficients are obtained from the Biot-Savart law and depend only on the shape of the vortex filament and its relative position to the control point. The formula for the influence coefficient for some common vortex filament shapes can be found in \citep{katz1991low}. Then, the total induced velocities are found by considering all $N$ vortex systems created by the $N$ segments:
\begin{equation}
  \begin{bmatrix}
    u_{y}^{vi}(\mathbf{x}_1)  \\
    u_{y}^{vi}(\mathbf{x}_2)  \\
    \vdots  \\
    u_{y}^{vi}(\mathbf{x}_N)  \\
  \end{bmatrix} =
  \begin{bmatrix}
    a_{y,11} & a_{y,12} & \cdots & a_{y,1N} \\
    a_{y,21} & a_{y,22} & \cdots & a_{y,2N} \\
    \vdots   &          & \ddots & \vdots \\
    a_{y,N1} & a_{y,N2} & \cdots & a_{y,NN} \\
  \end{bmatrix}
  \begin{bmatrix}
    \Gamma_1 \\
    \Gamma_2 \\
    \vdots  \\
    \Gamma_N
  \end{bmatrix}
\end{equation}
that can be written in matrix form as
\begin{equation}
  \mathsfbi{u_y}^{vi} = \mathsfbi{A_y} \boldsymbol{\Gamma}
  \label{eq:u_yLLL}
\end{equation}
and analogously for $u_{z}^{v}$:
\begin{equation}
  \mathsfbi{u_z}^{vi} = \mathsfbi{A_z} \boldsymbol{\Gamma} .
  \label{eq:u_zLLL}
\end{equation}

Equation \eqref{eq:LNLL} can also be written in matrix form:
\begin{equation}
  \boldsymbol{\Gamma} = \boldsymbol{\Gamma}_0 + \mathsfbi{b_y} \circ \mathsfbi{u_y}^{vi} + \mathsfbi{b_z} \circ \mathsfbi{u_z}^{vi} = \boldsymbol{\Gamma}_0 + (diag(\mathsfbi{b_y}) \mathsfbi{A_y}) \boldsymbol{\Gamma} + (diag(\mathsfbi{b_z}) \mathsfbi{A_z}) \boldsymbol{\Gamma}
\end{equation}
where $\circ$ denotes the element-wise product and $diag(\mathsfbi{b})$ is the square diagonal matrix formed by the elements of column vector $\mathsfbi{b}$. The circulation is finally found by solving the linear system
\begin{equation}
  \left(\mathsfbi{I} - diag(\mathsfbi{b_y}) \mathsfbi{A_y} - diag(\mathsfbi{b_z}) \mathsfbi{A_z} \right) \boldsymbol{\Gamma} = \boldsymbol{\Gamma}_0 ,
  \label{eq:circLLL}
\end{equation}
where $\mathsfbi{I}$ is the identity matrix. Therefore, equation \eqref{eq:circLLL} provides a direct way to calculate the circulation along the wing, instead of the iteration of Section \ref{sec:NLLL}. If needed, the induced velocities can be found using relations \eqref{eq:u_yLLL} and \eqref{eq:u_zLLL}.

It is worth remembering that the current reference system is related to the local reference system of figure \ref{fig:2dref}. Equation \eqref{eq:Gamma_jLLL2}, for example, depends only on the local variables. The only term that takes into account the contributions of other sections of the blades are the influence coefficients, hence, these are defined first in an absolute system of reference and then transformed into the local system of reference. Also, there is no constraint regarding the values or direction of the velocity (as opposed to the classical linearization of the lifting line method, which requires constant $U_z$ and $U_z \gg U_y$). Therefore, this method is directly applicable to any system of reference, including the case of rotating blades, both in a fixed frame of reference and a rotating frame of reference.

This formulation is reduced to one of the classical implementations of the numerical lifting line for $U_y=0$ and constant $U_z=U_{\infty}$, when an ideal airfoil and horseshoe vortices are employed. In this case, equation \eqref{eq:circLLL} becomes
\begin{equation}
  - \frac{1}{2 \pi} \left(diag(\mathsfbi{c/2})\right)^{-1} \boldsymbol{\Gamma} + \mathsfbi{A_y}  \boldsymbol{\Gamma} = - U_{\infty} \boldsymbol{\alpha_{g}} ,
  \label{eq:classicalLLapp}
\end{equation}
in which each term of equation \eqref{eq:classicalLLapp} can be directly compared to a term in equation 8.11 of \citep{katz1991low}:
\begin{equation}
  - \frac{\Gamma (x)}{2 \pi \, \frac{c(x)}{2}} - \frac{1}{4 \pi}\int_{x_{min}}^{x_{max}}\frac{\frac{-\partial \Gamma(x')}{dx}}{x-x'}dx' = - U_{\infty} \alpha_g
  \label{eq:katzLL}
\end{equation}
where the notation of \citep{katz1991low} was modified to be consistent with our notation.

\section{Resolution of the Gaussian vortex core} \label{app:grid}

Usually, it is desirable to employ a low smearing parameter. The minimum smearing parameter is commonly taken as $\varepsilon \approx 2 \Delta x$~\citep{troldborg2009actuator,martinez2015large}. However, for $\varepsilon = 2 \Delta x$, the Gaussian vortex core may not be well represented in the numerical grid. As mentioned in Section~\ref{sec:res_wing}, the low resolution of the vortex core may introduce errors, because the theory of Section~\ref{sec:vorticity} is based on a perfect representation of the Gaussian distribution of vorticity in the vortex core.

\begin{figure}
  \centering
%
%
\definecolor{mycolor1}{rgb}{1.00000,1.00000,0.00000}%
\begin{tikzpicture}

\begin{axis}[%
width=0.35\textwidth,
height=0.25\textwidth,
scale only axis,
xmin=0.0,
xmax=0.5,
xtick={0.0, 0.1, 0.2, 0.3, 0.4, 0.5},
xlabel style={font=\color{white!15!black}},
xlabel={$x/R$},
ymin=-0.002,
ymax=0.0005,
ytick={-0.002, -0.0015,..., 0.0005},
ylabel style={font=\color{white!15!black}},
ylabel={$(u^c_y - u^c_{yLL})/U_z$},
xmajorgrids,
ymajorgrids,
tick label style={font=\scriptsize},
axis background/.style={fill=white},
axis x line*=bottom,
axis y line*=left,
]
\addplot [color=olive, dashed, line width=1.0pt, mark size=1.5pt, mark=diamond, mark options={solid, olive}]
  table[row sep=crcr]{%
-0.49	-0.000527432278229184\\
-0.47	-0.000252560377386871\\
-0.45	-0.000279768197336897\\
-0.43	-0.000483907229174761\\
-0.41	-0.000648193299741584\\
-0.39	-0.000823823647044267\\
-0.37	-0.000923066681447739\\
-0.35	-0.00101898447629247\\
-0.33	-0.00109094916994872\\
-0.31	-0.00115188964171492\\
-0.29	-0.00119874342827684\\
-0.27	-0.00124004291095235\\
-0.25	-0.00127519761462496\\
-0.23	-0.0013000254732498\\
-0.21	-0.0013235723299468\\
-0.19	-0.00134367805560848\\
-0.17	-0.00136126506797021\\
-0.15	-0.00137632324179219\\
-0.13	-0.00138896093813202\\
-0.11	-0.00139870617757521\\
-0.09	-0.00140594068343628\\
-0.07	-0.00141090175120697\\
-0.05	-0.00141389571297143\\
-0.03	-0.00141571490872847\\
-0.01	-0.00141716265906225\\
0.01	-0.00141716402892953\\
0.03	-0.00141571685940481\\
0.05	-0.00141389812140897\\
0.07	-0.00141090549660821\\
0.09	-0.00140594566775464\\
0.11	-0.00139871090298554\\
0.13	-0.00138896469420273\\
0.15	-0.00137632662304337\\
0.17	-0.00136126700796181\\
0.19	-0.00134367830567811\\
0.21	-0.00132357174169295\\
0.23	-0.00130002449415838\\
0.25	-0.00127519504595059\\
0.27	-0.00124004112118318\\
0.29	-0.00119874100982394\\
0.31	-0.00115188559994912\\
0.33	-0.00109094351657467\\
0.35	-0.00101897806369091\\
0.37	-0.000923060183014753\\
0.39	-0.000823816105075384\\
0.41	-0.000648186047816912\\
0.43	-0.000483901011798328\\
0.45	-0.000279762549240517\\
0.47	-0.00025255533795332\\
0.49	-0.000527428814431755\\
};
\addlegendentry{ALM$^d$ ($\varepsilon = R/28 = 2 \Delta x$)}

\addplot [color=black, line width=0.5pt, mark size=1.5pt, mark=o, mark options={solid, black}]
  table[row sep=crcr]{%
-0.49	-0.000221916735371974\\
-0.47	0.000309313061101435\\
-0.45	0.000430305176602301\\
-0.43	0.000355075252081435\\
-0.41	0.000240053064902816\\
-0.39	0.000139170007239335\\
-0.37	6.17972934612646e-05\\
-0.35	2.81162093073231e-06\\
-0.33	-4.15472784554179e-05\\
-0.31	-7.52424502827205e-05\\
-0.29	-0.000101173994076841\\
-0.27	-0.00012134518783485\\
-0.25	-0.000137200009651964\\
-0.23	-0.000149726106862001\\
-0.21	-0.000159735895131199\\
-0.19	-0.000167765193323979\\
-0.17	-0.000174186914015113\\
-0.15	-0.000179358613345691\\
-0.13	-0.000183430568036321\\
-0.11	-0.000186701721692415\\
-0.09	-0.000189292621699779\\
-0.07	-0.000191288912918169\\
-0.05	-0.000192762701600624\\
-0.03	-0.000193748983756768\\
-0.01	-0.000194251793500553\\
0.01	-0.000194248591923525\\
0.03	-0.00019373857675601\\
0.05	-0.00019274880530817\\
0.07	-0.000191275855506306\\
0.09	-0.000189282066441337\\
0.11	-0.000186697479318747\\
0.13	-0.00018343238526513\\
0.15	-0.000179366122183167\\
0.17	-0.000174198001377412\\
0.19	-0.000167770131997411\\
0.21	-0.000159743022870855\\
0.23	-0.000149730472580385\\
0.25	-0.000137201847418791\\
0.27	-0.000121342438737182\\
0.29	-0.000101167437060536\\
0.31	-7.52394700173145e-05\\
0.33	-4.15369102478649e-05\\
0.35	2.81784269478419e-06\\
0.37	6.18015027537486e-05\\
0.39	0.000139173821861113\\
0.41	0.000240057991871885\\
0.43	0.000355076423856272\\
0.45	0.000430307901993582\\
0.47	0.000309317601410786\\
0.49	-0.000221916989775556\\
};
\addlegendentry{ALM$^d$ ($\varepsilon = R/28 = 4 \Delta x$)}

\legend{}
\end{axis}
\node [left] at (-0.5,-0.5) {(a)};
\end{tikzpicture}%
%
%
\definecolor{mycolor1}{rgb}{1.00000,1.00000,0.00000}%
\begin{tikzpicture}

\begin{axis}[%
width=0.35\textwidth,
height=0.25\textwidth,
scale only axis,
xmin=0.0,
xmax=0.5,
xtick={0.0, 0.1, 0.2, 0.3, 0.4, 0.5},
xlabel style={font=\color{white!15!black}},
xlabel={$x/R$},
ymin=-0.008,
ymax=0.000,
ylabel style={font=\color{white!15!black}},
ylabel={$(u^c_z - u^c_{zLL})/U_z$},
xmajorgrids,
ymajorgrids,
tick label style={font=\scriptsize},
axis background/.style={fill=white},
axis x line*=bottom,
axis y line*=left,
]
\addplot [color=olive, dashed, line width=1.0pt, mark size=1.5pt, mark=diamond, mark options={solid, olive}]
  table[row sep=crcr]{%
-0.49	-0.00110789556054813\\
-0.47	-0.00588917766356101\\
-0.45	-0.00732200879067757\\
-0.43	-0.00714266278781595\\
-0.41	-0.00666584929034209\\
-0.39	-0.00619565208362616\\
-0.37	-0.00589610011766095\\
-0.35	-0.00563382729432971\\
-0.33	-0.0054528983504547\\
-0.31	-0.00531409683540252\\
-0.29	-0.00520977754818697\\
-0.27	-0.00512202364116332\\
-0.25	-0.00505485587570875\\
-0.23	-0.00500653317880684\\
-0.21	-0.00496307260806483\\
-0.19	-0.00493006582928845\\
-0.17	-0.00490383827801411\\
-0.15	-0.00488201394616916\\
-0.13	-0.00486469033042647\\
-0.11	-0.00485349445215422\\
-0.09	-0.00484495935685403\\
-0.07	-0.00483711397823972\\
-0.05	-0.0048314588145576\\
-0.03	-0.00482922034136446\\
-0.01	-0.00482862203097778\\
0.01	-0.00482861680248892\\
0.03	-0.00482922130549746\\
0.05	-0.00483146484267505\\
0.07	-0.00483712157182548\\
0.09	-0.00484497068499323\\
0.11	-0.0048535092390164\\
0.13	-0.00486469930040145\\
0.15	-0.00488202519048719\\
0.17	-0.00490384468831384\\
0.19	-0.00493006732763379\\
0.21	-0.00496306958037518\\
0.23	-0.00500652741554619\\
0.25	-0.00505485887373147\\
0.27	-0.0051220172839006\\
0.29	-0.00520977130001077\\
0.31	-0.00531409567839669\\
0.33	-0.00545289963900564\\
0.35	-0.00563382581320371\\
0.37	-0.00589610064910995\\
0.39	-0.00619565993252202\\
0.41	-0.00666585697266053\\
0.43	-0.0071426665403248\\
0.45	-0.00732201041196501\\
0.47	-0.00588917690614577\\
0.49	-0.00110789421635005\\
};
\addlegendentry{ALM$^d$ ($\varepsilon = R/28 = 2 \Delta x$)}

\addplot [color=black, line width=0.5pt, mark size=1.5pt, mark=o, mark options={solid, black}]
  table[row sep=crcr]{%
-0.49	-0.000462210451076417\\
-0.47	-0.00501988164939204\\
-0.45	-0.00614332384683569\\
-0.43	-0.00564706677840976\\
-0.41	-0.00483462849824723\\
-0.39	-0.00412483875979797\\
-0.37	-0.00356945304440692\\
-0.35	-0.00313181162287079\\
-0.33	-0.0027853618439363\\
-0.31	-0.00250745671330832\\
-0.29	-0.00228196023791138\\
-0.27	-0.0020974048489476\\
-0.25	-0.00194490419372795\\
-0.23	-0.00181886820285382\\
-0.21	-0.00171391991289493\\
-0.19	-0.00162635993535798\\
-0.17	-0.00155354146068709\\
-0.15	-0.0014929528861245\\
-0.13	-0.00144345475555485\\
-0.11	-0.00140297272176793\\
-0.09	-0.00137051495813256\\
-0.07	-0.00134545883989967\\
-0.05	-0.00132717731451426\\
-0.03	-0.0013151537367927\\
-0.01	-0.00130916713858953\\
0.01	-0.00130917411721521\\
0.03	-0.00131519166532468\\
0.05	-0.00132720988920609\\
0.07	-0.00134548834053305\\
0.09	-0.00137055775208417\\
0.11	-0.00140298579407594\\
0.13	-0.00144345461649109\\
0.15	-0.00149294014976975\\
0.17	-0.00155349855793396\\
0.19	-0.00162633633455889\\
0.21	-0.00171388269618389\\
0.23	-0.00181885408730043\\
0.25	-0.00194490365001165\\
0.27	-0.00209742297019722\\
0.29	-0.00228200011501134\\
0.31	-0.00250746473322438\\
0.33	-0.0027853982810756\\
0.35	-0.00313180904286758\\
0.37	-0.00356944154018757\\
0.39	-0.0041248234209087\\
0.41	-0.00483462684688551\\
0.43	-0.00564705222320294\\
0.45	-0.00614332588137623\\
0.47	-0.00501990369063376\\
0.49	-0.000462211276036628\\
};
\addlegendentry{ALM$^d$ ($\varepsilon = R/28 = 4 \Delta x$)}

\legend{}
\end{axis}
\node [left] at (-0.5,-0.5) {(b)};
\end{tikzpicture}%
%
%
\definecolor{mycolor1}{rgb}{1.00000,1.00000,0.00000}%
\begin{tikzpicture}

\begin{axis}[%
width=0.35\textwidth,
height=0.25\textwidth,
scale only axis,
xmin=0.0,
xmax=0.5,
xtick={0.0, 0.1, 0.2, 0.3, 0.4, 0.5},
xlabel style={font=\color{white!15!black}},
xlabel={$x/R$},
ymin=-0.02,
ymax=0.00,
ylabel style={font=\color{white!15!black}},
ylabel={$(\Gamma-\Gamma_{LL})/\Gamma_{0}$},
xmajorgrids,
ymajorgrids,
tick label style={font=\scriptsize},
axis background/.style={fill=white},
axis x line*=bottom,
axis y line*=left,
legend image post style={scale=0.8},
legend style={legend cell align=left, align=left, font=\scriptsize},
legend pos=outer north east
]
\addplot [color=olive, dashed, line width=1.0pt, mark size=1.5pt, mark=diamond, mark options={solid, olive}]
  table[row sep=crcr]{%
-0.49	-0.00443507886640593\\
-0.47	-0.00745777638749335\\
-0.45	-0.00903391267983344\\
-0.43	-0.0101286788803439\\
-0.41	-0.0106847444873899\\
-0.39	-0.0113196509821185\\
-0.37	-0.0116474913964535\\
-0.35	-0.0119910232493258\\
-0.33	-0.0122652757692833\\
-0.31	-0.0125119530714997\\
-0.29	-0.0127043373072392\\
-0.27	-0.0128779967366291\\
-0.25	-0.0130333388883044\\
-0.23	-0.013142503361454\\
-0.21	-0.0132481987812206\\
-0.19	-0.0133425377503452\\
-0.17	-0.013427659765888\\
-0.15	-0.0135011552471181\\
-0.13	-0.0135638234081422\\
-0.11	-0.0136143465748793\\
-0.09	-0.013651685035811\\
-0.07	-0.0136753520622324\\
-0.05	-0.0136887725219126\\
-0.03	-0.0136981409000377\\
-0.01	-0.0137067167868712\\
0.01	-0.0137067201416493\\
0.03	-0.0136981540851414\\
0.05	-0.0136887936378206\\
0.07	-0.0136753831188632\\
0.09	-0.0136517275869996\\
0.11	-0.0136143909610092\\
0.13	-0.0135638559049258\\
0.15	-0.0135011876685079\\
0.17	-0.0134276783255848\\
0.19	-0.0133425408144858\\
0.21	-0.0132481920705149\\
0.23	-0.0131424914691506\\
0.25	-0.0130333258043037\\
0.27	-0.0128779791780877\\
0.29	-0.0127043159254498\\
0.31	-0.0125119266254078\\
0.33	-0.0122652416927786\\
0.35	-0.0119909816668177\\
0.37	-0.0116474513029522\\
0.39	-0.0113196116975105\\
0.41	-0.0106847068701721\\
0.43	-0.0101286438203441\\
0.45	-0.00903387906774061\\
0.47	-0.00745774422803916\\
0.49	-0.00443505598915264\\
};
\addlegendentry{ALM$^d$ ($\varepsilon = R/28 = 2 \Delta x$)}

\addplot [color=black, line width=0.5pt, mark size=1.5pt, mark=o, mark options={solid, black}]
  table[row sep=crcr]{%
-0.49	-0.00189149524267488\\
-0.47	-0.0030884008942064\\
-0.45	-0.00342715992850282\\
-0.43	-0.00339724664219436\\
-0.41	-0.0033067946740969\\
-0.39	-0.00323212312260947\\
-0.37	-0.00316479611305511\\
-0.35	-0.00309971710381388\\
-0.33	-0.00303376064596356\\
-0.31	-0.0029691063252569\\
-0.29	-0.00290783521174826\\
-0.27	-0.00285109031006955\\
-0.25	-0.00279908885645324\\
-0.23	-0.00275247769997145\\
-0.21	-0.00271101057952342\\
-0.19	-0.00267437836519347\\
-0.17	-0.00264229233525609\\
-0.15	-0.0026145363231443\\
-0.13	-0.00259091245173052\\
-0.11	-0.0025712254848238\\
-0.09	-0.00255524381259398\\
-0.07	-0.0025428853023536\\
-0.05	-0.00253397769253851\\
-0.03	-0.00252822578794727\\
-0.01	-0.00252543527948218\\
0.01	-0.00252542219514232\\
0.03	-0.002528198498273\\
0.05	-0.00253392318639045\\
0.07	-0.00254283298090838\\
0.09	-0.00255522046276455\\
0.11	-0.0025712119742238\\
0.13	-0.00259092369807812\\
0.15	-0.002614570630986\\
0.17	-0.00264231889650498\\
0.19	-0.00267438570436205\\
0.21	-0.00271101801335572\\
0.23	-0.00275249092787411\\
0.25	-0.00279909982027656\\
0.27	-0.00285109121554011\\
0.29	-0.00290783403411901\\
0.31	-0.00296909569194281\\
0.33	-0.00303373220173479\\
0.35	-0.00309967561074976\\
0.37	-0.00316475829689199\\
0.39	-0.00323208395290144\\
0.41	-0.00330676224805713\\
0.43	-0.00339722478500262\\
0.45	-0.00342714495949609\\
0.47	-0.00308839460770814\\
0.49	-0.00189149764728958\\
};
\addlegendentry{ALM$^d$ ($\varepsilon = R/28 = 4 \Delta x$)}

\end{axis}
\node [left] at (-0.5,-0.5) {(c)};
\end{tikzpicture}%
  \caption{Influence of grid resolution for the same value of smearing parameter. (a) Difference between $y$-velocities obtained by the ALM and LL. (b) Difference between $z$-velocities obtained by the ALM and LL. (c) Relative difference between circulation obtained by the ALM and LL.}
  \label{fig:res_LLACLstraightgrid}
\end{figure}
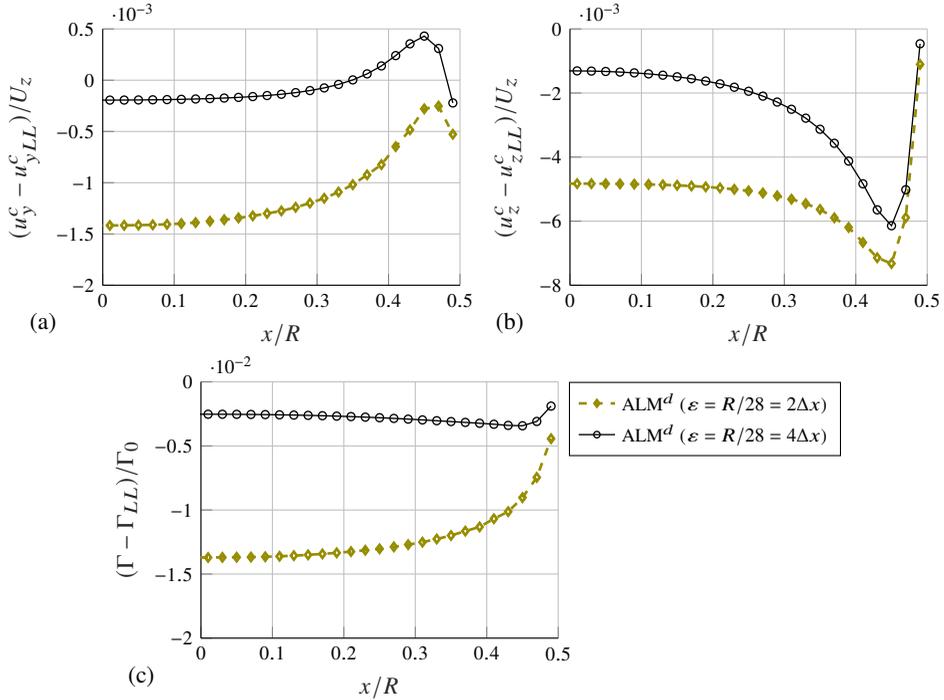

In order to exemplify this error, the case of Section~\ref{sec:res_wing} was simulated with the standard grid ($\Delta x = R/56$ in the region of the actuator line) for a low smearing parameter of $\varepsilon=2 \Delta x = R/28$. This case is compared to a simulation with the same smearing parameter, but a higher grid resolution in figure~\ref{fig:res_LLACLstraightgrid}.

 For $\varepsilon=2 \Delta x$, the difference in the induced velocity $u_y$ is in the order of $10^{-3}$ ($0.1\%$ of the undisturbed velocity). It is still considerably better than the results found in the literature, and a generally acceptable error for an ALM method. However, it is a clear deterioration from the agreement found for larger smearing parameters in Section~\ref{sec:res_wing}. By increasing the grid resolution to $\Delta x=R/112$, so that $\varepsilon=R/28=4 \Delta x$, the differences are back to the order of magnitude discussed in Section~\ref{sec:res_wing}.
 
\begin{figure}
    \centering
        \sbox0{\includegraphics[width=1.0\textwidth]{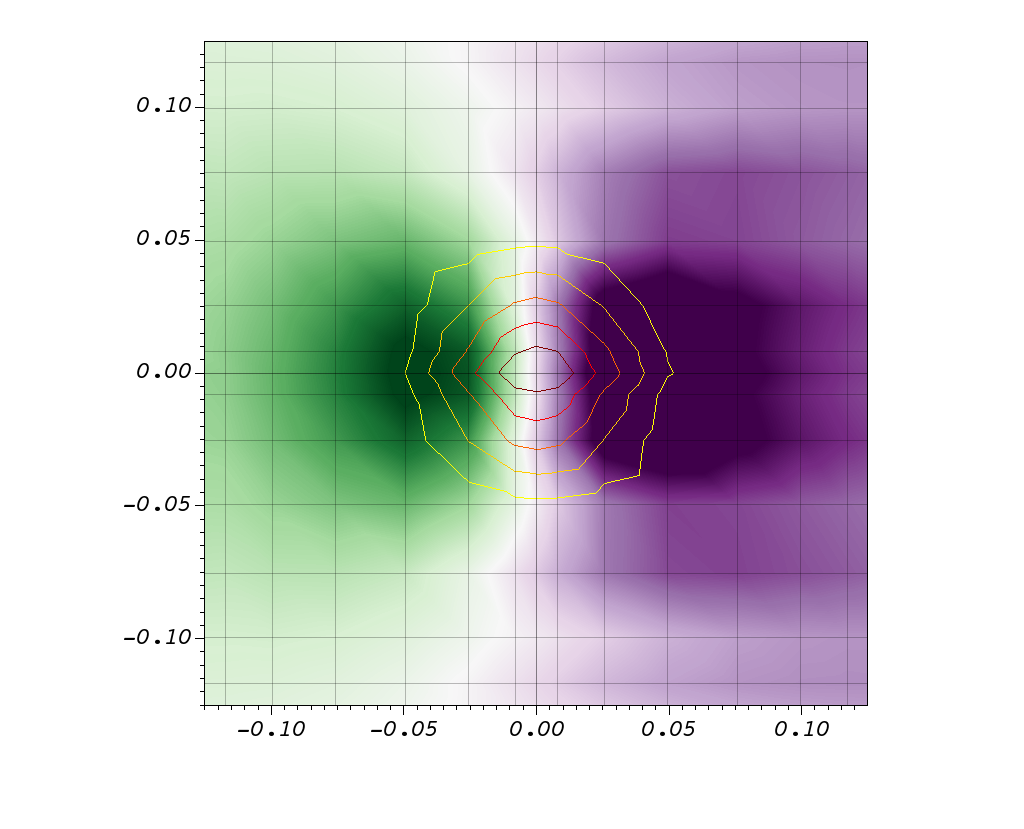}}
    \begin{tikzpicture}
      \node[anchor=south west,inner sep=0] (image) at (0,0) {\includegraphics[clip,trim={.22\wd0} {.15\ht0} {.35\wd0} {.1\ht0}, width=0.45\textwidth]{Figures/vcore_eps2.png}};
      \sbox0{{\includegraphics[clip,trim={.22\wd0} {.15\ht0} {.35\wd0} {.1\ht0}, width=0.45\textwidth]{Figures/vcore_eps2.png}}}
      \node at (0.05\wd0,0.05\ht0) {(a)};
      \node at (0.56\wd0,0.02\ht0) {\small{$z/R$}};
      \node at (0.02\wd0,0.54\ht0) {\small{$\frac{y}{R}$}};
    \sbox0{\includegraphics[width=1.0\textwidth]{Figures/vcore_eps2.png}}
      \node[anchor=south west,inner sep=0] (image) at (0.45\wd0,0) {\includegraphics[clip,trim={.22\wd0} {.15\ht0} {.35\wd0} {.1\ht0}, width=0.45\textwidth]{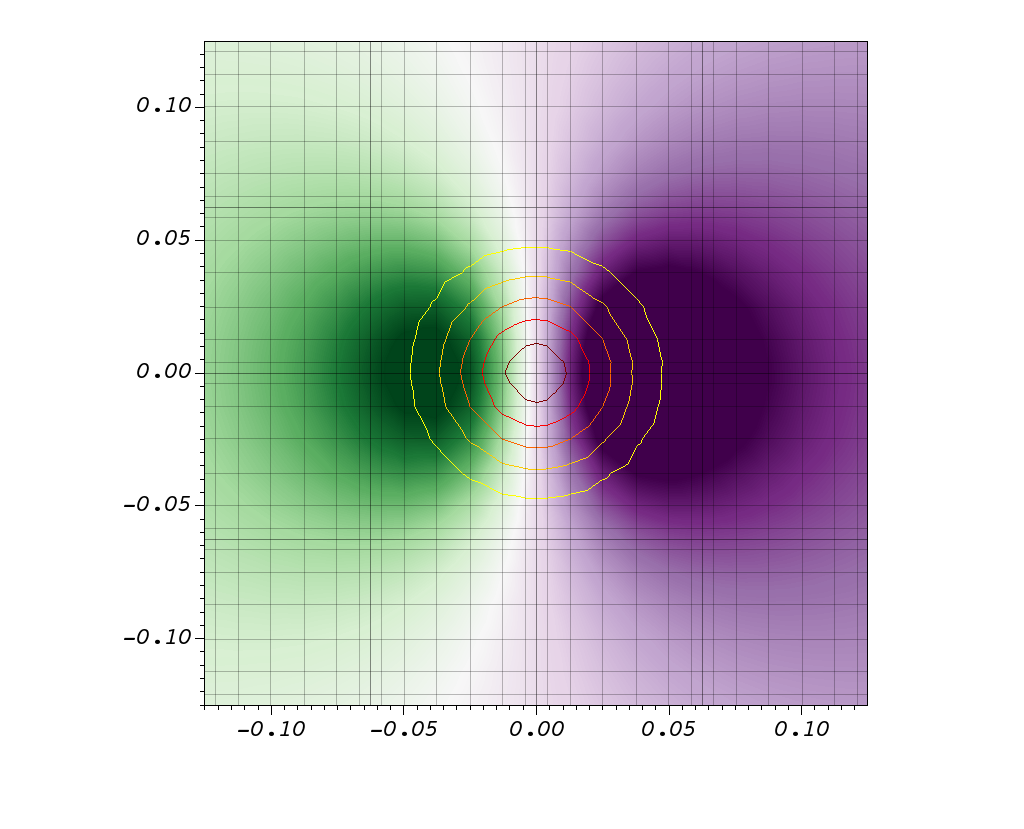}};
      \sbox0{{\includegraphics[clip,trim={.22\wd0} {.15\ht0} {.35\wd0} {.1\ht0}, width=0.45\textwidth]{Figures/vcore_eps4.png}}}
      \node at (1.05\wd0,0.05\ht0) {(b)};
      \node at (1.55\wd0,0.02\ht0) {\small{$z/R$}};
      \node[anchor=south west,inner sep=0] (image) at (1.98\wd0,0.09\ht0) {\includegraphics[clip,trim={.15\wd0} {.0\ht0} {.03\wd0} {.0\ht0}, height=0.9\ht0]{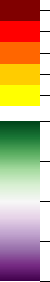}};
      \node at (2.09\wd0,0.095\ht0) {\scriptsize{$-0.10$}};
      \node at (2.09\wd0,0.222\ht0) {\scriptsize{$-0.05$}};
      \node at (2.09\wd0,0.350\ht0) {\scriptsize{$0.0$}};
      \node at (2.09\wd0,0.477\ht0) {\scriptsize{$0.05$}};
      \node at (2.09\wd0,0.605\ht0) {\scriptsize{$0.10$}};
      \node at (2.07\wd0,0.69\ht0) {\scriptsize{$2$}};
      \node at (2.07\wd0,0.76\ht0) {\scriptsize{$4$}};
      \node at (2.07\wd0,0.825\ht0) {\scriptsize{$6$}};
      \node at (2.07\wd0,0.895\ht0) {\scriptsize{$8$}};
      \node at (2.08\wd0,0.96\ht0) {\scriptsize{$10$}};
    \end{tikzpicture}
    \caption{Comparison of the influence of grid resolution for the same smearing parameter ($\varepsilon=R/28$). Pseudocolor of $u_y$ and contours of total vorticity in the vortex core. (a) $\Delta x = R/56 = \varepsilon/2$.  (b) $\Delta x = R/112 = \varepsilon/4$.}
    \label{fig:vortexcore}
\end{figure}

 Observing the velocity $u_y$ and the vorticity in the region of the bound vortex in the plane of symmetry ($x=0$), in figure~\ref{fig:vortexcore}, it is possible to note that the poorer grid does not have adequate resolution to represent the Gaussian vortex core. We conclude that this is the main cause of the high values of error in $u_y$ for the coarser grid. This error has already been identified by~\cite{shives2013mesh}, where a smearing parameter of $\varepsilon=4 \Delta x$ was recommended to reduce errors in the angle of attack, and is further discussed by~\citet{forsting2020generalised}. \citet{meyer2019vortex} also employed a larger smearing parameter than the recommended minimum in order to avoid this error.

Hence, for $\varepsilon=2 \Delta x$, it can be concluded that the error in $u_y$ is mostly dependent on the grid resolution, not the smearing parameter. The differences in $u_z$, however, have a component related to the discretization error, but also a relevant component related to the smearing parameter, especially near the tip. This is due to the differences in the formulation of the ALM and LL, as discussed in Section~\ref{sec:res_wing}.

\bibliographystyle{jfm}
\bibliography{references}

\listofchanges

\end{document}